\documentclass[onecolumn]{svjour3}

\usepackage{amsmath}
\usepackage{amssymb}
\usepackage{graphicx}

\title {Quantum walks: a comprehensive review}
\dedication{To my dear friend Carlos Fuentes. In Memoriam.\\To my beloved daughter Renata, welcome to my life and to Planet Earth!}
\author {Salvador El\'{\i}as Venegas-Andraca}
\institute {Quantum Information Processing Group at Tecnol\'ogico de Monterrey Campus Estado de M\'exico and Texia, SA de CV\\
\email {sva@mindsofmexico.org, sva@texia.mx, salvador.venegas-andraca@keble.oxon.org}}
\date{}

\begin{document}

\maketitle

\begin{abstract}

Quantum walks, the quantum mechanical counterpart of classical random walks, is an advanced tool for building quantum algorithms that has been recently shown to constitute a universal model of quantum computation. Quantum walks is now a solid field of research of quantum computation full of exciting open problems for physicists, computer scientists and engineers.

In this paper we review theoretical advances on the foundations of both discrete- and continuous-time quantum walks, together with  the role that randomness plays in quantum walks, the connections between the mathematical models of coined discrete quantum walks and continuous quantum walks, the quantumness of quantum walks, a summary of papers published on discrete quantum walks and entanglement as well as a succinct review of experimental proposals and realizations of discrete-time quantum walks. Furthermore, we have reviewed several algorithms based on both discrete- and continuous-time quantum walks as well as a most important result: the computational universality of both continuous- and discrete-time quantum walks.

\keywords{quantum walks \and quantum algorithms \and quantum computing \and quantum and classical simulation of quantum systems}

\end{abstract}

\section{Introduction}

Computer science and computer engineering are disciplines that have transformed every aspect of modern society. In these fields, cutting-edge research is about new models of computation, new materials and techniques for building computer hardware, novel methods for speeding-up algorithms, and building bridges between computer science and several other scientific fields that allow scientists to both think of natural phenomena as computational procedures as well as to employ novel models of computation to simulate natural processes (e.g. \cite{mukanata07,robinett07,kong07,stadler07,reif07,bacon07,abdeldayem07,marr07,aono07}.) In particular, quantifying the resources required to process information and/or to compute a solution, i.e. to assess the complexity of a computational process, is a prioritized research area as it allows us to estimate implementation costs as well as to compare problems by comparing the complexity of their solutions. Among the mathematical tools employed in advanced algorithm development, classical random walks, a subset of stochastic processes (that is, processes whose evolution involves chance),  have proved to be a very powerful technique for the development of stochastic algorithms \cite{motwani95,schoning99}. In addition to the key role they play in algorithmics, classical random walks are ubiquitous in many areas of knowledge as physics, biology, finance theory, computer vision, and earthquake modelling  \cite{landau80,berg93,hull05,grady06,helmstetter02,ward04}, to name a few.

Theoretical computer science, in its canonical form, does not take into account the physical properties of those devices used for performing computational or information processing tasks. As this characteristic could be perceived as a drawback  because the behavior of any {\it physical} device used for computation or  information processing must ultimately be predicted by the laws of physics, several research approaches have therefore concentrated on thinking of computation in a physical context (e.g \cite{benioff80,benioff82a,benioff82b,benioff82c,feynman82,feynman86,feynman_lectures_computation,deutsch85,deutsch00,galindo02,margolus03}.) Among those physical theories that could be used for this purpose, quantum mechanics stands in first place.

Quantum computation can be defined as the interdisciplinary scientific field devoted to build quantum computers and quantum information processing systems, i.e. computers and information processing systems that use the quantum mechanical properties of Nature. Research on quantum computation heavily focuses on building and running algorithms which exploit the physical properties of quantum computers. Among the theoretical discoveries and promising conjectures that have positioned  quantum computation as a key element in modern science, we find: 

\begin{enumerate}

\item
The development of novel and powerful methods of computation that may allow us to significantly increase our processing power  for solving certain problems (e.g. \cite{nielsen00,kitaev99,aharonov07,briegel09}.)

\item
The increasing number of  quantum computing applications in several branches of science and technology (e.g. image processing and computational geometry \cite{sva03a,sva03b,sva10,lanzagorta10,le10,le11,sun11,le11b,iliyasu11,iliyasu12}, pattern recognition \cite{trugenberger01,trugenberger02a,trugenberger02b,horn02}, quantum games \cite{abal08_iqg}, and warfare \cite{lanzagorta11}.)

\item
 The simulation of complex physical systems and mathematical problems for which we know no classical digital computer algorithm that could efficiently simulate them \cite{feynman82,kassal08,perdomo08,kassal09,kassal11,shor97}. A detailed summary of scientific and technological applications of quantum computers can be found in \cite{roadmapusa04,qipc07}.

\end{enumerate}

Building good quantum algorithms is a difficult task as quantum mechanics is a counterintuitive theory and intuition plays a major role in algorithm design and, for a quantum algorithm to be good, it is not enough to perform the task it is intended to: it must also do better, i.e. be more efficient, than any classical algorithm (at least better than those classical algorithms known at the time of developing corresponding quantum algorithms.) Examples of successful results in quantum computation can be found in \cite{deutschjosza92,shor97,grover96,childs03,horn02,aspuru05,kassal08,perdomo08,kassal09,kassal11}. Good introductions and reviews of quantum algorithms can be found in \cite{kitaev99,gruska99,nielsen00,bouwmeester01,mermin03,imre05,kempe06,vedral06,wootters06,mermin07,santha08,lanzagorta09,mosca09,ambainis10_mfcs,childs10,bacon10,williams11,rieffel11}.

Quantum walks, the quantum mechanical counterpart of classical random walks, is an advanced tool for building quantum algorithms (e.g. \cite{shenvi02,ambainis03,ambainis04a,childs03,ambainis08_sofsem,mohseni08}) that has been recently shown to constitute a universal model of quantum computation \cite{childs09,lovett10,underwood10}. There are two kinds of quantum walks: discrete and continuous quantum  walks. The main difference between these two sets is the timing used to apply corresponding evolution operators. In the case of discrete quantum walks, the corresponding evolution operator of the system is applied only in discrete time steps, while in the continuous quantum walk case, the evolution operator can be applied at any time. 

Our approach in the development of this work has been to study those concepts of quantum mechanics and quantum computation  relevant to the computational aspects of quantum walks.  Thus, in the history of cross-fertilization between physics  and computation, this review is meant to be situated as a contribution within the field of quantum walks from the perspective of a computer scientist. In addition to this paper, the reader may also find the scientific documents written by Kempe \cite{kempe04},  Kendon \cite{kendon07}, Konno \cite{konno08}, Ambainis \cite{ambainis04,ambainis04a,ambainis08_sofsem,ambainis10_mfcs}, Santha \cite{santha08}, and Venegas-Andraca \cite{sva08} relevant to deepening into the mathematical, physical and algorithmic properties of quantum walks.

The following lines provide a summary of the main ideas and contributions of this review article.

Section \ref{quantum_walks_intro}. {\bf Fundamentals of Quantum Walks}. In this section I offer a comprehensive yet concise introduction to the main concepts and results of discrete and continuous quantum walks on a line and other graphs. This section starts with  a short and rigorous introduction to those properties of classical discrete random walks on undirected graphs relevant to algorithm development, including definitions for hitting time, mixing time and mixing rate, as well as mathematical expressions for hitting time on an unrestricted line and on a circle. I then introduce the basic components of a discrete-time quantum walk on a line, followed by a detailed analysis of the Hadamard quantum walk on an infinite line, using a method based on the Discrete Time Fourier Transform known as the Schr\"{o}dinger approach. This analysis includes the enunciation of relevant theorems, as well as the advantages of the Hadamard quantum walk on an infinite line with respect to its closest classical counterpart. In particular, I explore the context in which the properties of the Hadamard quantum walk on an infinite line are compared with classical random walks on an infinite line and with two reflecting barriers. Also, I briefly review another method for studying the Hadamard walk on an infinite line: path counting approach. I then proceed to study a quantum walk on an infinite line with an arbitrary  coin operator and explain why the study of the Hadamard quantum walk on an infinite line is enough as for the analysis of arbitrary quantum walks on an infinite line. Then, I present several results of quantum walks on a line with one and two absorbing barriers, followed by an analysis on the behavior of discrete-time coined quantum walks using many coins and a study of the effects of decoherence, a detailed review on limit theorems for discrete-time quantum walks, a subsection devoted to the recently founded subfield of localization on discrete-time quantum walks, and a summary of other relevant results. 

I then focus on the properties of discrete-time quantum walks on graphs: we study discrete-time quantum walks on a circle, on the hypercube and some general properties of this kind of quantum walks on Cayley graphs, including  a limit theorem of averaged probability distributions for quantum walks on graphs. I continue this section with a general introduction to continuous quantum walks together with several relevant results published in this field. Then, I present  an analysis of the role that randomness plays in quantum walks and the connections between the mathematical models of coined discrete quantum walks and continuous quantum walks. The last part of this section focuses on issues about the quantumness of quantum walks that includes a brief summary of reports on discrete quantum walks and entanglement, Finally, I briefly summarize several experimental proposals and realizations of discrete-time quantum walks.

Section \ref{qw_based_algorithms}. {\bf Algorithms based on quantum walks and classical simulation of quantum algorithms-quantum walks}. We review several links between computer science and quantum walks. We start by introducing the notions of oracle and hitting time, followed by a detailed analysis of  quantum algorithms developed to solve the following problems: searching in an unordered list and in a hypercube, the element ditinctness problem, and the triangle problem. I then provide an introduction to a seminal paper written by M. Szegedy in which a new definiton of quantum walks based on quantizing a stochastic matrix is proposed.  The second part of this section is devoted to analyzing continuous quantum walks. We start by reviewing the most successful quantum algorithm based on a continuous quantum walk known so far, which consists of traversing, in polynomial time, a family of graphs of trees with an exponential number of vertices (the same family of graphs would be traversed only in exponential time by any classical algorithm). We then briefly review a generalization of a continuous quantum walk, now allowed to perform non-unitary evolution, in order to simulate photosynthetic processes, and we finish by reviewing the state of the art on classical digital computer simulation of quantum algorithms and, particularly, quantum walks.

Section \ref{qw_computational_universality}. {\bf Universality of quantum walks}. I review in this last section a very recent and most important contribution in the field of quantum walks: computational universality of both continuous- and discrete-time quantum walks.

\section{\label{quantum_walks_intro}Fundamentals of Quantum Walks}

Quantum walks are quantum counterparts of classical random walks. Since classical random walks have been successfully adopted to develop classical algorithms and one of the main topics in quantum computation is the creation of quantum algorithms which are faster than their classical counterparts, there has been a huge interest in understanding the properties of quantum walks over the last few years. In addition to their usage in computer science, the study of quantum walks is relevant to building methods in order to test the \lq \lq quantumness'' of emerging technologies for the creation of quantum computers as well as to model natural phenomena.

Quantum walks is a relatively new research topic. Although some authors have selected the name \lq \lq quantum random walk'' to refer to quantum phenomena \cite{godoy92,gudder88} and, in fact, in a seminal work by R.P. Feynman about quantum mechanical computers \cite{feynman86} we find a proposal that could be interpreted as a continuous quantum walk \cite{chase08}, it is generally accepted that the first paper with quantum walks as its main topic was published in 1993 by Aharonov {\it et al} \cite{aharonov93}. Thus, the links between classical random walks and quantum walks as well as the utility of quantum walks in computer science, are two fresh and open areas of research (among scientific contributions on the links between classical and quantum walks,  Konno has proposed in \cite{konno09_correlated} solid mathematical connections between correlated random walks and quantum walks using the $PQRS$ matrix method introduced in \cite{konno02,konno02a}.) Two models of quantum walks have been suggested:
\\\\
- The first model, called {\bf discrete quantum walks}, consists of two quantum mechanical systems, named a walker and a coin, as well as an evolution operator which is applied to both systems only in discrete time steps. The mathematical structure of this model is evolution via unitary operator, i.e. $|\psi\rangle_{t_2} = {\hat U}|\psi\rangle_{t_1}$.
\\
- The second model, named {\bf continuous quantum walks}, consists of a walker and an evolution (Hamiltonian) operator of the system that can be applied  with no timing restrictions at all, i.e. the walker walks any time. The mathematical structure of this model is evolution via the Schr\"odinger equation.
\\

In both discrete and continuous models, the topology on which quantum walks have been performed and their properties computed are discrete graphs. This is mainly because graphs are widely used in computer science and building up quantum algorithms based on quantum walks has been a prioritized activity in this field.

The original idea behind the construction of quantum algorithms was to start by initializing a set of qubits and then to apply (one of more) evolution operators several times  {\it without making intermediate measurements}, as measurements were meant to be performed only at the end of the computational process (for example, see the quantum algorithms reported in \cite{bouwmeester01,nielsen00}.) Not surprisingly, the first quantum algorithms based on quantum walks were designed using the same strategy: initialize qubits, apply evolution operators and measure only to calculate the final outcome of the algorithm. Indeed, this method has proved itself very useful for building several remarkable algorithms (e.g. \cite{ambainis04a,kempe04}.) However, as the field has matured, it has been reported that performing (partial) measurements on a quantum walk may lead to interesting mathematical properties for algorithm development, like the \lq top hat'  probability distribution (e.g. \cite{maloyer07,kendon07}.) Moreover and expanding on the idea of using more sophisticated tools from the repertoire of quantum mechanics, recent reports have shown the effect of using weak measurements on the walker probability distribution of discrete quantum walks \cite{ghoshal11}.

The rest of this section is organized as follows. I begin with  a short introduction to those properties of classical discrete random walks on undirected graphs relevant to algorithm development, including definitions for hitting time, mixing time and mixing rate, as well as mathematical expressions for hitting time on an unrestricted line and on a circle. I then introduce the basic components of a discrete-time quantum walk on a line, followed by a detailed analysis of the Hadamard quantum walk on an infinite line, using a method based on the Discrete Time Fourier Transform known as the Schr\"{o}dinger approach. This analysis includes the enunciation of relevant theorems, as well as the advantages of the Hadamard quantum walk on an infinite line with respect to its closest classical counterpart. In particular, I explore the context in which the properties of the Hadamard quantum walk on an infinite line are compared with classical random walks on an infinite line and with two reflecting barriers. Also, I briefly review another method for studying the Hadamard walk on an infinite line: path counting approach. I then proceed to study a quantum walk on an infinite line with an arbitrary  coin operator and explain why the study of the Hadamard quantum walk on an infinite line is enough as for the analysis of arbitrary quantum walks on an infinite line. Then, I present several results of quantum walks on a line with one and two absorbing barriers, followed by an analysis on the behavior of discrete-time coined quantum walks using many coins and a study of the effects of decoherence, a detailed review on limit theorems for discrete-time quantum walks, a subsection devoted to the recently founded subfield of localization on discrete-time quantum walks, and a summary of other relevant results. 

In addition to this review paper, the reader may also find the scientific documents written by Kempe \cite{kempe04},  Kendon \cite{kendon07}, Konno \cite{konno08}, Ambainis \cite{ambainis04,ambainis04a,ambainis08_sofsem,ambainis10_mfcs}, Santha \cite{santha08},  and Venegas-Andraca \cite{sva08} relevant to deepening into the mathematical, physical and algorithmic properties of quantum walks.  Finally, readers who are not yet acquainted with the mathematical and/or physical foundations of quantum computation may find the following references useful:  \cite{gruska99,nielsen00,rieffel00,mermin03,imre05,mermin07,sva_dphil,lanzagorta09,rieffel11}.

\subsection{Classical random walk on an unrestricted line}

Classical discrete random walks were first thought as stochastic processes with no straightforward relation to algorithm development. Thus, in addition to references like \cite{polya21,spitzer64,coleman74,doyle84,grinstead97,norris99,woess00,rudnick04}  in which the mathematical foundations of  random walks can be found, references \cite{motwani95,lovasz96,lovasz98,rantanen04}  are highly recommendable for a deeper understanding of algorithm development based on classical random walks.

\begin{figure}
\begin{center}
\scalebox{0.3}{\includegraphics{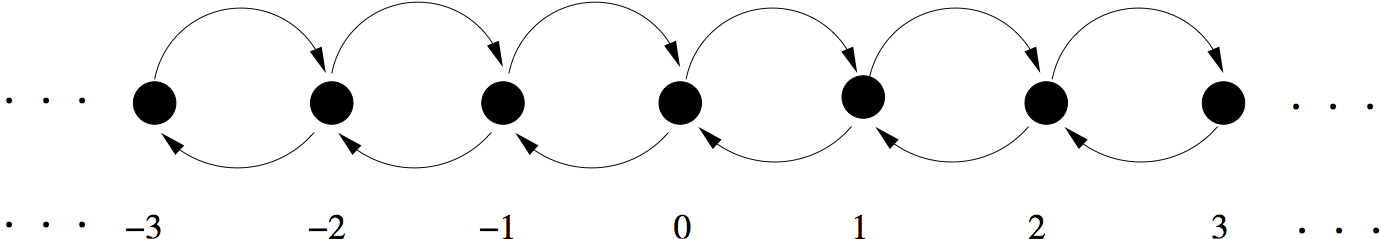}}
\end{center}
\caption{{\small An unrestricted classical discrete random walk on a line. The probability of going to the right is $p$ and the probability of going to the left is $q=1-p$.}
}\label{ucrw_line}
\end{figure}

A classical discrete random walk on a line is a particular kind of stochastic process. The simplest classical random walk on a line consists of a particle (\lq \lq the walker'') jumping to either left or right depending on the outcomes of a probability system (\lq \lq the coin'') with (at least) two mutually exclusive results, i.e. the particle moves according to a probability distribution (Fig. (\ref{ucrw_line}).) The generalization to discrete random walks on spaces of higher dimensions (graphs) is straightforward. An example of a discrete random walk on a graph is a particle moving on a lattice where each node has $6$ vertices, and the particle moves according to the outcomes produced by tossing a dice. Classical random walks on graphs can be seen as Markov chains (\cite{motwani95,norris99}.) 

Now, let $\{ Z_n \}$ be a stochastic process which consists of the path of a particle which moves along an axis with steps of one unit at time intervals also of one unit (Fig. (\ref{ucrw_line}).) At any step,  the particle has a probability $p$ of going to the right and $q=1-p$  of going to the left. Each step is modelled by a Bernoulli-distributed random variable \cite{coleman74,sva_dphil} and the probability of finding the particle  in position $k$ after $n$ steps and having as initial position $Z_0 = 0$ is given by the binomial distribution
$T_n = \sum_{k=1}^{n} Y_i = \frac{1}{2}(Z_n + n)$ $\Rightarrow$

\begin{equation}
\text{pr}(Z_n = k | Z_0 = 0)=
\begin{cases}
\binom{n}{{1 \over 2}(k+n)} p^{{1 \over 2}(k+n)}q^{{1 \over 2}(n-k)},&
\frac{1}{2}(k+n) \in \mathbb{N}\cup\{0\}; \\
0,& \text{otherwise}
\end{cases}
\label{unrestricted_classical_random_walk}
\end{equation}

\begin{figure}
\begin{center}
\scalebox{0.3}{\includegraphics{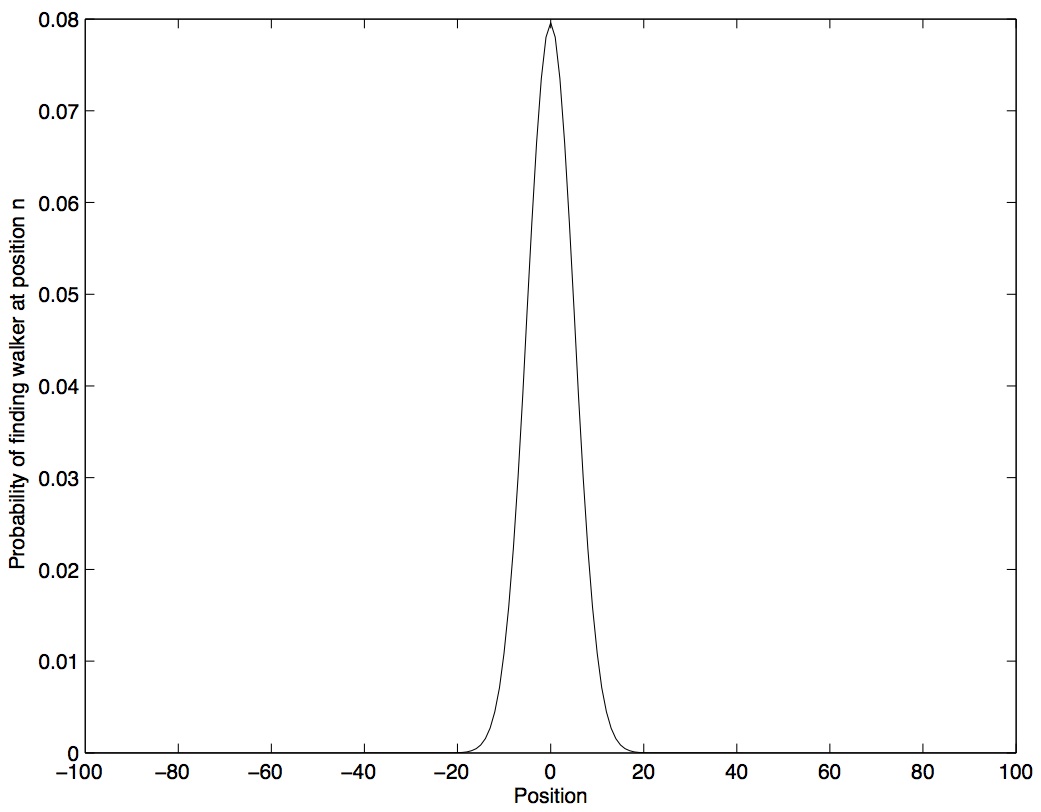}}
\end{center}
\caption{ {\small Plot of $P_{ok}^{(n)} = \binom{n}{\frac{1}{2}(k+n)}p^{\frac{1}{2}(k+n)}q^{\frac{1}{2}(n-k)}$
for $n=100$ and $p=\frac{1}{2}$. The probability of finding the walker in position $k=0$ is equal to 0.0795. Only probabilities corresponding to even positions are shown, as odd positions have probability equal to zero.}}
\label{binomial_25}
\end{figure}

Fig. (\ref{binomial_25}) shows a plot of Eq. (\ref{unrestricted_classical_random_walk}) with number of steps $n=100$ and $p = \frac{1}{2}$.

Since $T_n$ is Bin$(n,p)$ then the expected value is given by $E[T_n] = np$ and the variance is computed as $V[T_n] = npq$. Thus,

\begin{equation}
V[Z_n] = V[2T_n - n] = 4npq. \text{ In other words, } V[Z_n] = O(n)
\label{variance_unrestricted_rw_line}
\end{equation}

Eq. (\ref{variance_unrestricted_rw_line}) will be used in the following sections to show one of the earliest results on comparing classical random walks to quantum walks.

Graphs that encode the structure of a group are called {\bf Cayley graphs}. Cayley graphs are a vehicle for translating mathematical structures of scientific and engineering problems into forms amenable to algorithm development for scientific computing.

\begin{definition}{\bf Cayley graph}. Let $G$ be a finite group, and  let $S=\{s_1, s_2, \ldots , s_k \}$ be a generating set for G.  The Cayley graph of $G$ with respect to $S$ has a vertex for every element of $G$, with an edge from $g$ to $gs$ $\forall$  $g \in G$ and $s \in S$.
\label{cayley_graph}
\end{definition}

Cayley graphs are $k$-regular, that is, each vertex has degree $k$. Cayley graphs have more structure than arbitrary Markov graphs and their properties can be used for algorithm development  \cite{kempe_phd_01}.  Graphs and Markov chains can be put in an elegant framework which turns out to be very useful for the development of algorithmic applications:

Let $G=(V,E)$ be a connected, undirected graph with $|V|=n$ and $|E|=m$. $G$ induces a Markov chain $M_G$ if the states of $M_G$ are the vertices of $G$, and $\forall$
$u,v \in V$
\[ p_{uv} = \left\{ \begin{array}{ll}
        {1 \over d(u)} & \mbox{if $(u,v) \in E$};\\
        0 & \mbox{otherwise}.\end{array} \right. \]

where $d(u)$ is the degree of vertex $u$. Since $G$ is connected, then $M_G$ is irreducible and aperiodic \cite{motwani95}. Moreover, $M_G$ has a unique stationary distribution. 

\begin{theorem}
Let $G$ be a connected, undirected graph with $n$ nodes and
$m$ edges, and let $M_G$ be its corresponding Markov chain. 
Then, $M_G$ has a unique distribution 
$${\overrightarrow \pi} = (d(v_i)/2m)$$
for all components $v_i$ of ${\overrightarrow \pi}$.
\label{theorem_stationary_distribution_undirected_graph}
\end{theorem}

Note that Theorem \ref{theorem_stationary_distribution_undirected_graph} holds even when the distribution $\{ d(v_i) \}$ is not uniform.  In particular, the stationary distribution of  an undirected and connected graph with $n$ nodes,  $m$ edges and constant degree $d(v_i) = r$ $\forall$ $v_i \in G$, i.e. a Cayley graph, is ${\overrightarrow \pi} = (r/2m)$,  the uniform distribution.

We have established the relationship between Markov chains and graphs. We now proceed to define the concepts that make discrete random walks on graphs useful in computer science.  We shall begin by formally describing a random walk on a graph: let  $G$ be a graph. A random walk, starting from a vertex $u \in V$ is the random process defined by 
\\\\
s=u  \\
{\bf repeat} \\
\indent choose a neighbor $v$ of $u$ according to a certain probability distribution $P$\\
\indent u = v \\
{\bf until} (stop condition)

So, we start at a node $v_0$ and, if at $\text{t}^\text{th}$
step we are at a node $v_t$, we move to a neighbour of $v_t$ with probability
given by probability distribution $P$. It is common practice to make 
$P_{uv} = {1 \over d(v_t)}$, where $d(v_t)$ is the degree of vertex $v_t$.
Examples of discrete random walks on graphs are a classical random walk on a circle or on a 3-dimensional mesh.

We now introduce several measures to quantify the performance of discrete random walks
on graphs. These measures play an important role in the quantitative
theory of random walks, as well as in the application of this kind of Markov chains
in computer science.

\begin{definition}{\bf Hitting time}.
The hitting time $H_{ij}$ is the expected number of steps
before node $j$ is visited, starting from node $i$. 
\label{hitting_time_classical}
\end{definition}

\begin{definition}{\bf Mixing rate}. The mixing rate is a measure of how fast the discrete random walk converges to its limiting distribution. The mixing rate can be defined in many ways, depending on the type of graph we want to work with. We use the definition given in \cite{lovasz96}.
\\
If the graph is non-bipartite then $p_{ij}^t \rightarrow d_j/2m$ as $t \rightarrow \infty$, and the mixing rate is given by

$$
\mu = \lim_{t \rightarrow \infty} \sup \max \left|p_{ij}^{(t)} - {d_j \over 2m} \right | ^{1/ t}
$$
\end{definition}

As it is the case with the mixing rate, the {\bf mixing time} can be defined in several ways. Basically, the notion of  mixing time comprises the number of steps one must perform a classical discrete random walk before its distribution is close to its limiting distribution. 
\\

\begin{definition}{\bf Mixing time} {\it \cite{ambainis01}}. Let $M_G$ be an ergodic Markov chain which induces a probability distribution $P_u(t)$ on the states at time $t$. Also, let $\overrightarrow{\pi}$ denote the limiting distribution of $M_G$. The mixing time $\tau_\epsilon$ is then defined as

$$
\tau_\epsilon = \max_u \min_t \{ t | t \geq T \Rightarrow || P_u(t) - \overrightarrow{\pi} || < \epsilon \}
$$
\\
where $|| P_u(t) - \overrightarrow{\pi} ||$ is a standard distance measure. For example, we could use the total variation distance, defined as $|| P_u(t) - \overrightarrow{\pi} || = {1 \over 2}\sum_i |P_{u_i}(t) - \pi_i |$. Thus, the mixing time is defined as the first time $t$ such that $P_u(t)$ is within distance $\epsilon$ of $\overrightarrow{\pi}$ at all subsequent time steps $t \geq T$, irrespective of the initial state.
\label{mixing_time}
\end{definition}

Let us now provide two examples of hitting times on graphs.

\subsubsection{Hitting time of an unrestricted classical discrete random walk on a line}

It has been shown in Eq. (\ref{unrestricted_classical_random_walk}) that,
for an unrestricted classical discrete random walk on a line with $p=q={1 \over 2}$, the probability of
finding the walker in position $k$ after $n$ steps is given by

$$
\text{pr}(Z_n = k | Z_0 = 0)=
\begin{cases}
\binom{n}{{1 \over 2}(k+n)} {1 \over 2^n},&
{1 \over 2}(k+n) \in \mathbb{N}\cup\{0\}; \\
0,& \text{otherwise}
\end{cases}
$$

Using Stirling's approximation $n! \approx \sqrt{2\pi n} \left ( {n \over e} \right )^n $
and after some algebra, we find 

\begin{equation}
\text{pr}(Z_n = k | Z_0 = 0)=
{1 \over 2^n} \binom{n}{{1 \over 2}(k+n)} \approx
\sqrt{2n \over {\pi^2 (n^2 - k^2)} } \text{  } { n^n \over { (n+k)^{{n+k} \over 2}   (n-k)^{{n-k} \over 2} }   }
\label{binomial_approx_stirling}
\end{equation}

We know that Eq. (\ref{unrestricted_classical_random_walk}) is a binomial distribution, thus it makes sense to study the mixing time in two different vertex populations:  $k << n$ and $k \approx n$ (the first population is mainly contained under the bell-shape part of the distribution, while the second can be found along the tails of the distribution.)  In both cases, we shall find the expected hitting time by calculating the inverse of  Eq. (\ref{binomial_approx_stirling}) (i.e., the expected time of the geometric distribution): 

{\bf Case} ${\mathbf k} {\mathbf \ll} {\mathbf n}$. Since
$
\sqrt{2n \over {\pi^2 (n^2 - k^2)} } \text{  } { n^n \over { (n+k)^{{n+k} \over 2}   (n-k)^{{n-k} \over 2} }   }
\approx \sqrt{2n \over {\pi^2 n^2}}  \text{  } { n^n \over {n^{n/2} n^{n/2}} } = {c \over \sqrt{n}} \Rightarrow
$

\begin{equation}
\text{Hitting time  } H_{0,k}= O(\sqrt{n})
\label{hitting_time_k_smaller_n}
\end{equation}

{\bf Case} ${\mathbf k} {\mathbf \approx} {\mathbf n}$. Let $n-k=C_1$ and $n^2 - k^2 = C_2$, where $C_1$ and $C_2$ are small integer numbers. 
Since
$
\sqrt{2n \over {\pi^2 (n^2 - k^2)} } \text{  } { n^n \over { (n+k)^{{n+k} \over 2}   (n-k)^{{n-k} \over 2} }   }
\approx \sqrt{2n \over {\pi C_2}}  \text{  } { n^n \over { 2^n n^n C_1^{C_1/2}} } = 
{1 \over 2^n}\sqrt{{2n} \over {\pi C_1^{C_1} C_2}} \Rightarrow
$

\begin{equation}
\text{Hitting time  }  H_{0,k}= O(2^{n})
\label{hitting_time_k_close_n}
\end{equation}

Thus, the hitting time for a given vertex $k$ of an $n$-step unrestricted classical discrete random walk on a line
depends on which region vertex $k$ is located in. If $k<<n$ then it will take $\sqrt{n}$ steps to reach
$k$, in average. However, if $k \approx n$ then it will take an exponential number of steps to
reach $k$, as one would expect from the properties of the binomial distribution.

\subsubsection{Hitting time of a classical discrete random walk on a circle}

The definitions of discrete random walks on a circle and on a line with
two reflecting barriers are very similar. In fact, the only difference
is the behavior of the extreme nodes. 

\begin{figure}
\begin{center}
\scalebox{0.3}{\includegraphics{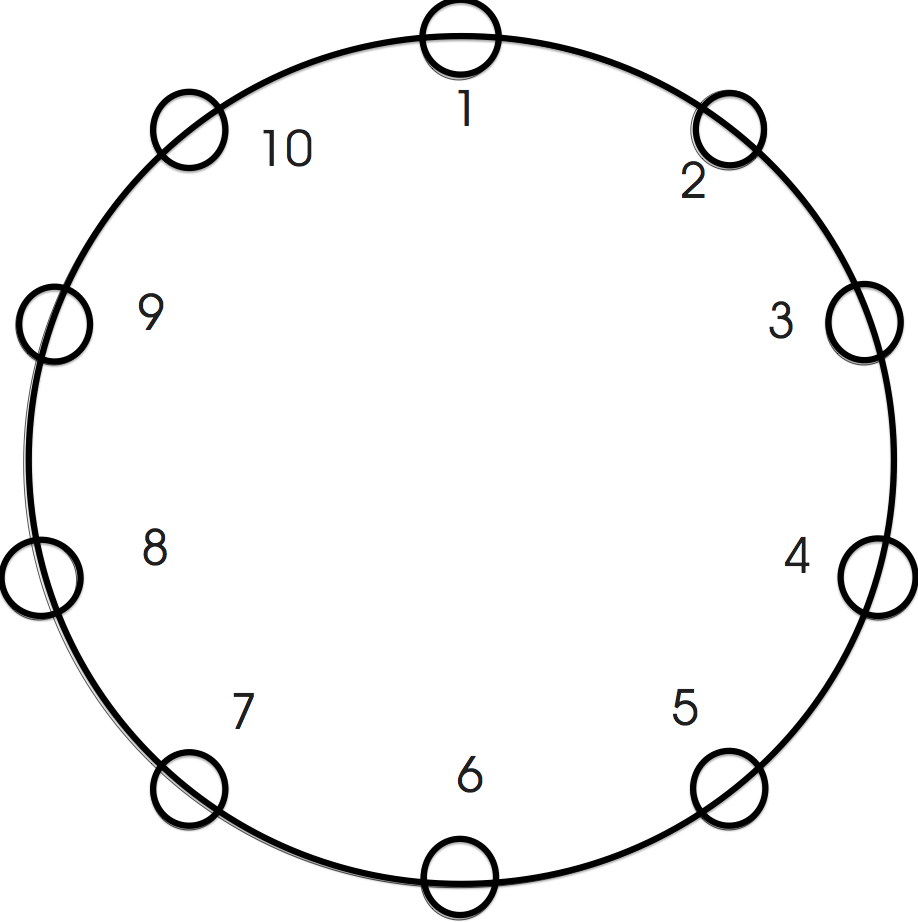}}
\end{center}
\caption{{\small Classical discrete random walk on a 10 nodes circle.}
}\label{circle}
\end{figure}

Let $\{ Z_n \}$ be a stochastic process which consists of the path of a particle which moves along a circle with steps of one unit at time intervals also of  one unit. The circle has $n$ different position sites (for an example with 10 nodes, see Fig. (\ref{circle})). At any step, the particle has a probability $p$ of going to the right and $q = 1-p$ of going to the left. If the particle is on $Z_t = 0$ at time $t$ then the particle will move to $Z_{t+1} = 1$ with probability $p$ and to $Z_{t+1} = n-1$ with probability $q$. Similarly, if the particle is on $Z_t = n-1$  at time $t$ then at time $t+1$ the particle will go to $Z_{t+1} = 0$ with probability $p$ and  to $Z_{t+1} = n-2$ with probability $q$.

According to Theorem \ref{theorem_stationary_distribution_undirected_graph}, the Markov chain defined by $\{ Z_n \}$ has a stationary distribution given by 

\begin{equation}
\overrightarrow{\pi} = {1 \over n}
\label{stationary_distribution_circle}
\end{equation}

And a hitting time $H_{0,n}$ given by (\cite{lovasz96})

\begin{equation}
H_{0,n} = O(n^2)
\label{hitting_time_circle}
\end{equation}

\subsection{Discrete quantum walk on a line}

Discrete quantum walks on a line (DQWL) is the most studied model of discrete quantum walks. As its name suggests, this kind of quantum walks are performed on graphs $G$ composed of a set of vertices $V$ and a set of edges $E$ (i.e., $G=(V,E)$), and having each vertex two edges, i.e. $|V|=2$. Studying DQWL is important in quantum computation for several reasons, including:
\\\\
1. DQWL can be used to build quantum walks on more sophisticated structures like circles or general graphs.
\\
2. DQWL is a simple model that can be exploited to explore, find and understand relevant properties of quantum walks for the development of quantum algorithms.
\\
3. DQWL can be employed to test the quantumness of experimental realizations of quantum computers.
\\

In \cite{meyer95}, Meyer made two contributions to the study of DQWL while working on models of Quantum Cellular Automata (QCA) and Quantum Lattice Gases:
\\
1. He proposed a model of quantum dynamics that would be used later on to analytically characterize DQWL.
\\
2. He showed that a quantum process in which, at each time step, a quantum particle (the walker) moves in superposition both to left and right with equal amplitudes, is physically impossible in general, the only exception being the trivial motion in a single direction.
\\
\\
In order to perform a discrete DQWL with non-trivial evolution, it was proposed in  \cite{ambainis01} and \cite{nayak00} to use an additional quantum system: a coin. Thus, a DQWL comprises two quantum systems, {\bf coin} and {\bf walker}, along with a unitary coin operator (\lq \lq to toss a coin'') and a conditional shift operator (to displace the walker to either left or right depending on the accompanying coin state component.)

In a different perspective, Patel {\it et al}  proposed in \cite{patel05}  to eliminate the use of coins by rearranging the Hamiltonian operator  associated with the evolution operator of the quantum walk (however,  there is a price to be paid on the translation invariance of the quantum walk.) Moreover, Hines and Stamp have proposed the development of quantum walk Hamiltonians \cite{hines07} in order to reflect the properties of potential experimental realizations of quantum walks in their mathematical structure.

Motivated by \cite{patel05}, Hamada {\it et al} \cite{hamada05} wrote a general  setting for QCA, developed a correspondence between DQWL and QCA, and applied this connection to show that the quantum walk proposed in \cite{patel05} could be  modelled as a QCA. The relationship between QCA and quantum walks has been indirectly explored by Meyer \cite{meyer95}. Additionally, Konno {\it et al} \cite{konno04a} have studied the relationship between quantum walks and cellular automata, Van Dam has shown \cite{dam96} that it is possible to build a quantum cellular automaton capable of universal computation, and Gross {\it et al} have introduced a comprehensive mathematical setting for developing index theory of one-dimensional automata and cellular automata \cite{gross11}.

We now review the mathematical structure of a basic coined DQWL.

\subsubsection{Structure of a  basic coined DQWL}

The main components of a coined DQWL are a walker, a coin, evolution operators for both walker and coin, and a set of observables:
\\
\noindent\emph{\textbf{Walker and Coin:}} The walker is a quantum system living in a Hilbert space of infinite but countable dimension ${\cal H}_p$. It is customary to use vectors from the canonical (computational) basis of ${\cal H}_p$ as \lq\lq position sites" for the walker. So, we denote the walker as $|\text{position}\rangle \in {\cal H}_p$ and affirm that the canonical basis states $|i\rangle_p$ that span ${\cal H}_p$, as well as any superposition of the form $\sum_{i} \alpha_i|i\rangle_p$ subject to $\sum_i|\alpha_i|^2 = 1$, are valid states for $|\text{position}\rangle$. The walker is usually initialized at the \lq origin', i.e. $|\text{position}\rangle_{\text{initial}} = |0\rangle_p$.

The coin is a quantum system living in a 2-dimensional Hilbert space ${\cal H}_c$. The coin may take the canonical basis states $|0\rangle$ and $|1\rangle$ as well as any superposition of these basis states. Therefore $|$coin$\rangle$ $\in {\cal H}_c$ and a general normalized state of the coin may be written as $|$coin$\rangle$ $= a |0\rangle_c + b |1\rangle_c$, where $|a|^2 + |b|^2 = 1$.

The total state of the quantum walk resides in ${\cal H}_t = {\cal H}_p \otimes {\cal H}_c$. It is customary to use product states of ${\cal H}_t$ as initial states, that is, $|\psi\rangle_{\text{initial}} = |\text{position}\rangle_{\text{initial}} \otimes |\text{coin}\rangle_{\text{initial}}$.
\\
\noindent\emph{\textbf{Evolution Operators:}}  The evolution of a quantum walk is divided into two parts that closely resemble the behavior of a classical random walk. In the classical case, chance plays a key role in the evolution of the system. In the quantum case, the equivalent of the previous process is to apply an evolution operator to the coin state followed by a conditional shift operator to the total quantum system. The purpose of the coin operator is to render the coin state in a superposition, and the randomness is introduced by performing a measurement on the system after both evolution operators have been applied to the total quantum system several times.

Among coin operators, customarily denoted by $\hat{C}$, the Hadamard operator has been extensively employed:

\begin{equation}
{\hat H} = {1 \over \sqrt{2}} (|0\rangle_{c} \langle 0| + |0
\rangle_{c} \langle 1| + |1\rangle_{c} \langle 0| - |1\rangle_{c}
\langle 1|)
\label{hadamard_single}
\end{equation}

For the conditional shift operator use is made of a unitary operator that allows the walker to go one step forward if the accompanying coin state is one of the two basis states (e.g. $|0\rangle$), or one step backwards if the accompanying coin state is the other basis state (e.g. $|1\rangle$). A suitable conditional shift operator has the form

\begin{equation}
{\hat S} = |0\rangle_{c} \langle 0| \otimes \sum_i |i+1 \rangle_{p} \langle i|
+ |1\rangle_{c} \langle 1| \otimes \sum_i |i-1 \rangle_{p} \langle i|.
\label{shift_single}
\end{equation}

Consequently, the operator on the total Hilbert space is ${\hat U} = {\hat S} \cdot ({\hat C} \otimes \mathbb{{\hat I}}_p)$ and a succinct mathematical representation of a discrete quantum walk after $t$ steps is

\begin{equation}
|\psi \rangle_t = ({\hat U})^t |\psi\rangle_\text{initial},
\label{succint_quantum_walk}
\end{equation}

where $|\psi\rangle_{\text{initial}} = |\text{position}
\rangle_{\text{initial}} \otimes |\text{coin}
\rangle_{\text{initial}}$.
\\
\noindent\emph{\textbf{Observables:}}  Several advantages of quantum walks over classical random walks are a consequence of interference effects between coin and walker after several applications of ${\hat U}$ (other advantages come from quantum entanglement between walker(s) and coin(s) as well as partial measurement and/or interaction of coins and walkers with the environment.) However, we must perform a measurement at some point in order to know the outcome of our walk. To do so, we define a set of observables according to the basis states that have been used to define coin and walker.

There are several ways to extract information from the composite quantum system. For example, we may first perform a measurement on the coin using the observable

\begin{equation}
{\hat M}_c = \alpha_0 |0\rangle_{c}\langle 0| + \alpha_1
|1\rangle_{c}\langle 1|.
\end{equation}

A measurement must then be performed on the position states of the
walker by using the operator

\begin{equation}
{\hat M}_p = \sum_i a_i |i\rangle_{p}\langle i|.
\end{equation}

\begin{figure}
\begin{center}
(a) \scalebox{0.5}{\includegraphics{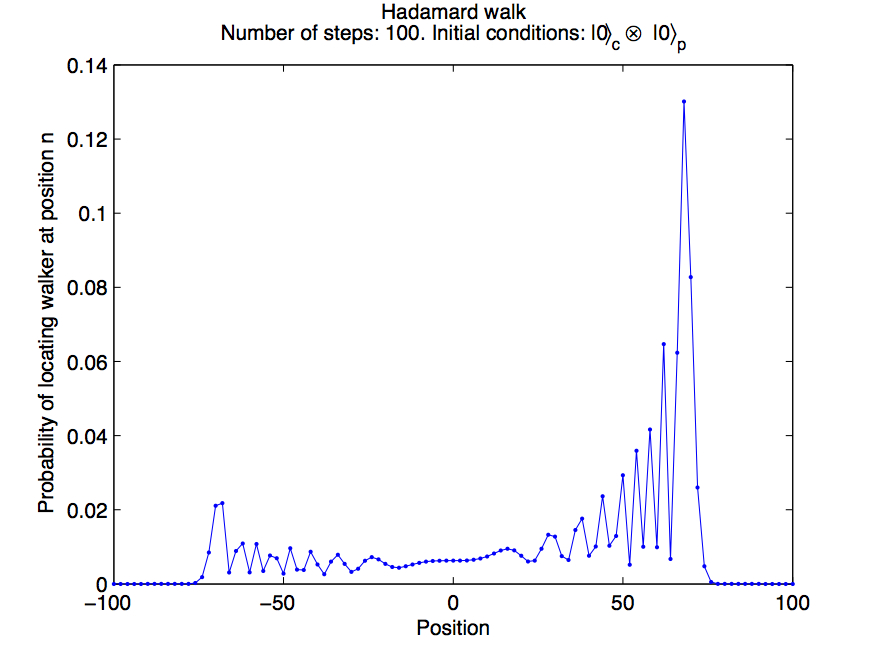}}\\
(b) \scalebox{0.5}{\includegraphics{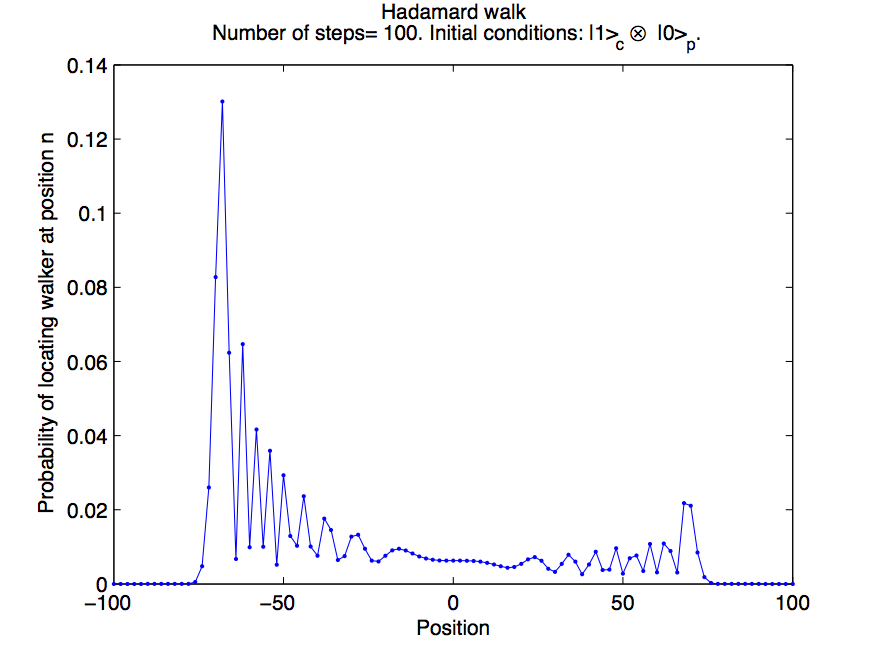}}
\end{center}
\caption{{\small Probability distributions of $100$ steps DQWLs using coin and shift operators given by Eqs. (\ref{hadamard_single}) and (\ref{shift_single}) respectively. Plot (a) corresponds to a DQWL with total initial quantum state $|0\rangle_c \otimes |0\rangle_p$, while plot (b) had total initial quantum state $|1\rangle_c \otimes |0\rangle_p$. Two interesting properties of these quantum walks is the skewness of corresponding probability distributions, along with the dependance of the symmetry of such skewness from the coin initial  state.}
}\label{hadamard_skewed}
\end{figure}

We show in Fig. (\ref{hadamard_skewed}) the probability distributions of two 100-steps DQWL. Coin and shift operators for both quantum walks are given by Eqs. (\ref{hadamard_single}) and (\ref{shift_single}) respectively. The DQWLs from plots (a) and (b) have corresponding initial quantum states $|0\rangle_c \otimes |0\rangle_p$ and $|1\rangle_c \otimes |0\rangle_p$. The first evident property of these quantum walks is the skewness of their probability distributions, as well as the dependance of the symmetry of such a skewness from the coin initial quantum state ($|0\rangle$ for plot (a) and $|1\rangle$ for plot (b).) This skewness comes from constructive and destructive interference due to the minus sign included in  Eq. (\ref{hadamard_single}). Also, we notice a quasi-uniform behavior in the central area of both probability distributions, approximately in the interval $[-70, 70]$. Finally, we notice that regardless their skewness, both probability distributions cover the same number of positions (in this case, even positions from -100 to 100. If the quantum walk had been performed an odd number of times, then only odd position sites could have non-zero probability.) 

Two approaches have been extensively used to study DQWL:
\\
{\bf 1. Schr\"{o}dinger approach}. In this case, we take an arbitrary component $|\psi \rangle_n = (\alpha |1\rangle_c + \beta |0\rangle_c) \otimes|n\rangle_p  \text{ }$ of the quantum walk, the tensor product of coin and position components for a certain walker position. $|\psi\rangle_n$ is then Fourier-transformed in order to get a closed form of the coin amplitudes. Then, standard tools of complex analysis are used to calculate the statistical properties of the probability distribution computed from corresponding coin amplitudes.
\\
{\bf 2. Combinatorial approach}.  In this method we compute the amplitude for a particular position component $|n\rangle_p$ by summing up the amplitudes of all the paths which begin in the given initial condition and end up in $|n\rangle_p \text{. }$ This approach can be seen as using a discrete version of path integrals.

In addition, Fuss {\it et al} have proposed an analytic description of probability densities and moments for the one-dimensional quantum walk on a line \cite{fuss07}, Bressler and Pemantle \cite{bressler07} as well as Zhang \cite{zhang08} have employed generating functions to asimptotically analize position probability distributions in one-dimensional quantum walks, and Feldman and Hillery \cite{feldman07} have proposed an alternative formulation of discrete quantum walks based on  scattering theory. In particular, \cite{feldman07} plays an increasingly important role on the foundations of the field of quantum walks for being an alternative formulation for discrete quantum walks as well as a key tool to describe and understand the proof of computational universality delivered by Childs in \cite{childs09}, this latter paper is to be reviewed in section \ref{qw_computational_universality}.

In the following lines we review both Schr\"{o}dinger and combinatorial approaches to analyze the Hadamard walk, a specific but very powerful DQWL with coin and shift operators given by Eqs. (\ref{hadamard_single}) and (\ref{shift_single}) respectively. Later on we show how the Hadamard walk is related to the more general case of a DQWL with arbitrary coin operator.

\subsubsection{Schr\"odinger approach for the Hadamard walk}

The analysis of DQWL properties using the Discrete Time Fourier Transform (DTFT) and methods from complex analysis was first made by Nayak and Vishwanath \cite{nayak00}, followed by Ambainis {\it et al} \cite{ambainis01}, Ko\v{s}\'{\i}k \cite{kosik03} and Carteret {\it et al} \cite{carteret05a,carteret05b}. Following \cite{ambainis01,nayak00}, a quantum walk on an infinite line after $t$ steps can be written as $|\psi \rangle_t = {\hat U}^t |\psi\rangle_\text{initial}$ (Eq. (\ref{succint_quantum_walk})) or, alternatively, as

\begin{equation}
\sum_k [a_k |0\rangle_c + b_k |1\rangle_c] |k\rangle_p
\end{equation}

where $|0\rangle_c$, $|1\rangle_c$ are the coin state components and $|k\rangle_p$ are the walker state components. For example, let us suppose we have

\begin{equation}
|\psi\rangle_0 = |0\rangle_c \otimes |0\rangle_p
\label{initial_state}
\end{equation}

as the quantum walk initial state, with Eq.(\ref{hadamard_single}) and Eq.(\ref{shift_single}) as coin and shift operators. Then, the first three steps of this quantum walk can be written as:

$$
|\psi\rangle_1 = {1 \over \sqrt{2}}|0\rangle_c |1\rangle_p + {1 \over
\sqrt{2}}|1\rangle_c |-1\rangle_p \text{ ,}
$$

$$
|\psi\rangle_2 = ({1 \over 2}|0\rangle_c + 0|1\rangle_c)           |2\rangle_p +
                ({1 \over 2}|0\rangle_c + {1 \over 2}|1\rangle_c) |0\rangle_p +
                (          0|0\rangle_c - {1 \over 2}|1\rangle_c)
|-2\rangle_p \text{ ,}
$$
and

\begin{eqnarray}
|\psi\rangle_3 = ( {1 \over 2\sqrt{2}}|0\rangle_c + 0
 |1\rangle_c )   |3\rangle_p +
                ( {1 \over  \sqrt{2}}|0\rangle_c + {1 \over
2\sqrt{2}}  |1\rangle_c )   |1\rangle_p + \nonumber \\
                ( {-1 \over  2\sqrt{2}}|0\rangle_c + 0
 |1\rangle_c )   |-1\rangle_p +
                ( 0                    |0\rangle_c + {1 \over
2\sqrt{2}}|1\rangle_c )   |-3\rangle_p  \text{ .}\nonumber
\end{eqnarray}

We now define

\begin{equation}
\Psi(n,t) = \left(\begin{matrix} \Psi_R (n,t) \cr \Psi_L (n,t) \cr
\end{matrix}\right)
\label{component_amplitude_position_n_time_t}
\end{equation}

as the two component vector of amplitudes of the particle being at point $n$ and time $t$ or, in operator notation

\begin{equation}
|\Psi(n,t) \rangle = \Psi_L(n,t) |1\rangle +  \Psi_R(n,t) |0\rangle
\label{initial_state_fourier}
\end{equation}

We shall now analyze the behavior of a Hadamard walk at point $n$ after $t+1$ steps. We begin by applying the Hadamard operator given by Eq. (\ref{hadamard_single}) to those coin state components in position $n-1$, $n$ and $n+1$:

\begin{eqnarray}
\lefteqn{\hat{H}(|\Psi(n-1,t) \rangle + |\Psi(n,t) \rangle +
|\Psi(n+1,t) \rangle) =} \nonumber \\
& & {1 \over \sqrt{2}}( |\Psi_L(n-1,t) \rangle |0\rangle +
|\Psi_R(n-1,t) \rangle |0\rangle
                      -|\Psi_L(n+1,t) \rangle |1\rangle +
|\Psi_R(n+1,t) \rangle |1\rangle  \nonumber \\
& &                    -|\Psi_L(n-1,t) \rangle |1\rangle +
|\Psi_R(n-1,t) \rangle |1\rangle +
                       |\Psi_L(n+1,t) \rangle |0\rangle +
|\Psi_R(n+1,t) \rangle |0\rangle   \nonumber \\
& &                    +|\Psi_L(n,t)   \rangle |0\rangle +
|\Psi_R(n,t)   \rangle |0\rangle
                      -|\Psi_L(n,t)   \rangle |1\rangle +
|\Psi_R(n,t)   \rangle |1\rangle)
\label{intermediate_state}
\end{eqnarray}

Now, we apply the shift operator given by Eq. (\ref{shift_single}) to Eq. (\ref{intermediate_state})

\begin{eqnarray}
   \lefteqn{\hat{U}(\hat{H}(|\Psi(n-1,t) \rangle + |\Psi(n,t) \rangle
+ |\Psi(n+1,t) \rangle)) = } \nonumber \\
& & {1 \over \sqrt{2}} (\mathbf{|\Psi_L(n,t) \rangle |0\rangle +
|\Psi_R(n,t) \rangle |0\rangle }
   \mathbf{                  -|\Psi_L(n,t) \rangle |1\rangle +
|\Psi_R(n,t) \rangle |1\rangle} \nonumber \\
& &                           -|\Psi_L(n-2,t) \rangle |1\rangle +
|\Psi_R(n-2,t) \rangle |1\rangle +
                              |\Psi_L(n+2,t) \rangle |0\rangle +
|\Psi_R(n+2,t) \rangle |0\rangle \nonumber \\
& &                           -|\Psi_L(n-1,t) \rangle |1\rangle +
|\Psi_R(n-1,t) \rangle |1\rangle +
                              |\Psi_L(n+1,t) \rangle |0\rangle +
|\Psi_R(n+1,t) \rangle |0\rangle ) \nonumber \\
\label{final_state}
\end{eqnarray}

The bold font amplitude components of Eq. (\ref{final_state}) are the amplitude components of $|\Psi(n,t+1)\rangle$, which can be written in matrix notation as

\begin{equation}
\Psi(n,t+1) = \left(\begin{matrix} {-1 \over \sqrt{2}} & {1 \over
\sqrt{2}} \cr  0 & 0 \cr \end{matrix}\right) \Psi(n+1,t)
+ \left(\begin{matrix} 0 & 0 \cr {1 \over \sqrt{2}} & {1 \over
\sqrt{2}} \cr \end{matrix}\right) \Psi(n-1,t)
\label{component_amplitude_position_n+1_time_t}
\end{equation}

Let us label

$$
M_{\_} = \left(\begin{matrix} {-1 \over \sqrt{2}} & {1 \over \sqrt{2}}
\cr  0 & 0 \cr \end{matrix}\right) \text{ and }
M_{+} = \left(\begin{matrix} 0 & 0 \cr {1 \over \sqrt{2}} & {1 \over
\sqrt{2}} \cr \end{matrix}\right)
$$

Thus

\begin{equation}
\Psi(n,t+1) = M_{\_} \Psi(n+1,t) + M_{+} \Psi(n-1,t)
\label{difference_recurrence}
\end{equation}

Eq. (\ref{difference_recurrence}) is a difference equation with $\Psi(0,0) = \left(\begin{matrix} 1 \cr 0 \cr \end{matrix}\right)$ and $\Psi(n,0) = \left(\begin{matrix} 0 \cr 0 \cr \end{matrix}\right)$, $\forall \text{ }n \neq 0$ as initial conditions (Eq. (\ref{initial_state}).)

Our objective is to find analytical expressions for $\Psi_L(n,t)$ and $\Psi_R(n,t)$. To do so, we compute the Discrete Time Fourier transform of Eq. (\ref{difference_recurrence}). The Discrete Time Fourier Transform is given by

\begin{definition}{\bf Discrete Time Fourier Transform}. 
Let $f: \mathbb{Z} \rightarrow \mathbb{C}$ be a complex function over the integers $\Rightarrow$ its Discrete Time Fourier Transform (DTFT) $\tilde{f}: [-\pi,\pi] \rightarrow \mathbb{C}$ is given by

$$
\tilde{f} = \tilde{f}(e^{i\omega}) = \sum_{n=-\infty}^{\infty} f(n) e^{-in\omega} \text{ ,}
$$
and its inverse is given by
$$
f(n) = {1 \over 2\pi }\int_{-\pi}^{\pi} F(e^{i\omega}) e^{in\omega} d\omega
$$
\label{dtft}
\end{definition}

Ambainis {\it et al} \cite{ambainis01} employ the following slight variant of the DTFT:

\begin{equation}
\tilde{f}(k) = \sum_n f(n) e^{ik} \text{ ,}
\label{ambainis_dtft}
\end{equation}
where $f: \mathbb{Z} \rightarrow \mathbb{C}$ and $\tilde{f}:
[-\pi,\pi] \rightarrow \mathbb{C}$.
Corresponding inverse DTFT is given by

\begin{equation}
f(n) = {1 \over 2\pi} \int_{-\pi}^{\pi} \tilde{f}(k) e^{-ik} dk
\label{inverse_ambainis_dtft}
\end{equation}

So, using Eq. (\ref{ambainis_dtft}) we have

\begin{equation}
\tilde{\Psi}(k,t) = \sum_n \Psi(n,t) e^{ikn}
\end{equation}
\\

Using Eq. (\ref{difference_recurrence}) we obtain

\begin{equation}
\tilde{\Psi}(k,t+1) = \sum_n (M_{\_} \Psi(n+1,t) + M_{+} \Psi(n-1,t)) e^{ikn}
\end{equation}

After some algebra we get

\begin{equation}
\tilde{\Psi}(k,t+1) = M_k \tilde{\Psi}(k,t) \text{, where }
M_k = e^{-ik} M_{\_} + e^{ik} M_{+} = {1 \over \sqrt{2}}
\left(\begin{matrix} -e^{-ik}  & e^{-ik} \cr e^{ik} & e^{ik} \cr
\end{matrix}\right)
\end{equation}

Thus

\begin{equation}
\tilde{\Psi}(k,t) = \left(\begin{matrix} \tilde{\Psi}_L(k,t) \cr
\tilde{\Psi}_R(k,t) \cr \end{matrix}\right)
= M_k^t \tilde{\Psi}(k,0) \text{ , where }
\tilde{\Psi}(k,0) = \left(\begin{matrix} 1 \cr 0\cr \end{matrix}\right)
\end{equation}

Our problem now consists in diagonalizing the (unitary) matrix $M_k$ in order to calculate $M_k^t$. If $M_k$ has eigenvalues $\{\lambda_k^1, \lambda_k^2  \}$ and eigenvectors $|\Phi_k^1 \rangle, |\Phi_k^2 \rangle$ then

\begin{equation}
M_k= \lambda_k^1 |\Phi_k^1 \rangle \langle  \Phi_k^1| + \lambda_k^2
|\Phi_k^2 \rangle \langle \Phi_k^2|
\label{matrix_mk}
\end{equation}

Using the mathematical properties of linear operators, we then find:

\begin{equation}
M_k^t= (\lambda_k^1)^t |\Phi_k^1 \rangle \langle  \Phi_k^1| +
(\lambda_k^2)^t |\Phi_k^2 \rangle \langle \Phi_k^2|
\label{matrix_mkt}
\end{equation}

It is shown in \cite{nayak00} and \cite{ambainis01} that

\begin{equation}
\lambda_k^1 = e^{i\omega_k} \text{, } \lambda_k^2 = e^{i(\pi -
\omega_k)} \text{, where }
\omega_k \in [-{\pi \over 2}, {\pi \over 2}] \text{ and }
\sin(\omega_k) = {{\sin k} \over \sqrt{2}}
\label{eigenvalues_dqwl}
\end{equation}

and

\begin{subequations}
\begin{equation}
\Phi_k^1 = { 1 \over \sqrt{2[(1+ \cos^2(k)) + \cos(k)\sqrt{1+\cos^2 k}]} }
\left(\begin{matrix} e^{-ik} \cr \sqrt{2}e^{i\omega_k} + e^{-ik} \cr
\end{matrix}\right)
\label{first_eigenvector_dqwl}
\end{equation}
\begin{equation}
\Phi_k^2 = { 1 \over \sqrt{2[(1+ \cos^2(\pi - k)) + \cos(\pi -
k)\sqrt{1+\cos^2 (\pi - k)}]} }
\left(\begin{matrix} e^{-ik} \cr -\sqrt{2}e^{-i\omega_k} + e^{-ik} \cr
\end{matrix}\right)
\label{second_eigenvector_dqwl}
\end{equation}
\end{subequations}

From Eqs. (\ref{eigenvalues_dqwl}), (\ref{first_eigenvector_dqwl}) and (\ref{second_eigenvector_dqwl}) we compute the Fourier-transformed amplitudes $\tilde{\Psi}_L(n,t)$ and $\tilde{\Psi}_R(n,t)$

\begin{subequations}
\begin{equation}
\tilde{\Psi}_L(n,t) = {e^{-ik} \over {2\sqrt{1 + \cos^2 k}}}
(e^{i\omega_k t} - (-1)^t e^{-i\omega_k t})
\label{left_amplitude_dwql}
\end{equation}
\begin{equation}
\tilde{\Psi}_R(n,t) = {1 \over 2} (1 + {{\cos k} \over \sqrt{1+\cos^2
k}}) e^{i\omega_k t}
+ {(-1)^t \over 2} (1 - {{\cos k} \over \sqrt{1+\cos^2 k}}) e^{-i\omega_k t}
\label{right_amplitude_dwql}
\end{equation}
\end{subequations}

Using  Eq. (\ref{dtft}) on Eqs. (\ref{left_amplitude_dwql}) and (\ref{right_amplitude_dwql}), it is possible to prove the following theorem:
\\

\begin{theorem}
Let $|\Psi\rangle_0 = |0\rangle_p \otimes |0\rangle_c$ be the initial state of a discrete quantum walk on an infinite line with coin and shift operators given by Eqs. (\ref{hadamard_single}) and (\ref{shift_single}) respectively $\Rightarrow$

$$
\Psi_L(n,t) = {1 \over 2\pi} \int_{-\pi}^{\pi} {-ie^{ik} \over
{2\sqrt{1 + \cos^2 k}}} (e^{-i(\omega_k t - kn)}) dk
$$

$$
\Psi_R(n,t) = {1 \over 2\pi} \int_{-\pi}^{\pi} (1 + {{\cos k} \over
\sqrt{1+\cos^2 k}})  (e^{-i(\omega_k t - kn)}) dk
$$

where $\omega_k = \sin^{-1} ({\sin k \over \sqrt{2}})$ and $\omega_k
\in [{-\pi \over 2}, {\pi \over 2}]$.
\label{amplitudes_dqwl}
\end{theorem}

The amplitudes for even $n$ (odd $n$) at odd $t$ (even $t$) are zero, as it can be inferred from the definition of the quantum walk. Now we have an analytical expression for $\Psi_L(n,t)$ and $\Psi_R(n,t)$, and taking into account that $P(n,t) = |\Psi_L(n,t)|^2 + |\Psi_R(n,t)|^2$, we are interested in studying the asymptotical behavior of $\Psi(n,t)$ and $P(n,t)$. Integrals in Theorem \ref{amplitudes_dqwl} are of the form

$$
I(\alpha,t) = {1 \over {2\pi}}\int_{-\pi}^{\pi} g(k)e^{i\phi(k,\alpha)t} dk
\text{ , where } \alpha = n/t (\text{ }= \text{position} /
\text{number of steps})
$$

The asymptotical properties of this kind of integral can be studied using the method of stationary phase (\cite{bender78} and \cite{bleistein75}), a standard method in complex analysis. Using such a method, the authors of \cite{ambainis01} and \cite{nayak00} reported the following theorems and conclusions:

\begin{theorem}
Let $\epsilon > 0$ be any constant, and $\alpha$ be in the interval
$({{-1} \over \sqrt{2}} + \epsilon, {{{1} \over \sqrt{2}} - \epsilon})$.
Then, as $t \rightarrow \infty$, we have (uniformly in $n$)

$$
p_L(n,t) \backsim {2 \over {\pi\sqrt{1-2\alpha^2 t}}} \cos^2(-\omega t
+ {\pi \over 4} - \rho) \text{ ,}
$$

$$
p_R(n,t) \backsim {{2(1+\alpha) } \over {\pi (1-\alpha )
\sqrt{1-2\alpha^2 t}}} \cos^2(-\omega t + {\pi \over 4})
$$

where $\omega = \alpha \rho + \theta$, $\rho = \arg(-B + \sqrt{\Delta})$,
$\theta = \arg (B + 2 + \sqrt{\Delta})$, $B = { {2 \alpha} \over {1-\alpha}}$
and $\Delta = B^2 - 4(B+1)$.
\label{theorem_probabilities_dqwl}
\end{theorem}

\begin{theorem}
Let $n = \alpha t \rightarrow \infty$ with $\alpha$ fixed. In case
$\alpha \in (-1, -1/\sqrt{2}) \cup (1/\sqrt{2}, 1)$ $\Rightarrow$
$\exists \text{ } c>1$
for which $p_L(n,t) = O(c^{-n})$ and $p_R(n,t) = O(c^{-n})$.
\label{asymptotics_amplitudes}
\end{theorem}

{\bf Conclusions}
\\
1. {\bf Quasi-uniform behavior and standard deviation}. The wave function $\Psi_L(n,t)$ and $\Psi_R(n,t)$ (Theorem \ref{amplitudes_dqwl}) is almost uniformily spread over the region for which $\alpha$ is in the interval $[-1 / \sqrt{2}, 1 / \sqrt{2}]$ (Theorem \ref{theorem_probabilities_dqwl}), and shrinks quickly outside this region (Theorem \ref{asymptotics_amplitudes}). Furthermore, by integrating the probability functions from Theorem \ref{theorem_probabilities_dqwl}, it is possible to see that almost all of the probability is concentrated in the interval $[(-1/\sqrt{2}+\epsilon)t, (1/ \sqrt{2}-\epsilon)t]$. In fact, the exact probability value in that interval is $P= 1 - {{2\epsilon}\over \pi} - {O(1) \over t}$.

Furthermore, the position probability distribution spreads as a function of $t$, i.e.  $[-t / \sqrt{2}, t / \sqrt{2}]$, hence an evidence of 

\begin{equation}
\sigma_{\hat H} = O(t)
\end{equation}

Konno \cite{konno02} as well as Kendon and Tregenna \cite{kendon03}   have computed the actual variance of the probability distribution given in Theorem \ref{theorem_probabilities_dqwl}. Furthermore, by introducing a novel method to compute the probability distribution $X$ of the unrestricted DQWL, it was shown in \cite{konno02} that ${\sigma(X) \over t} \rightarrow \sqrt{{\sqrt{2}-1 \over 2}}$ as $t \rightarrow \infty$. In any case, the standard deviation of the unrestricted Hadamard DQWL is $O(t)$ and that result is in contrast with the standard deviation of an unrestricted classical random walk on a line, which is $O(\sqrt{t})$ ({\it cf.} Eq. (\ref{variance_unrestricted_rw_line}).)
\\  
2. {\bf Mixing time}.  It was shown in \cite{ambainis01} and \cite{nayak00} that an unrestricted Hadamard DQWL has a linear mixing time $\tau_\epsilon^{(q)} = O(t)$, where $t$ is the number of steps. Furthermore, $\tau_\epsilon^{(q)}$ was compared with the corresponding mixing time of a classical random walk on a line, which is quadratic, i.e. $\tau_\epsilon^{(c)} = O(t^2)$.

In order to properly bound and evaluate the impact of this result in the fields of quantum walks and quantum computation, a few clarifications are needed.

a) The mixing time measure used in this case is not the same as Eq. (\ref{mixing_time}), the reason being that {\it unitary} Markov chains in {\bf finite} state space (such as finite graph analogues of quantum walks) have no stationary distribution (section 2 of \cite{ambainis01}.) Instead, the mixing time measure proposed is given by

\begin{definition}{\bf Instantaneous Mixing Time.}
$\tau_\epsilon = \max_u \min_t \{  t | \text{  } ||P_u(t) - \pi||
\leq \epsilon\}$
\label{instantaneous_mixing_time}
\end{definition}

which is a more relaxed definition in the sense that it measures the first time that the current probability distribution $P_u(t)$ is $\epsilon$-close to the stationary distribution, {\it without the requirement of continuing being $\epsilon$-close for all future steps}.

b) The stationary distribution of an unrestricted classical random walk on a line is the binomial distribution, spread all over $\mathbb{Z}$. The only difference between $P_t$, the probability distribution of an unrestricted classical random walk on a line at step $t$, and its limiting distribution $P$ is the numerical value of the probability assigned to each node, as the shape of the distribution is the same. Although the binomial distribution can be {\it roughly} approximated by a uniform distribution for large values of $t$, depending on the precision we need for a certain task, that comparison is not accurate: as shown in our previous subsection on classical random walks, the hitting time of an unrestricted classical random walk on a line depends on the region we are looking into. Specifically, the hitting time is $O(\sqrt{t})$ for $k \ll t$ and $O(2^t)$ for $k \approx t$ (Eqs. (\ref{hitting_time_k_smaller_n}) and (\ref{hitting_time_k_close_n}).) Thus, to hit node $k$ with equal probabilities $P_{t_k} = P_k$ may depend on the region where $k$ is located. For example, it may take $O(\sqrt{t})$ if $k \ll t$ and $O(2^t)$ if $k \approx t$.

So, comparing mixing and hitting times for quantum and classical unrestricted walks on a line is not necessarily clear and straightforward. In order to reduce complexity in the analysis of algorithms, the infiniteness property of unrestricted classical random walks can sometimes be relaxed and properties of classical random walks on finite lines could be used instead, as proposed by  Rantanen in \cite{rantanen04}.

\subsubsection{Discrete Path Integral Analysis of the Hadamard Walk}

A different proposal to study the properties of quantum walks, based on combinatorics and the method to quantify quantum state amplitudes given by Meyer in \cite{meyer95}, has been delivered by Ambainis {\it et al} in \cite{ambainis01} as well as Carteret {\it et al} in \cite{carteret05a,carteret05b}.) The main idea behind this approach is to count the number of paths that take a quantum walker from point $a$ to point $b$. Thus, this approach can also be seen as a discrete path-integral method. Let us begin by stating the following lemma:

\begin{lemma} \cite{ambainis01} and \cite{meyer95}. Let $t \in [-n,n) \cap  \mathbb{Z}$ and $l = {{t-n} \over 2}$. The amplitudes of position $n$ after $t$ steps of the Hadamard walk are:

\begin{subequations}
\begin{equation}
\psi_L(n,t) = {1 \over \sqrt{2^t}} \sum_k \binom{l-1}{k}
\binom{t-l}{k} (-1)^{l-k-1}
\end{equation}
\begin{equation}
\psi_R(n,t) = {1 \over \sqrt{2^t}} \sum_k \binom{l-1}{k-1}
\binom{t-l}{k} (-1)^{l-k}
\end{equation}
\end{subequations}
\label{amplitudes_path_integral}
\end{lemma}

It was shown in \cite{ambainis01} that the probabilities computed from those amplitudes of Lemma (\ref{amplitudes_path_integral}) can be expressed using Jacobi polynomials. Furthermore, it was shown in \cite{carteret05b} that both Schr\"{o}dinger and combinatorial approaches are equivalent.

\begin{theorem}
Let $n \in \mathbb{N} \cup \{0\}$ and $J_\nu^{(a,b)} (z)$
be the normalised degree $\nu$ Jacobi polynomial with $J_\nu^{(a,b)}$ as its constant term. Let us
also define $\nu = {(t-n) \over 2} - 1$. Then

\begin{subequations}
\begin{equation}
P_l(n,t) = 2^{-n-2} (J_\nu^{(0,n+1)})^2
\end{equation}
\begin{equation}
P_R(n,t) = \left (  {{t+n}  \over {t-n} }\right )^2 2^{-n-2} (J_\nu^{(1,n)})^2, 
\end{equation}
\end{subequations}
$$ \text{with } p_L(-n,t) = p_L(n-2,t) \text{ and } p_R(-n,t) = \left (  {{t-n}
\over {t+n} }\right )^2 p_R(n,t)$$

\label{probabilities_path_integral}
\end{theorem}

A slight variation of this approach has been given by Brun {\it et al} in \cite{brunetal103}. An alternative and powerful method for building quantum walks, based on combinatorics and decompositions of unitary matrices, has been proposed by Konno in \cite{konno02,konno04b,konno05b,konno05a}. Also, Katori {\it et al} proposed in \cite{katori05} to apply Group Theory to analyze symmetry properties of quantum walks on a line and, along the same line of thought, Chandrashekar {\it et al} have proposed a generalized version of  the discrete quantum walk with coins living in {\it SU(2)} \cite{chandrashekar07}.

\subsubsection{Unrestricted DQWL with a general coin}

The study of the Hadamard walk is relevant to the field of quantum walks not only as an example but also because of the fact that some important properties shown by the Hadamard walk (for example, its standard deviation and mixing time) are shared by any quantum walk on the line. In \cite{tregenna03}, Tregenna {\it et al} showed that, for a general unbiased initial coin state

\begin{equation}
|\psi(x,0)\rangle = \sqrt{\eta} (|0 \rangle_c + e^{i\alpha}
\sqrt{1-\eta}|1\rangle_c) \otimes |0\rangle_p
\label{general_coin_initial_state}
\end{equation}
and a single step (in Fourier space) of the quantum walk
$$
|\tilde{\psi}(k,t+1)\rangle = \tilde{C}_k |\tilde{\psi}(k,t)\rangle
$$
where

\begin{equation}
\tilde{C}_k  = \left( \begin{array}{cc}
\sqrt{\rho}e^{ik}          & \sqrt{1-\rho}e^{i(\theta+k)} \\
\sqrt{1-\rho}e^{i(-k+\phi)}  & -\sqrt{\rho}e^{i(-k+\theta+\phi)}
\end{array} \right)
\label{fourier_general_coin_operator}
\end{equation}
is the Fourier transformed version of the most general 2-dimensional
coin operator

$$\mathbf{C_2} = \left( \begin{array}{cc}
\sqrt{\rho}               & \sqrt{1-\rho}e^{i\theta} \\
\sqrt{1-\rho}e^{i\phi}  & -\sqrt{\rho}e^{i(\theta+\phi)}
\end{array} \right)
$$
with $\theta,\phi \in [0,\pi]$ and $\rho \in [0,1]$,
we can express a $t$-step quantum walk on a line as

\begin{equation}
|\tilde{\psi}(k,t+1)\rangle = \tilde{C}_k^t |\tilde{\psi}(k,0)\rangle
\text{, where } 
|\tilde{\psi}(k,0)\rangle = \left( \begin{array}{c}             
\sqrt{\eta}    \\
e^{i\alpha}\sqrt{1-\eta} 
\end{array} \right) \otimes |k\rangle
\label{quantum_walk_general}
\end{equation}

If $\tilde{C}_k$ is expressed in terms of its eigenvalues $\lambda_k^{\pm}$ 
and eigenvectors $|\lambda_k^{\pm}\rangle$ 
then
$\tilde{C}_k^t = (\lambda^+_k)^t | \lambda^+_k \rangle \langle \lambda^+_k |
             + (\lambda^-_k)^t | \lambda^-_k \rangle \langle \lambda^-_k |$,
and Eq. (\ref{quantum_walk_general}) can be written as

\begin{equation}
|\tilde{\psi}(k,t+1)\rangle =
 (\lambda^+_k)^t | \lambda^+_k \rangle \langle \lambda^+_k |
\tilde{\psi}(k,0)\rangle
+ (\lambda^-_k)^t | \lambda^-_k \rangle \langle \lambda^-_k |
\tilde{\psi}(k,0)\rangle
\end{equation}
with

\begin{equation}
(\lambda^\pm_k)^t \langle \lambda^\pm_k | \tilde{\psi}(k,0)\rangle =
{(\lambda^\pm_k)^t \over n_k^\pm} e^{-ik} \left [  \sqrt{\eta} -
\sqrt{ {1-\eta} \over {1-\rho} }e^{i(\theta + \alpha)}
 (\sqrt{\rho} \mp e^{i(k-\delta)} e^{\mp i\omega_k}) \right ]\text{ , }
\label{general_quantum_walk}
\end{equation}
where  
$\delta = (\theta + \phi) /2$, $\sin(\omega_k) = \sqrt{\rho} \sin(k-\delta)$,
$\lambda_k^\pm = \pm e^{i\delta}e^{\pm i\omega_k}$, 
$n_k = \sqrt{ { {2[1 \mp \sqrt{\rho} \cos(k-\delta \mp \omega_k)]}
\over {1-\rho} } } $,
 $\lambda^{\pm}=\pm e^{i\delta}e^{\pm i\omega_k}$ and
$|\lambda^{\pm}\rangle = {1 \over n_k^{\pm}} 
\left( \begin{array}{c}
e^{ik}    \\
e^{i\theta}(\lambda^{\pm} - \sqrt{\rho} e^{ik})/\sqrt{1-\rho} 
\end{array} \right)$.

\begin{figure}
\begin{center}
(a)\scalebox{0.5}{\includegraphics{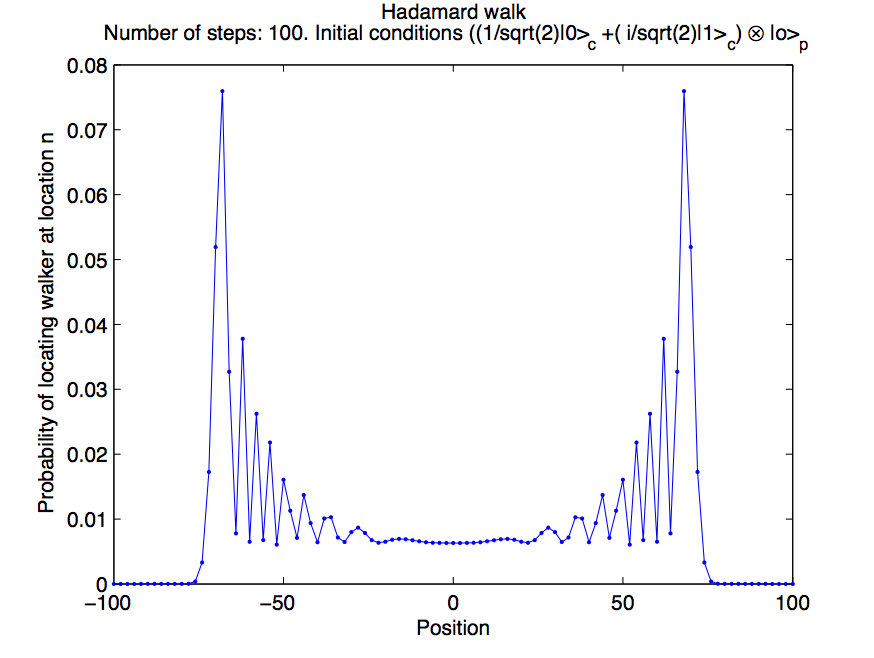}}
(b)\scalebox{0.5}{\includegraphics{hadamard_walk_00.jpg}}
(c)\scalebox{0.5}{\includegraphics{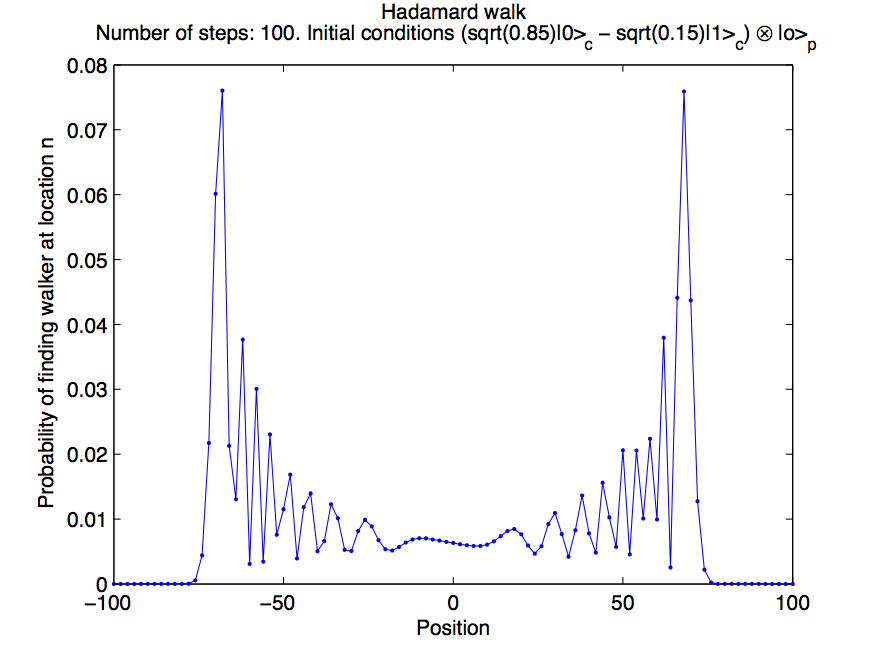}}
\end{center}
\caption{{\small Graph (a) was computed using initial state $|\psi\rangle_0 = {1 \over \sqrt{2}}(|0\rangle_c + i |1\rangle_c) \otimes |0\rangle_p$. Graphs (b) and (c) had $| \psi \rangle_0 = | 0 \rangle_c \otimes | 0 \rangle_p$ and $|\psi\rangle_0 = \sqrt{0.85}|0\rangle_c - \sqrt{0.15} |1\rangle_c) \otimes |0\rangle_p$ as initial states, respectively. Notice that symmetry in the probability distribution can be achieved by using coin initial states with either complex or real relative phase factors \cite{tregenna03}. All graphs were computed from $100$-step Hadamard quantum walks on a line with Eq. (\ref{shift_single}) as shift operator.}}
\label{hadamard_symmetry}
\end{figure}

As in the Hadamard walk case, the properties of the quantum walk defined by Eqs. (\ref{general_quantum_walk},\ref{quantum_walk_general}) may be studied by inverting the Fourier transform and using methods of complex analysis. Let us concentrate on the phase factors $\alpha \in \mathbb{R}$  of the coin initial state (Eq. (\ref{general_coin_initial_state})) and $\theta \in \mathbb{R}$ of the coin operator (Eq. (\ref{fourier_general_coin_operator}).) Note that we can choose many pairs of values ($\alpha, \theta$) for any phase factor $r = \alpha + \theta$. So, if we fix a value for $\theta$ (i.e. if we use only one coin operator) we can always vary the initial coin state $|\psi(x,0)\rangle$ (Eq. (\ref{general_coin_initial_state})) to get a value for $\alpha$ so that we can compute a quantum walk with a certain phase factor value $r$. It is in this sense that we say that the study of a Hadamard walk suffices to analyze the properties of all unrestricted quantum walks on a line. In Fig. (\ref{hadamard_symmetry}) we show the probability distributions of three Hadamard walks with different initial coin states.

On further studies of coined quantum walks on a line, Villagra {\it et al} \cite{villagra10} present a closed-form of the probability that a quantum walk arrives at a given vertex after $n$ steps, for a general symmetric {\it SU(2)} coin operator.

\subsubsection{Discrete Quantum walk with boundaries}

The properties of discrete quantum walks on a line with one and two absorbing barriers were first studied in \cite{ambainis01}. For the semi-infinite discrete quantum walk on a line, Theorem \ref{quantum_walk_one_barrier} was reported

\begin{theorem}
Let us denote by $p_\infty$ the probability that the measurement of whether the particle is at the location of the absorbing boundary (location $0$ in \cite{ambainis01}) $\Rightarrow$ $ p_\infty = {2 \over \pi}$.
\label{quantum_walk_one_barrier}
\end{theorem}

Theorem \ref{quantum_walk_one_barrier} is in stark contrast with its classical counterpart (Theorem 8 of \cite{ambainis01}), as the probability of eventually being absorbed (in the classical case) is equal to unity. Furthermore, Yang, Liu and Zhang have introduced  an interesting and relevant result in \cite{liu07_path}: the absorbing probability of Theorem \ref{quantum_walk_one_barrier} decays faster than the classical case and, consequently, the conditional expectation of the quantum-mechanical case is finite (as opposed to the classical case in which the corresponding conditional expectation is infinite.)

The case of a quantum walk on a line with two absorbing boundaries was also studied in \cite{ambainis01}, and their main result is given in Theorem \ref{quantum_walk_two_barriers}.

\begin{theorem}
For each $n > 1$, let $p_n$ be the probability that the process eventually
exits to the left. Also define $q_n$ to be the probability that the process
exits to the right. Then
$$
i) \forall \text{ } n>1 \text{ } \Rightarrow \text{ } p_n + q_n = 1
$$

$$
ii) \lim_{n \rightarrow \infty} p_n = {1 \over \sqrt{2}}
$$
\label{quantum_walk_two_barriers}
\end{theorem}

In \cite{bach04}, Bach {\it et al} revisit  Theorems \ref{quantum_walk_one_barrier} and \ref{quantum_walk_two_barriers} with detailed corresponding proofs using both Fourier transform and path counting approaches as well as prove some conjectures given in \cite{yamasaki03}. Moreover, in \cite{bach09},  Bach and Borisov further study the absorption probabilities of the two-barrier quantum walk. Finally, Konno studied the properties of quantum walks with boundaries using a set of matrices derived from a general unitary matrix together with a path counting method (\cite{konno02a,konno03}.)

\subsubsection{Unrestricted quantum walks on a line with several coins}

The effect of different and multiple coins has been studied by several authors. In  \cite{konno04}, Inui and Konno have analyzed the localization phenomena due to eigenvalue degeneracies in one-dimensional quantum walks with 4-state coins (the results shown in \cite{konno04} have some similarities with the quantum walks with maximally entangled coins reported by Venegas-Andraca {\it et al} in \cite{sva05} in the sense that both quantum walks tend to concentrate most of their probability distributions about the origin of the walk, i.e. the localization phenomenon is present.) Moreover, in \cite{konno05c}, Konno, Inui and Segawa have derived an analytical expression for the stationary distribution of one-dimensional quantum walks with 3-state coins that make the walker go either right or left or, alternatively, rest in the same position. Additionally, Ribeiro {\it et al} \cite{ribeiro04} have considered quantum walks with several biased coins applied aperiodically, D'Alessandro {\it et al} \cite{dalessandro07} have studied non-stationary quantum walks on a cycle using different coin operators at each computational step, and Feinsilver and Kocik \cite{feinsilver02} have proposed the use of Krawtchouk matrices (via tensor powers of the Hadamard matrix) for calculating quantum amplitudes.

Linden and Sharam have formally introduced  a family of quantum walks, inhomogeneous quantum walks, being their main characteristic to allow  coin operators to depend on both position and coin registers \cite{linden09}. Shikano and Katsura \cite{shikano10} have studied the properties of self-duality, localization and fractality on a generalization of the inhomogeneous quantum walk model defined in \cite{linden09}, Konno has presented and proved a theorem on return probability for inhomogeneous walks which are periodic in position \cite{konno10_inhomogeneous}, Machida \cite{machida10_combination} has found that combining the action of two unitary operators in an inhomogenenous quantum walk will result in a limit distribution for $X_t/t$ that can be expressed as a $\delta$ function and a combination of density functions (for a detailed analisys of weak convergence $X_t/t$ please go to subsection \ref{qw_limit_theorems}), and Konno has proved that the return probability of a one-dimensional discrete-time quantum walk can be written in terms of elliptic integrals \cite{konno10_elliptic}.

In \cite{brunetal103}, Brun {\it et al} analyzed the behavior of a quantum walk on the line using both $M$ 2-dimensional coins and single coins of $2^M$ dimension, and Sewaga {\it et al} \cite{segawa08} have computed analytical expressions for limit distributions of quantum walks driven by $M$ 2-dimensional coins as well as analyzed the conditions upon which applying $M$ 2-dimensional coins to a quantum walk leads to classical behavior. Furthermore, Ba\~{n}uls {\it et al} \cite{banulus05} have studied the behavior of quantum walks with a time-dependent coin and Machida and Konno \cite{machida10_time_dependent} have produced limit distributions for such quantum walks with ${\hat C}={\hat C(t)}$, Chandrashekar \cite{chandrashekar08} has proposed a generic model of quantum walk whose dynamics is described by means of a Hamiltonian with an embedded coin, and Romanelli \cite{romanelli09a} has generalized the standard definition of a discrete quantum walk and shown that appropriate choices of quantum coin lead to obtaining a variety of wave-function spreading. Finally, Ahlbrecht {\it et al} have produced a comprehensive analysis of  asymptotical behavior of ballistic and diffusive spreading, using Fourier methods together with perturbed and unperturbed operators \cite{ahlbrecht11}.

\subsubsection{Decoherence and other considerations on classical and quantum walks}

The links between classical and quantum versions of random walks have been studied by several authors under different perspectives:
\\
1) Simulating classical random walks using quantum walks. Studies on this area (e.g. \cite{watrous01}) would provide us not only with interesting computational properties of both  types of walks, but also with a deeper insight of the correspondences between  the laws that govern computational processes in both classical and quantum physical systems.
\\
2) Transitions from quantum walks into classical random walks. This area of research is interesting not only for exploring computational properties of both kinds of walks, but also because we would provide quantum computer builders (i.e. experimental physicists and engineers) with some criteria and thresholds for testing the quantumness of a quantum computer.  Moreover, these studies have allowed the scientific community to reflect on the quantum nature of quantum walks and some of their implications in algorithm development (in fact, we shall discuss the quantum nature of quantum walks in subsection \ref{quantumness}.) 

Decoherence is a physical phenomenon that typically arises from the interaction of quantum systems and their environment. Decoherence used to be thought of as an annoyance as it used to be equated with loss of quantum information. However, it has been found that decoherence can indeed play a beneficial role in natural processes (e.g. \cite{mohseni08}) as well as produce interesting results for quantum information processing (e.g. \cite{kendon02,romanelli05,brunetal203}.) In addition to these properties, decoherence via measurement or free interaction with a classical environment is a typical framework for studying transitions of quantum walks into classical random walks. Thus, for the sake of getting a deeper understanding of the physical and mathematical relations between quantum systems and their environment, together with searching for new paradigms for building quantum algorithms, studying decoherence properties and effects on quantum walks is an important field in quantum computation.

Tregenna and Kendon \cite{kendon02} have studied the impact of decoherence in quantum walks on a line, cycle and the hypercube, and have found that some of those decoherence effects could be useful for building quantum algorithms, Strauch \cite{strauch09_dhypercube} has also studied the effects of decoherence on {\it continuous-time} quantum walks on the hypercube, and Fan {\it et al} \cite{fan11_convergence_decoherence} have proposed a convergent rescaled limit distribution for quantum walks subject to decoherence. Brun {\it et al} \cite{brun03} have shown that the quantum-classical walk transition could be achieved via two possible methods, in addition to performing measurements: decoherence in the quantum coin and the use of higher-dimensional coins, Ampadu  \cite{ampadu12} has focused on generalizing the method of decoherent quantum walk proposed in \cite{brun03} for two-dimensional quantum walks, and Annabestani {\it et al} have generalized the results of \cite{brun03}   by providing analytical expressions for different kinds of decoherence \cite{annabestani10b}. Moreover, by using a discrete path approach, it was shown by Konno  that introducing a random selection of coins (that is, amplitude components for coin operators are chosen randomly, being under the unitarity constraint) makes quantum walks behave classically \cite{konno05a}. In \cite{childs02}, Childs {\it et al} make use of a family of graphs (e.g. Fig. (\ref{trees}(a)) to exemplify the different behavior of (continuous) quantum walks and classical random walks. 

Several authors have addressed the physical and computational properties of decoherence in quantum walks: Ermann {\it et al} \cite{ermann05} have inspected the decoherence of quantum walks  with a complex coin, where the coin is part of a larger quantum system, Chandrashekar et al \cite{chandrashekar0702} have studied symmetries and noise effects  on coined discrete quantum walks, and Obuse and Kawakami \cite{obuse11_topological}  have studied one-dimensional quantum walks with spatial or temporal random defects as a consequence of interactions with randome environments, having found that this kind of quantum walks can avoid complete localization. Also, Kendon {\it et al} \cite{kendon02,kendon02a,kendon03} have extensively studied the computational consequences of coin decoherence  in quantum walks, Alagi\'c and Russell \cite{alagic05} have studied the effects of independent measurements on a quantum walker travelling along the hypercube (please see Def. \ref{hypercube_1} and Fig. \ref{hypercube_3d}), Ko\v s\'{\i}k {\it et al} \cite{kosik06} have studied the quantum to classical transition of a quantum walk by introducing randoms phase shifts in the coin particle, Romanelli \cite{romanelli07} has studied one-dimensional quantum walks subjected to decoherence induced by measurements perfomed with timing provided by the L{\'e}vi waiting time distribution, P{\'e}rez and Romanelli \cite{romanelli11_brokenlinks} have analyzed a one-dimensional discrete quantum walk under decoherence, on the coin degree of freedom, with a strong spatial dependence (decoherence acts only when the walker moves on one half of the line), and Oliveira {\it et al} \cite{oliveira06_decoherence2dqw} have analyzed two-dimensional quantum walks under a decoherence regime due to random broken links on the lattice. Furthermore and taking as basis a global chirality probability distribution (GCD) independent of the walker's position proposed in \cite{romanelli10}, Romanelli has studied the behavior of one-dimensional quantum walks under two models of decoherence: periodic measurements of position and chirality as well as randomly broken links on the one-dimensional lattice \cite{romanelli11}. Additionally, Chisaki {\it et al} \cite{chisaki11_crossovers} have studied both quantum to classical and classical to quantum transitions using discrete-time and classical random walks, and have also introduced a new kind of quantum walk entitled final-time-dependent discrete-time quantum walk (FD-DTQW) together with a limit theorem for FD-DTQW.

In \cite{zhang08}, Zhang studied the effect of increasing decoherence (caused by measurements probabilistically performed on both walker and coin) in coined quantum walks and derived analytical expressions for position-related probability distributions, Annabestani {\it et al} have studied the impact of decoherence on the walker in one-dimensional quantum walks \cite{annabestani10c}, Srikanth {\it et al} \cite{srikanth10} have quantified the degree of \lq quantumness'  in decoherent quantum walks using measurement-induced disturbance, G{\"{o}}n{\"{u}}lol {\it et al} \cite{gonulol09_dtdqw} have studied decoherence phenomena in two-dimensional quantum walks with traps, and Rao {\it et al} have analyzed noisy quantum walks using measurement-induced disturbance and quantum discord \cite{rao11}. Moreover, Liu and Petulante have proposed a model for decoherence in an $n$-site cycle together with a definition for decoherence time \cite{liu10}, as well as derived analytical expressions for i) the asymptotic dynamics of discrete quantum walks under decoherence on the coin degree of freedom \cite{liu_10_cycle} and on both coin and walker degrees of freedom running on n-site cycles \cite{liu11}, ii) the order ({\it big O}) of the mixing time for the time-averaged probability of a quantum walk subject to decoherence on the coin quantum system \cite{liu_10_cycle}, and iii) the limiting behavior of quantum entanglement between coin and walker under the same decoherence regime \cite{liu11}.

Schreiber {\it et al} \cite{schreiber11} have analyzed the effect of decoherence and disorder in a photonic implementation of a quantum walk, and have shown how to use dynamic and static disorder to produce diffusive spread and Anderson localization, respectively. In addition, Ahlbrecht {\it et al}  have produced a detailed manuscript in which several topics from the field of discrete quantum walks are analyzed, including ballistic and diffusive behavior, decoherent and invariance on translation, asymptotic behavior with perturbation, together with several examples \cite{ahlbrecht11}.

\subsubsection{\label{qw_limit_theorems}Limit theorems for quantum walks}

The central limit theorem plays a key role in determining many properties of statistical estimates. This key role has been a crucial motivation for members of the quantum computing community to derive limit distributions for quantum walks. Among the scientific contributions produced in this field, the seminal papers produced by Norio Konno and collaborators have been central to the effort of deriving analytical results and establishing solid grounds for quantum walk limit distributions.

Let us start this summary with a fundamental result for quantum walks on a line: Konno's weak limit theorem \cite{konno02,konno02a,konno05b,konno08} (following mathematical statements are taken verbatim from corresponding papers.) 
\\
Let $\Phi=\{ \varphi = (\alpha,\beta)^t \in \mathbb{C}^2  \text{ :  }|\alpha|^2 + |\beta|^2 = 1 \}$ be the set of initial qubit states of a one-dimensional quantum walk, and let $X^{\varphi} _n$  denote a one-dimensional quantum walk at time $n$ starting from initial qubit state $\varphi \in \Phi$  with evolution operator given by  a $2 \times 2$ unitary matrix

\begin{equation}
U=\left[ \begin{array}{cc} a & b \\ c & d \end{array} \right]
\label{u_unitary_konno}
\end{equation}

Using a path integral approach, Konno proves the following theorem:

\begin{theorem}
\cite{konno02,konno05b,konno08}
We assume $abcd \not= 0$. If $n \to \infty$, then 
\[
{X_n ^{\varphi} \over n} \quad \Rightarrow \quad Z^{\varphi}
\]
where $Z^{\varphi}$ has the following density, known as {\bf Konno's density function} 
\[
f(x; {}^t[\alpha, \beta])
= { \sqrt{1 - |a|^2} \over \pi (1 - x^2) \sqrt{|a|^2 - x^2}} 
 \left\{ 1- \left( |\alpha|^2 - |\beta|^2 + {a \alpha \overline{b \beta} + \overline{a \alpha} b \beta \over |a|^2 } \right) x \right\} 
\]
for $x \in (- |a|, |a|)$ with 
\begin{eqnarray*}
&& E(Z^{\varphi}) 
=
- \left( |\alpha|^2 - |\beta|^2 + { a \alpha \overline{b \beta} 
+ \overline{a \alpha} b \beta \over |a|^2 } \right) 
\times (1 - \sqrt{1 - |a|^2}) 
\\
&& E ((Z^{\varphi})^2) = 1 - \sqrt{1 - |a|^2}
\end{eqnarray*}
and $ Y_n \Rightarrow Y$ means that $Y_n$ converges in distribution to a limit $Y$. 
\label{weak_limit_theorem_konno}
\end{theorem}

That is, the quantity  ${X_n ^{\varphi} \over n}$, later on named a {\it pseudovelocity}, does converge to the limit distribution $Z$. In \cite{hamada09_orthogonal}, Hamada {\it et al} study the symmetric $\left[\left( |\alpha|^2 - |\beta|^2 + {a \alpha \overline{b \beta} + \overline{a \alpha} b \beta \over |a|^2 } \right)=0\right]$ and asymmetric $\left[\left( |\alpha|^2 - |\beta|^2 + {a \alpha \overline{b \beta} + \overline{a \alpha} b \beta \over |a|^2 } \right) \in [\frac{-1}{r}, \frac{1}{r}], \text{ where } r \in (0,1) \right]$ cases of Konno's density function.
\\

A plethora of central results are published in \cite{konno02,konno02a,konno05b,konno08}. Among them, I mention the following:

\begin {itemize}

\item {\bf Symmetry of probability distribution $P({X_n ^{\varphi}})$}.
\\

Let us define the following sets:

\begin{definition}
\begin{eqnarray*}
\Phi_s &=&  \{ \varphi \in 
\Phi : \> 
P(X_n ^{\varphi}=k) = P(X_n ^{\varphi}=-k) \>\> 
\hbox{for any} \> n \in {\bf Z}_+ \> \hbox{and} \> k \in {\bf Z}
\},
\\
\Phi_0 &=& \left\{ \varphi \in 
\Phi : \> 
E(X_n ^{\varphi})=0 \>\> \hbox{for any} \> n \in {\bf Z}_+
\right\},
\\
\Phi_{\bot} &=& \left\{ \varphi = [\alpha, \beta]^t \in 
\Phi :
|\alpha|= |\beta|=1/\sqrt{2}, \> a \alpha \overline{b \beta} + \overline{a \alpha} b \beta =0 
\right\}, 
\end{eqnarray*}
\label{symmetry_qw}
\end{definition}
where ${\bf Z}_+$ is the set of the positive integers. Then,

\begin{theorem}
Let $\Phi_{s}, \Phi_0,$ and $\Phi_{\bot}$ be as in Def. (\ref{symmetry_qw}). Suppose $abcd \not= 0$. Then we have $\Phi_{s} = \Phi_0 = \Phi_{\bot}.$
\label{theorem_symmetry_qw_konno}
\end{theorem}

Theorem \ref{theorem_symmetry_qw_konno} is a generalization of the result given by \cite{konno04b} for the Hadamard walk, i.e. a one-dimensional quantum walk with the Hadamard operator (Def. \ref{hadamard_single}) as evolution operator. Also, Nayak and Vishwanath \cite{nayak00} discussed the symmetry of distribution and showed that $[1/\sqrt{2},\>$ $\pm i/\sqrt{2}]{}^t \in \Phi_s$ for the Hadamard walk. 
\\

\item {\bf $m^{th}$ moment of $ X^{\varphi} _{n}$}. A most interesting result from \cite{konno02,konno02a,konno05b,konno08} is the expected behavior of $(X_n ^{\varphi})^m$: for $m$ even, $E((X_n ^{\varphi})^m)$ is independent of  the initial qubit state ${\varphi}$. In contrast, for  $m$ odd, $E((X_n ^{\varphi})^m)$ does depend on  the initial qubit state ${\varphi}$.

\begin{theorem}
\label{cor:cor5} 
\par\noindent
\hbox{(i)} Suppose $abcd \not= 0$. When $m$ is odd, we have
\begin{eqnarray*}
E((X_n ^{\varphi}) ^m) 
&=& 
- |a|^{2(n-1)}
\Biggl[ \mu_{\alpha,\beta} \> n^m + \sum_{k=1}^{\left[{n-1 \over 2}\right]}
\sum_{\gamma =1} ^{k} \sum_{\delta =1} ^{k}
\left(-{|b|^2 \over |a|^2} \right)^{\gamma + \delta} \> {(n-2k)^{m+1} \> \kappa_{\gamma,\delta,n,k} \over  \gamma \delta} 
\\
&&
\qquad \qquad \qquad \qquad \qquad \qquad  \times
\biggl\{ 
\mu_{\alpha,\beta} \> n + {\gamma +\delta \over 2 |b|^2}  
(|\alpha|^2 - |\beta|^2 - \mu_{\alpha,\beta} ) 
\biggr\} \Biggr].
\end{eqnarray*}
\par\noindent
When $m$ is even, we have
\begin{eqnarray*}
E((X_n^{\varphi}) ^m) 
&=& |a|^{2(n-1)} 
\Biggl\{
n^m +
\sum_{k=1}^{\left[{n-1 \over 2}\right]}
\sum_{\gamma =1} ^{k} \sum_{\delta =1} ^{k}
\left(-{|b|^2 \over |a|^2} \right)^{\gamma + \delta} 
\> {(n-2k)^{m} \kappa_{\gamma,\delta,n,k} \> \nu_{\gamma,\delta,n,k} \over  \gamma \delta} 
\Biggr\}.
\end{eqnarray*}
\par\noindent
\hbox{(ii)}
Let $b=0$. Then we have 
\[
E((X_n^{\varphi}) ^m) =
\left\{
\begin{array}{cl}
n^m (|\beta|^2 - |\alpha|^2)
& \mbox{if $m$ is odd,} \\
n^m  
& \mbox{if $m$ is even.}
\end{array}
\right.
\]
\par\noindent
\mbox{(iii)}
Let $a=0$. Then we have
\[
E((X_n ^{\varphi})^m) =
\left\{
\begin{array}{cl}
|\alpha|^2 - |\beta|^2
& \mbox{if $n$ and $m$ are odd,} \\
1
& \mbox{if $n$ is odd and $m$ is even,} \\
0
& \mbox{if $n$ is even.}
\end{array}
\right.
\]
\end{theorem}

where $\mu_{\alpha,\beta} = \left(|a|^2 - |b|^2 \right)  \left(|\alpha|^2 - |\beta|^2 \right)  + 2 (a \alpha \overline{b \beta} + \overline{a \alpha} b \beta )$.
\\

\item {\bf Hadamard walk case}. Let the unitary matrix $U$ from Eq. (\ref{u_unitary_konno}) be the Hadamard operator given in Eq. (\ref{hadamard_single}). Then, the following result holds:

For any initial qubit state $\varphi = [\alpha, \beta]{}^t$, Theorem \ref{weak_limit_theorem_konno} implies  
\begin{eqnarray}
\lim_{n \to \infty} P(a \le X^{\varphi} _{n}/n \le b) = \int^b _a {  1-  (|\alpha|^2 - |\beta|^2 + \alpha \overline{\beta} + \overline{\alpha} \beta) x \over \pi (1-x^2) \sqrt{1-2x^2}} \> 1_{
(\frac {-1}{\sqrt{2}}, \frac {1}{\sqrt{2}})} (x) \> dx,
\label{probability_pseudovelocity_konno}
\end{eqnarray}
where $1_{(u,v)}(x)$ is the indicator function, that is, $1_{(u,v)}(x)=1 $ if $x \in (u,v)$, and $1_{(u,v)}(x) = 0$ if $x \notin (u,v).$

Compare Eq. (\ref{probability_pseudovelocity_konno}) with the corresponding result for the classical symmetric random walk $Y^o _n$ starting from the origin, 
Eq. (\ref{classical_moivre_laplace_probability_konno}):
 
\begin{eqnarray}
\lim_{n \to \infty}
P(a \le Y^o _{n}/ \sqrt{n} \le b) = \int^b _a {e^{-x^2/2} \over \sqrt{2 \pi}} dx.
\label{classical_moivre_laplace_probability_konno}
\end{eqnarray}

\end{itemize}

In addition to the scientific contributions already mentioned in previous sections, we now provide a summary of more results on limit distributions. Konno \cite{konno05_limit_cont} has proved the following weak limit theorem for continuous quantum walks:

\begin{theorem}
Let us denote a continuous-time quantum walk on $\mathbb{Z}$ by $X_t$ whose probability distribution is defined by $P(k,t)$ for any location $k \in \mathbb{Z} $ and time $t \ge 0.$  Then, the following result holds for a continuous-time quantum walk on a line:
\begin{eqnarray*}
P ( a \le X_t/t \le b) \quad \to \quad \int_a ^b {1 \over \pi \sqrt{1 - x^2}} \> dx \qquad \text{        as }  t \to \infty, \text{ for } -1 \le a < b \le 1. 
\end{eqnarray*}
\label{teorema_konno_limite_debil}
\end{theorem}

In \cite{grimmett04_weak}, Grimmett {\it et al} used Fourier transform methods to also rigorously prove weak convergence theorems for $one-$ and $d-$ dimensional quantum walks and, using the definition of pseudovelocities introduced by Konno \cite{konno05b}, the Fourier transform method proposed in \cite{grimmett04_weak} and the one-parameter family of quantum walks proposed by Inui {\it et al} in \cite{inui04}, Watabe {\it et al} \cite{watabe08} have derived analytical expressions for the limit and localization distributions of walker pseudovelocities in two-dimensional quantum walks, while Sato {\it et al} \cite{sato08_unpublished} have derived limit distributions for qudits in one-dimensional quantum walks, Liu and Petulante have presented limiting distributions for quantum Markov chains \cite{liupetulante11_markovchains}, and Chisaki {\it et al} have also deduced limit theorems for $X_t$ (localization) and $\frac{X_t}{n}$ (weak convergence) for quantum walks on Cayley trees \cite{chisaki09_trees}.

Furthermore and based on the Fourier transform approach developed by Grimmett {\it et al} \cite{grimmett04_weak}, Machida and Konno have deduced a limit theorem for discrete quantum walks with 2-dimensional time-dependent coins \cite{machida10_time_dependent}. In addition, Machida has produced analytical expressions for weak convergence as well as limit distributions for a localization model of a 2-state quantum walk \cite{machida11_localizationmodel}, Konno has derived limit theorems using path counting methods for discrete-time quantum walks in random (both quenched and annealed) environments \cite{konno09_random_environments}, and Liu \cite{liu11b} has derived a weak limit distribution as well as formulas for stationary probability distribution for quantum walks with two-entangled coins \cite{sva05}.

 Motivated by the properties of quantum walks with many coins published by Brun {\it et al} in \cite{brun03,brunetal103}, Segawa and Konno \cite{segawa08} have used the Wigner formula of rotation matrices for quantum walks published by Miyazaki {\it et al} in \cite{miyazaki07} to rigorously derive limit theorems for quantum walks driven by many coins. Also,  Sato and Katori \cite{sato10} have analyzed Konno's pseudovelocities within the context of relativistic quantum mechanics, di Molfetta and Debbasch have proposed a subset of quantum walks, named (1-jets), to study how continuous limits can be computed for discrete-time quantum walks \cite{molfetta11}. In addition, based on definitions and concepts found in \cite{konno05b,konno03,konno02}, Ampadu proposed a mathematical model for the localization and symmetrization phenomena in generalized Hadamard quantum walks as well as proposed conditions for the existence of localization \cite{ampadu11a}. Moreover, based on Mc Gettrick's model of discrete quantum walks with memory \cite{gettrick11_memory} and using the Fourier-based approach proposed by  Grimmett {\it et al} \cite{grimmett04_weak}, Konno and Machida \cite{konno10_qw_memory} have proved two new weak limit distribution theorems for that kind of quantum walk.

Finally, in \cite{konno11_inhomogeneous} Konno {\it et al}  have studied three kinds of measures (time averaged limit measure, weak limit measure and stationary measure) as well as studied conditions for localization in a family of inhomogeneous quantum walks, while Chisaki {\it et al} have produced limit theorems for discrete quantum walks running on joined half lines (i.e. lines with sites defined on $\mathbb{Z}^+ \cup \{0\})$ and (semi)homogeneous trees \cite{chisaki12}.

\subsubsection{Localization in discrete quantum walks}

In condensed-matter physics, localization is a well-studied physical phenomenon. According to Kramer and MacKinnon \cite{kramer93}, it is likely that the first paper in which localization was discussed within the context of quantum mechanical phenomena is \cite{anderson58} by P. W. Anderson. Since then, localization has been extensively studied (see the compilation of textbooks and reviews on localization provided in \cite{kramer93}) and, consequently, different cualitative and mathematical definitions have been provided for this concept. Nevertheless, the essential idea behind localization is {\it the absence of diffusion of a quantum mechanical state},  which could be caused by random or disordered environments that break the periodicity in the dynamics of the physical system. Moreover, localization could also be produced by evolution operators that mimic the behavior of disordered media, as shown by Chandrashekar in \cite{chandrashekar11}. As for quantum walks, localization phenomena has been detected as a result of either eigenvalue degeneracy (typically caused by using evolution operators that are all identical except for a few sites) or choosing coin operators that are site dependent \cite{joye12_pc}.

In order to have a precise and inclusive introduction to localization in quantum walks, we direct the reader's attention to \cite{joye12,joye11} by A. Joye, \cite{joye10} by A. Joye and M. Merkli, and \cite{hamza12} by E. Hamza and A. Joye, and references provided therein. In addition to these references and those presented in previous sections in which we have incidentally addressed the topic of localization, we also mention the numerical simulations of quantum walks on graphs shown by Tregenna {\it et al} \cite{tregenna03}, in which the localization phenomenon, due to the use of Grover's operator (Def. (\ref{grover_coin_operator})) in a 2-dimensional quantum walk, was detected. Inspired by this phenomenon, Inui {\it et al} proved in \cite{inui04} that the key factor behind this localization phenomenon is the degeneration of the eigenvectors of corresponding evolution operator, Inui and Konno \cite{konno04} have further studied the relationship between localization and eigenvalue degeneracy in the context of particle trapping in quantum walks on cycles, and  Ide {\it et al} have computed the return probability of final-time dependent quantum walks \cite{idekonno11_return}. Based on the study of aperiodic quantum walks given in \cite{ribeiro04}, Romanelli \cite{romanelli09} has proposed the computation of a trace map for Fibonacci quantum walks (this is a discrete quantum walk with two coin operators arranged in quasi-periodic sequences following a Fibonacci prescription) and Ampadu has shown that localization does not occur on Fibonacci quantum walks \cite{ampadu11b}. 

In \cite{grunbaum12_recurrence_dqw}, Gr{\"{u}}nbaum {\it et al} have studied recurrence processes on discrete-time quantum walks following a particle absorption monitoring approach (i.e. a projective measurement strategy), \ifmmode \check{S}\else \v{S}\fi{}tefa\ifmmode \check{n}\else \v{n}\fi{}\'ak {\it et al}  have analyzed the P{\'o}lya number (i.e. recurrence without monitoring particle absorption) for biased quantum walks on a line \cite{stefanak09_biased_polya_recurrence} as well as for $d$-dimensional quantum walks \cite{stefanak1108_polya_recurrence,stefanak08_ubiased_polya_recurrence}, and Dar{\'a}z and Kiss \cite{daraz10_polyacontinuousqw} have also proposed a P{\'o}lya number for continuous-time quantum walks. In \cite{stefanak1108_polya_recurrence}, \ifmmode \check{S}\else \v{S}\fi{}tefa\ifmmode \check{n}\else \v{n}\fi{}\'ak {\it et al} have proposed a criterion for localization and Koll\'ar {\it et al}   \cite{stefanak110_recurrence_three} found that, when executing a discrete-time quantum walk on a triangular lattice using a three-state Grover operator, there is no localization in the origin.

Furthermore, Chandrashekar has found that one-dimensional discrete coined quantum walks fail to fully satisfy the quantum recurrence theorem but suceed at exhibiting a fractional recurrence that can be characterized using the quantum P{\'o}lya number \cite{chandrashekar10_fractional_recurrence}, Ampadu has analyzed the motion of $M$ particles on a one-dimensional Hadamard walk and has presented a theoretical criterion for observing quantum walkers at an initial location with high probability \cite{ampadu11_mlocalization},  has also studied the conditions upon which a biased quantum walk on the plane is recurrent \cite{ampadu11_recurrence_hadamard}, as well as studied the localization phenomenon in two-dimensional five-state quantum walks \cite{ampadu11_two_five_localization}.

 In \cite{cantero10}, Cantero {\it et al} present an alternative method to formulate the theory of quantum walks based on matrix-valued Szeg{\"{o}} orthogonal polynomials, known as the CGMV method, associated with a particular kind of unitary matrices, named CMV matrices, and Hamada {\it et al} have independently introduce the idea of employing orthogonal polynomials for deriving analytical expressions for limit distributions of one-dimensional quantum  walks \cite{hamada09_orthogonal}.  Based on the mathematical formalism delivered in \cite{cantero10}, Konno and Segawa \cite{konnosegawa11} have studied quantum walks on a half line, focusing on analyzing the corresponding spectral measure  as well as on localization phenomena for this kind of quantum walks. Also based on the CMV method presented in \cite{cantero10}, Ampadu has studied both limit distributions and localization of quantum walks on the half plane \cite{ampadu11_halfplane}. Moreover, in \cite{cantero10a}, Cantero {\it et al} have produced an extensive analysis of the asymptotical behavior of quantum walks: starting with a definition for a quantum walk with one defect (i.e. a one-dimensional quantum walk with constant coins except for the origin) and using the CGMV method, Cantero {\it et al} have classified localization properties as well as derived analytical expressions for return probabilities to the origin. Finally, Gr{\"{u}}nbaum and Vel{\'a}zquez have studied models of quantum walks on the non-negative integers using Riez probability measures \cite{grunbaumvelazquez11_riez}.

On further studies, Konno \cite{konno10_inhomogeneous} has mathematically proved that inhomogenenous discrete-time quantum walks do exhibit localization, Shikano and Katsura \cite{shikano10} have proved that, for a class of inhomogenenous quantum walks, there is a  limit distribution that is localized at the origin, as well as found, through numerical studies, that the eigenvalue spectrum of such inhomogenenous walks exhibit a fractal structure similar to that of the Hofstadter butterfly. Also, Machida has proposed a localization model of quantum walks on a line \cite{machida11_localizationmodel} as well as computed a limit distribution for 2-state inhomogenenous quantum walks with different unitary operators applied in different times \cite{machida10_combination}, and Chandrashekar has proposed Hamiltonians for walking on different lattices as well as found links between localization and spatially static disordered operations \cite{chandrashekar11b}, and presented a scheme to induce localization in a Bose-Einsten condensate \cite{chandrashekar11}.  Finally, in \cite{ahlbrecht11_disordered}, Ahlbrecht {\it et al} have delivered a review on disordered one-dimensional quantum walks and dynamical localization.

\subsubsection{More results on discrete quantum walks}
 
A plethora of numerical, analytical and experimental results have made the field of quantum walks rich and solid. In addition to the results already mentioned in this review, I would like to direct the reader's attention to the following results:

In \cite{shikano11_arrow_time}, Shikano {\it et al} have proposed using discrete-time quantum walks to analyze problems in quantum foundations. Specifically, Shikano {\it et al} have derived an analytical expression for the limit distribution of a discrete-time quantum walk with periodic position measurements and analyzed the concepts of randomness and arrow of time. Also, G{\"{o}}n{\"{u}}lol {\it et al} have found that the quantum walker survival probability  in discrete-time quantum walks running of cycles with traps exhibits a piecewise stretched exponential character \cite{gonulol11_survival}, Kurzy{\'{n}}ski and W{\'{o}}jcik and shown that quantum state transfer is achievable in discrete-time quantum walks with position-dependent coins \cite{kurzynski11_statetransfer}, Stang {\it et al} have introduced a history-dependent discrete-time quantum walk (i.e. a quantum walk with memory) and proposed a correlation  function for measuring memory  effects on the evolution of discrete-time quantum walks \cite{stang09_correlatedqw}, Navarrete-Benlloch {\it et al} \cite{navarretebenlloch07} have introduced a nonlinear version of the optical Galton board, Whitfield {\it et al} \cite{whitfield10} have introduced an axiomatic approach for a generalization of both continuous and discrete quantum walks that evolve according to a quantum stochastic equation of motion (\cite{whitfield10} helps to realize why the behavior of some decoherent quantum walks is different from both classical and coherent quantum walks), Xu \cite{xu10} has derived analytical expressions for position probability distributions on unrestricted quantum walks on the line, together with an introduction to a quantum walk on infinite or even-numbered size of lattices which  is equivalent to the traditional quantum walk with symmetrical initial state and coin parameter,  Chandrashekar has introduced a quantum walk version of Parrondo's games \cite{chandrashekar11_parrondo} and, in \cite{chandrashekar10a}, Chandrashekar {\it et al} have introduced some mathematical relationships between quantum walks and relativistic quantum mechanics and have proposed Hamiltonian operators (that retain the coin degree of freedom) to run quantum walks on different lattices (e.g. cubic, kagome and honeycomb lattices)  as well as to study different kinds of disorder on quantum walks. Also, Feng {\it et al} have introduced the idea of using quantum walks to study waves \cite{feng10_qw_waves},  Cantero {\it et al} show how to use matrix valued orthogonal polynomials defined in the real line to build a large class of quantum walks \cite{cantero10}, and Jacobs  has analyzed quantum walks within the mathematical framework of coalgebras, monads and category theory \cite{jacobs09_coalgebraic,jacobs11_categories}.

Mc Gettrick \cite{gettrick11_memory} has proposed a model of discrete quantum walks with up to two memory steps and derived analytical expressions for corresponding quantum amplitudes. Based on \cite{gettrick11_memory}, Konno and Machida \cite{konno10_qw_memory} have proved two new weak limit distribution theorems. Moreover,  Romanelli \cite{romanelli11_temperature} has developed a thermodynamical approach to entanglement quantification between walker and coin and de Valc\'arcel {\it et al} \cite{devalcarcel10} have assigned extended probability distributions as initial walker position  in a discrete quantum walk, and have found a particular initial condition for producing a homogeneous position distribution (interestingly enough, a similar quasi-homogeneous position probability distribution has been shown in \cite{kendon07} as a result of a measurement-induced decoherent process in a discrete quantum walk.) Also, Goswani {\it et al} have extended the concept of persistence (i.e. the time during which a given site remains unvisited by the walker) \cite{goswani11_persistence},  Konno and Sato \cite{konno11_zetafunction} have presented a formula for the transition matrix of a discrete-time quantum walk in terms of the second weighted zeta function, and Konno {\it et al} have shown several relationships between the Heun and  Gauss differential equations with quantum walks \cite{konnomachidawasaka11_heun}.

In \cite{konno11_sojourn}, Konno has introduced the notion of sojourn time for Hadamard quantum walks and has also derived analytical expressions for corresponding probability distributions, while in \cite{ampadu11_sojourna}  Ampadu has shown the inexistence of sojourn time for {\it Grover} quantum walks. Brennen {\it et al}  have presented  foundational definitions and statistics of a family of discrete quantum walks with an anyonic walker \cite{brennen10} and Lehman {\it et al} have modelled the dynamics on a non-Abelian anyonic quantum walk and found that, asymptotically, the statistical dynamics of a non-Abelian Ising anyon reduce to that of a classical random walk (i.e. linear dispersion) \cite{lehman11}. In addition, Ghoshal {\it et al} have recently reported some effects of using weak measurements on the walker probability distribution of discrete quantum walks \cite{ghoshal11}, Konno \cite{konnoito11} has proposed an It{\^o}'s formula for discrete-time quantum walks, Endo {\it et al} \cite{endo09_ballistic} have studied the ballistic behavior of quantum walks having the walker initial state spread over $N$ neighboring sites, Venegas-Andraca and Bose have studied the behavior of quantum walks with walkers in superposition as initial condition \cite{sva09}, Xue and Sanders \cite{xuesanders11_sharing} have studied the joint position distribution of two independent quantum walks augmented by stepwise partial and full coin swapping,  and Chiang {\it et al}  \cite{chiang_nagaj_wocjan_10} have proposed a general method, based on \cite{szegedy04,rudolph_grover_02}, for realizing a quantum walk operator corresponding to an arbitrary sparse classical random walk.

\subsection{Discrete quantum walks on graphs}

Quantum walks on graphs is now an established active area of research in quantum computation. Among several scientific documents providing comprehensive introductions to quantum walks on graphs, we find a seminal paper by Aharonov et al \cite{aharonov01}, a rigorous mathematical analysis and description of quantum walks on different topologies and their limit distributions by Konno \cite{konno08}, as well as introductory reviews on discrete and continuous quantum walks on graphs by Kendon \cite{kendon03a} and  Venegas-Andraca \cite{sva08}. 

In \cite{aharonov01}, Aharonov {\it et al} studied several properties of quantum walks on graphs. Their first finding consisted in proving a counterintuitive theorem: if we adopt the classical definition of stationary distribution (see \cite{sva_dphil} and references cited therein for a concise introduction on mathematical properties of Markov chains), then quantum walks do not converge to any stationary state nor to any stationary distribution. In order to review the contributions of \cite{aharonov01} and other authors, let us begin by formally introducing the following elements:

Let $G=(V,E)$ be a $d$-regular graph with $|V|=n$ (note that graphs studied here are {\it finite}, as opposed to the unrestricted line we used in the beginning of this section) and ${\cal H}_v$ be the Hilbert space spanned by states $|v\rangle$ where $v \in V$. Also, we define ${\cal H}_A$, the coin space, as an auxiliary Hilbert space of dimension $d$ spanned by the basis states $\{ |i\rangle | i \in \{1, \ldots d \} \}$, and $\hat{C}$, the coin operator, as a unitary transformation on ${\cal H}_A$. Now, we define a shift operator $\hat{S}$ on ${\cal H}_v \otimes {\cal H}_A$ such that $\hat{S}|a,v\rangle = |a,u\rangle$, where $u$ is the $a^\text{th}$ neighbour of $v$ (since edge labeling is a permutation then $\hat{S}$ is unitary.) Finally, we define one step of the quantum walk on $G$ as $\hat{U}= \hat{S}( \hat{C} \otimes \hat{I})$.

As in the study of quantum walks on a line, if $|\psi\rangle_0$ is the quantum walk initial state then a quantum walk on a graph $G$ can be defined as

\begin{equation}
|\psi\rangle_t =  \hat{U}^t |\psi\rangle_0
\label{quantum_walk_graph}
\end{equation}

Now, we discuss the definition and properties of limiting distributions for quantum walks on graphs. Suppose we begin a quantum walk with initial state $|\psi\rangle_0$. Then, after $t$ steps, the probability distribution of the graph nodes induced by Eq. (\ref{quantum_walk_graph}) is given by

\begin{definition}{\bf Probability distribution on the nodes of $G$.} Let $v$ be a node of $G$ and ${\cal H}^d$ be the coin Hilbert space. Then $$P_t (v | \psi_0) = \sum_{i \in \{ 1, \ldots, d \}} |\langle
i,v|\psi\rangle_t|^2$$ \end{definition}

If probability distributions $P_0 , P_1$ at time $0$ and $1$ are different, it can be proved that $P_t$ does not converge \cite{aharonov01}. However, if we compute the {\it average} of distributions over time

\begin{definition}{\bf Averaged probability distribution}. $$\bar{P}_t (v | \psi_0) = {1 \over T}\sum_{t=0}^{T-1} P_t (v | \psi_0)$$
\label{averaged_prob_dist}
\end{definition}

we then obtain the following result

\begin{theorem}\cite{aharonov01}. Let $|k\rangle$, $\lambda_k$ denote the eigenvectors and corresponding eigenvalues of $\hat{U}$. Then, for an initial state $|\psi\rangle_0 = \sum_k a_k |k\rangle$
$$\lim_{t \rightarrow \infty} \bar{P}_t (v | \psi_0) = \sum_{i,j,a} a_i a_j^* \langle a,v|i\rangle \langle j |a,v\rangle$$ where the sum is only on pairs $i,j$ such that $\lambda_i = \lambda_j$.
\end{theorem}

If all the eigenvalues of ${\hat U}$ are distinct, the limiting averaged probability distribution takes a simple form. Let $p_i(v) = \sum_{i \in \{ 1, \ldots, d \}} |\langle i,v| k\rangle|^2$, i.e. $p_i(v)$ is the probability to measure node $v$ in the eigenstate $|k\rangle$. Then it is possible to prove \cite{aharonov01} that, for an initial state $|\psi\rangle_0 = \sum_k a_k |k\rangle$ $\Rightarrow$ $\lim_{T \rightarrow \infty} \bar{P}_t (v | \psi_0) = \sum_i |a_i|^2 p_i(v)$. Using this fact it is possible to prove the following theorem.

\begin{theorem}\cite{aharonov01}
Let ${\hat U}$ be a coined quantum walk on the Cayley graph of an Abelian group, such that all eigenvalues of ${\hat U}$ are distinct. Then the limiting distribution $\pi$ (Def. (\ref{averaged_prob_dist})) is uniform over the nodes of the graph, independent of the initial state $|\psi\rangle_0$.
\label{theorem_uniform_averaged_militing_distribution}
\end{theorem}

\begin{figure}
\begin{center}
\scalebox{0.3}{\includegraphics{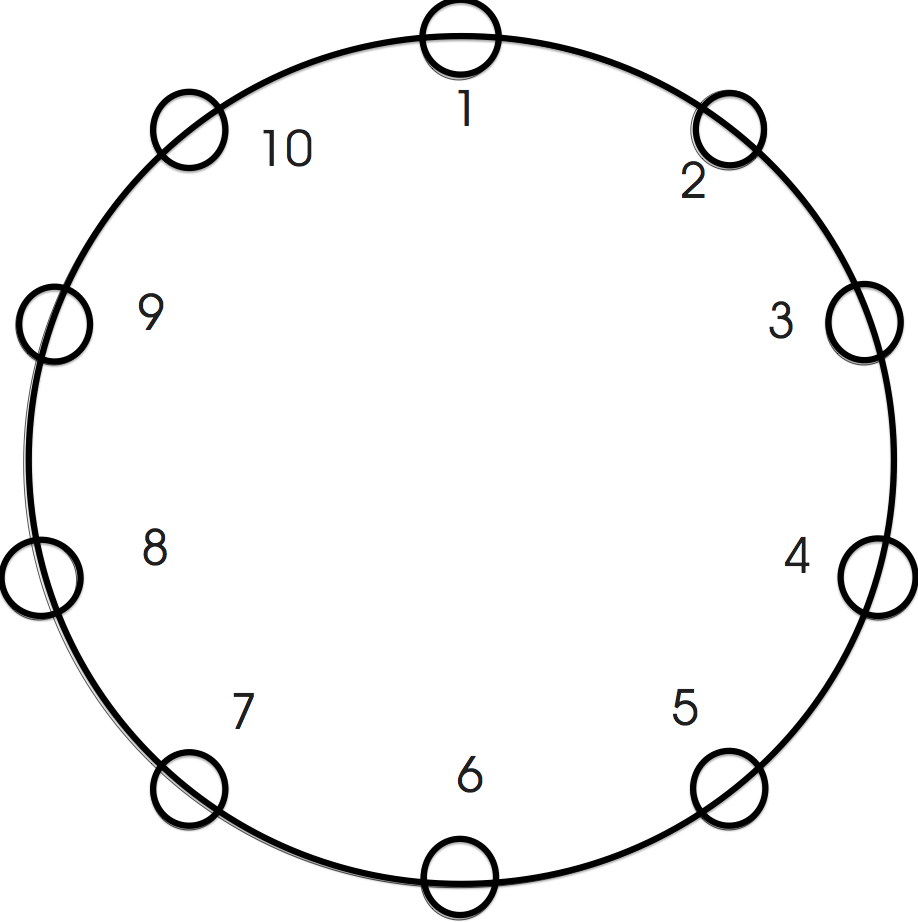}}
\end{center}
\caption{{\small Quantum walk on a cycle. A cycle is a 2-regular graph which can be viewed as a Cayley graph of the group $\mathbb{Z}_n$ with generators $1, -1$. The cycle shown in this figure has 10 vertices.}}
\label{cycle}
\end{figure}

Using Theorem \ref{theorem_uniform_averaged_militing_distribution} we compute the limiting distribution of a quantum walk on a cycle:

\begin{theorem}
Let $G_\text{cyc}$ be a cycle with $n$ nodes (see Fig. (\ref{cycle}).) A quantum walk on  $G_\text{cyc}$ acts on a total Hilbert space ${\cal H}^2 \otimes {\cal H}^n$. The limiting distribution $\pi$ for the coined quantum walk on the $n$-cycle, with $n$ odd, and with the Hadamard operator as coin, is uniform on the nodes, independent of the initial state $|\psi\rangle_0$.
\end{theorem}

Several other important results for quantum walks on a graph are delivered in \cite{aharonov01}. Among them, we mention some results on mixing times.

\begin{definition}{\bf Average Mixing time}. The mixing time $M_\epsilon$ of a quantum Markov chain with initial state $|k,v\rangle$  is given by
$$ M_\epsilon = \min \{T | \forall t \geq T \Rightarrow  || \bar{P}_t (k,v) - \pi(k,v)  || \leq \epsilon  \} $$
\label{average_mixing_time}
\end{definition}

\begin{theorem}
For the quantum walk on the $n$-cycle, with n odd, and the Hadamard operator as coin, we have
$$ M_\epsilon \leq O({ {n \log n} \over \epsilon^3}) $$
\end{theorem}

So, the mixing time of a quantum walk on a cycle is $O(n \log n)$. The mixing time of corresponding classical random walk on a circle is $O(n^2)$. Now we focus on a general property of mixing times.

\begin{theorem}
For a general quantum walk on a bounded degree graph, the mixing time is at most quadratically faster than the mixing time of the simple classical random walk on that graph.
\label{boundary_mixing_time}
\end{theorem}

So, according to Theorem \ref{boundary_mixing_time}, the speedup that can be provided by a quantum walk on a graph is not enough to exponentially outperform classical walks. Consequently, other parameters of quantum walks have been investigated, among them their {\it hitting time}. In \cite{kempe03}, Kempe offers an analysis of hitting time of discrete quantum walks on the hypercube (due to the potential service of hitting times in the construction of quantum algorithms, we shall analyze \cite{kempe03} in detail on Section \ref{qw_based_algorithms}.) Further studies on mixing time for discrete quantum walks on several graphs as well a convergence criterion for stationary distribution in {\it non-unitary} quantum walks are presented in \cite{kargin2010_mixing}.

The properties of the wave function of a quantum particle walking on a circle have been studied by Fjelds{\o}                               
{\it  et al} in \cite{fjelds88}, some details of limiting distributions of quantum walks on cycles are shown by Bednarska {\it et al} in \cite{bednarska03,bednarska04}, Liu and Petulante have presented limiting distributions for quantum Markov chains \cite{liupetulante11_markovchains}, the effect of using different coins on the behavior of quantum walks on an $n$-cycle as well as in graphs of higher degree has been studied by Tregenna {\it et al}  in \cite{tregenna03},  a standard deviation measure for quantum walks on circles is introduced by Inui {\it et al} in \cite{inui05}, and Banerjee {\it et al} have studied some effects of noise in the probability distribution symmetry of quantum walks on a cycle \cite{banerjee08}.

Another graph studied in quantum walks is the hypercube, defined by

\begin{definition}{\bf The hypercube}.  The hypercube is an undirected graph
with $2^n$ nodes, each of which is labeled by a binary string of $n$ bits. Two nodes $\vec{x},\vec{y}$ in the hypercube are connected by an edge if $\vec{x},\vec{y}$ differ only by a single bit flip, i.e. if $|\vec{x} - \vec{y}|=1$, where $|\vec{x} - \vec{y}|$ is the Hamming distance between $\vec{x}$ and $\vec{y}$.
\label{hypercube_1}
\end{definition}

In \cite{moore02}, Moore and Russell derived values for {\it the two notions} of mixing times we have studied (Defs. (\ref{instantaneous_mixing_time}) and (\ref{average_mixing_time})) for continuous and discrete quantum walks on the hypercube. As for the discrete quantum walk, \cite{moore02} begins by defining Grover's operator as coin operator.

\begin{definition}{\bf Grover's operator}.
Let ${\cal H}$ be an $n$-dimensional Hilbert space and $|i\rangle$
be the canonical basis for ${\cal H}$ and
$|\psi \rangle = {1 \over \sqrt{n}}\sum_{i=0}^{n-1} |i\rangle$.
Then we define Grover's operator as
$\hat {G} = 2|\psi \rangle \langle \psi | - \hat {I}$.
\label{grover_coin_operator}
\end{definition}
\paragraph{}

Additionally, their shift operator is given by

\begin{equation}
\hat {S} = \sum_{d=0}^{n-1} \sum_{\vec{x}} |d,\vec{x} \oplus \vec{e_d}
\rangle \langle d,\vec{x}|
\end{equation}
where $\vec{e_d}$ is the $i^\text{th}$ basis vector of the
$n$-dimensional hypercube.
So, the quantum walk on the hypercube proposed in \cite{moore02} can
be written as

\begin{equation}
|\psi\rangle_t = \hat{U}^t |\psi\rangle_0 = [\hat{S}(\hat{G} \otimes
\hat{I}_n)]^t |\psi\rangle_0
\label{qw_hypercube}
\end{equation}
for a given initial state $|\psi\rangle_0$. Using a Fourier transform approach
as in \cite{nayak00}, it was proved in \cite{moore02} that

\begin{theorem}
For the discrete quantum walk defined in Eq. (\ref{qw_hypercube}), its
instantaneous mixing time (Def. (\ref{instantaneous_mixing_time})) is
given by $t = { {k\pi} \over 4}n$, i.e. $t = O(n)$, with $\epsilon =
O(n^{-7/6})$
for all odd $k$.

\end{theorem}

Additionally, \cite{moore02} provides analytical expressions for eigenvalues and corresponding eigenvectors of the evolution operator defined in Eq. (\ref{qw_hypercube}) which were later used in \cite{shenvi02} for the design of a search algorithm based on a discrete quantum walk. 

In addition to the articles I have already mentioned, a substantial number of scientific papers has been published over the last few years. Please let me now provide a summary of more results on properties and developments on discrete quantum walks on graphs (we leave published algorithmic applications of quantum walks for section \ref{qw_based_algorithms}.)

\subsubsection{Several results on discrete quantum walks on graphs}

In \cite{mackay02}, MacKay {\it et al} present numerical simulations of quantum walks in higher dimensions using separable and non-separable coin operators,  Gottlieb {\it et al} \cite{gottlieb04} studied the convergence of coined quantum walks in $\mathbb{R}^d$, and Dimcovic {\it et al} have put forward a general framework for describing discrete quantum walks in which the coin operator is substituted by an interchange operator \cite{dimcovic11_framework}.

Kempf and Portugal \cite{kempf09} have introduced a new definition of hitting time for quantum walks that exhibit phase and group velocities, Marquezino {\it et al} \cite{marquezino10_torus} have studied and computed the mixing time and limiting distribution of a discrete quantum walk on a torus-like lattice, Leung {\it et al} \cite{leung10_percolation} have studied the behavior of coined quantum walks on 1- and 2-dimensional percolation graphs (i.e. graphs in which edges or sites are randomly missing) under two regimes: quantum tunneling employing general coin operators and the potential path redundancy present in 2-d grids, and Lovett {\it et al} \cite{lovett11_modified_search_factors} have presented a further numerical study on how dimensionality, tunneling and connectivity affect a discrete quantum-walk based search algorithm. In addition, $\check{\text{S}}$tefa$\check{\text{n}}${\'a}k {\it et al} have presented in \cite{stefanak10_revival} how eigenvalue independency from momenta imply a cyclic evolution that correspondingly leads to quantum state full revivals in two-dimensional discrete quantum walks.

On further studies on classical and quantum hitting times, in \cite{magniez10} Magniez {\it et al}: i) have presented mathematical definitions of hitting time according to Las Vegas and Monte Carlo algorithms for finding a marked element, ii) have introduced quantum analogues of such classical hitting times, and iii) have proved that, for any reversible ergodic Markov chain P, the corresponding quantum hitting time of the quantum analogue of P is of the same order as the square root of the classical hitting time of P.  Moreover, based on space-time generating functions and the mathematical methods introduced in \cite{pemantle04_asymptotics}, Baryshnikov {\it et al} have presented a mathematically rigorous and highly elegant treatment of quantum walks on two dimensions  in \cite{baryshnikov08_twoqw}, being this work followed by \cite{bressler10_qwilattice} in which Bressler {\it et al} have presented examples of results shown in \cite{baryshnikov08_twoqw} as well as derived asymptotic properties for 1-d quantum walk amplitudes. In addition, Gudder and Sorkin have presented a study of discrete quantum walks based on measure theory \cite{gudder_11} and Smith has studied graph invariants closely related to both continuous- and discrete-time quantum walks \cite{smith11_graph_invariants}.

Feldman and Hillery have studied the relationship between quantum walks on graphs and scattering theory in \cite{feldman04} as well as proposed a protocol for detecting graph anomalies using discrete quantum walks \cite{feldman10_anomalies}. Also,  Berry and Wang \cite{berry10_centrality} have analyzed, for a variety of graphs including Cayley trees, fractals and Husmi cactuses, the relationship betwen search success probability and the position of a marked vertex in such graphs, L\'{o}pez-Acevedo and Gobron \cite{lopez05} delivered an algebraic oriented analysis of quantum walks on Cayley graphs, Montanaro presented in \cite{montanaro05} a study on quantum walks on directed graphs, Krovi and Brun \cite{krovi07} have studied quantum walks (and their hitting times) on quotient graphs as well as links between those quantum walks and the group theory properties of Cayley graphs (for an extended work on this last topic, see \cite{krovi_phdthesis}.) Also, Hoyer and Meyer \cite{hoyer10_faster_transport} have presented a discrete quantum walk model for traversing a directed 1-d graph with self-loops and have found that, on this topology, the quantum walker proceeds an expected distance $\Theta (1)$ in constant time regardless the number of self-loops, Berry and Wang \cite{berry11_isomorphism} have presented a scheme for building  discrete quantum walks upon interacting and non-interacting particles and have produced two results: a numerical study of entanglement generation in such quantum walks together with a potential application on those quantum walks for testing graph isomorphism (in contrast to the results presented by Gamble {\it et al} in \cite{gamble_10_isomorphism} for continuous-time quantum walks also built upon interacting and no-interacting particles, the scheme proposed in \cite{berry11_isomorphism} can only detect some non-isomorphic strongly regular graphs.)

Resources for experimental realizations of quantum walks are costly. With this fact in mind, Di Franco {\it et al} have suggested a novel scheme for implementing a Grover discrete quantum walks on two dimensions, consisting of using a single qubit as coin (instead of using a four-dimensional quantum system) and alternating the use of such coin for motion on the $x$ and $y$ axes \cite{difranco11_mimicking}. As stated in \cite{difranco11_mimicking}, a step on this walk consists substituting the Grover operator for a sequence of two Hadamard operators on the qubit acting as coin system (one for the $x$ axis, the other for the $y$ axis), together with the movement on both $x$ and $y$ axes.  Moreover, Di Franco {\it et al} \cite{difranco11_alternate} have provided a proof of equivalence between the Grover walk and the alternate quantum walk introduced in \cite{difranco11_mimicking} as well as a limit theorem and a numerical study of entanglement generation for the alternate quantum walk, and Rohde {\it et al} have studied the dynamics of entanglement on discrete-time quantum walks running on bounded finite sized graphs \cite{rohde11_quasiperiodicity}.

Finally, Kitagawa {\it et al} \cite{kitagawa10} have shown that discrete time quantum walks can be useful for studying topological phases, Attal {\it et al} \cite{attal11_open} have proposed a formalism for modeling open quantum walk on graphs, based on completely positive maps and, in a fresh and most interesting potential application of quantum walks to engineering science, Albertini and D'Alessandro have devised the execution of quantum walks with coins allowed to change at every time step as control systems \cite{albertini09_control,albertini10_control}. In particular, Albertini and D'Alessandro have found in \cite{albertini10_control} that if the degree of of the graph $G=(V,E)$ is greater than $|V|/2$ then the quantum walk is always completely controllable.

\subsection{Continuous quantum walks}

We start by defining a continuous quantum walk so that we can use it in subsection (\ref{connection_discrete_continuous}) where we present recent advances about the mathematical bonds between discrete and continuous quantum walks, as well as in subsection \ref{qw_based_algorithms}, where we explore how this kind of quantum processes is utilized in algorithm development.

In addition to Feynman's celebrated contribution \cite{feynman86} about the simulation of quantum systems, continuous quantum walks were defined by Farhi and Gutmann \cite{farhi98}, being the latter the basis upon which Childs {\it et al} \cite{childs02} present the following formulation of a continuous classical random walk:

\begin{definition}
Let $G=(V,E)$ be a graph with $|V|=n$ then a continuous time random walk on $G$ can be described by the order $n$ infinitesimal generator matrix M given by

\begin{equation}
M_{ab}=
\begin{cases}
-\gamma, & a \neq b, (a,b) \in G  \\
0, & a \neq b, (a,b) \notin G  \\
k\gamma,& a=b\text{ and $k$ is the valence of vertex $a$}
\end{cases}
\label{generator_matrix}
\end{equation}

\end{definition}

Following \cite{childs02} and \cite{farhi98}, the probability  of being at vertex $a$ at time $t$ is given by

\begin{equation}
{{\rm d} p_a(t) \over {\rm d}t} = - \sum_{b} M_{ab} p_b(t)
\end{equation}

Now, let us define a Hamiltonian  (\cite{childs02,farhi98}) that closely follows Eq. (\ref{generator_matrix})

\begin{definition}
Let ${\hat H}$ be a Hamiltonian with matrix elements given by


\begin{equation}
\langle a | H | b \rangle= M_{ab}
\label{hamiltonian_ctqw_chapter5}
\end{equation}
\end{definition}

We can then employ Hamiltonian ${\hat H}$ as given in Eq. (\ref{hamiltonian_ctqw_chapter5}), defined in a Hilbert space ${\cal H}$ with basis $\{|1\rangle, |2\rangle, \ldots, |n\rangle   \}$, for constructing the following Schr\"odinger equation of a quantum state $|\psi\rangle \in {\cal H}$

\begin{equation}
i {{\rm d} \langle a | \psi (t) \rangle \over {\rm d}t} =  \sum_{b} \langle a | H | b \rangle \langle b | \psi(t) \rangle
\label{schrodinger_equation_ctqw_chapter5}
\end{equation}

Finally, taking Eqs. (\ref{hamiltonian_ctqw_chapter5}) and (\ref{schrodinger_equation_ctqw_chapter5}) the unitary operator ${\hat U}$

\begin{equation}
 {\hat U} = exp(-i{\hat H}t)
\label{unitary_operator_ctqw_chapter5}
\end{equation}

defines a {\bf continuous quantum walk} on graph $G$. Note that the continuous quantum walk given by Eq. (\ref{unitary_operator_ctqw_chapter5}) defines a process on continuous time and discrete space.

Since the publication of \cite{farhi98,childs02}, there has been an increasing number of publications with relevant results of continuous quantum walks. We now provide a summary of more results on this area.

In \cite{konno05_limit_cont}, Konno has proved the weak limit theorem for continuous quantum walks presented on Theorem \ref{teorema_konno_limite_debil}. Also, in \cite{varbanov08} Varbanov {\it et al} present a definition of hitting time for continuous quantum walks, based on performing measurements on the walker at Poisson-distributed random times; moreover, they have proved that, depending on the measurement rate, continuous quantum walks may or may not have infinite hitting times. Xu \cite{xu08_cqw_regular_networks} has derived transition probabilities and computed transport velocity in continuous quantum walks on ring lattices, Xu and Liu \cite{xu08_cqw_erdos_networks} have studied quantum and classical transport on both finite and infinite versions of  Erd{\H{o}}s-R{\'e}nyi networks while Agliari {\it et al}, motivated by recent advances on quantum transport phenomena on photosynthesis, have studied trapping processes in rings and shown that carrying trap configuration leads to changes in quantal mean survival probability \cite{agliarimulkenblumen10_trapping}. Also, Agliari {\it et al} \cite{agliariblumenmulken08_restricted} have studied the average displacement of quantum walker on Gasket, Cayley tree and square torus graphs, Agliari \cite{agliari11_erdos_cqw} has studied coherent transport models with traps on Erd{\H{o}}s-R{\'e}nyi graphs, Tsomokos has investigated the properties of continuous quantum walks on complex networks with community structure \cite{tsomokos11_complex_networks_cqw}, and Salimi and Jafarizadeh have studied both classical and continuous quantum walks on several Cayle graphs \cite{salimi09_classicalquantum}  and spidernet graphs \cite{salimiqip10_spidernetcqw}. A review on models for coherent transport on complex networks has been recently  published by O. M{\"{u}}ken and A. Blumen in \cite{mulken11_review_cqw}. Furthermore, Kargin \cite{kargin10_homogeneous_trees_cqw} has calculated the limit of average probability distribution for nearest-neighbor walks on $\mathbb{Z}^d$ and infinite homogeneous trees, Rosmanis \cite{rosmanis11_snake} has introduced quantum snake walks (i.e. continuous quantum walks with fixed-length paths) on graphs, Godsil and Guo \cite{godsil11_eigengraph} have analyzed the properties of transition matrix of continuous quantum walks on regular graphs, and Kieferov\'a and Nagaj have analyzed the evolution of continuous quantum walks on necklaces \cite{kieferova11_necklace}.

Mixing and hitting times as well as the structure of probability distributions and transitions probabilities have been analyzed in this field. Analytical expressions of transition probabilities on star graphs have been presented by Xu in \cite{xu09_star_analytical} and Godsil has proposed some properties of average mixing of continuous quantum walks \cite{godsil11_averagecqw}, while Salimi \cite{salimi08_star} has produced a version of the central limit theorem for  continuous quantum walks also on star graphs, Inui {\it et al} have proposed both instantaneous uniform mixing property and temporal standard deviation for continuous-time quantum random walks on circles \cite{inui04_cqw_circle_mix}, Best {\it et al} have studied instantaneous and uniform mixing of continuous quantum walks on generalized hypercubes \cite{best08_mixcqw}, Drezgich {\it et al} \cite{drezgich09_characterization} have characterized the mixing time of continuous quantum walks on the hypercube under a Markovian decoherence model, Salimi and Radgohar have also analyzed effects of decoherence on mixing time in cycles \cite{salimiradgohar10_mixingtimecqw}, and Anishchenko {\it et al} have studied how highly degenerate eigenvalue spectra impact the quantum walk spreading on a star graph  \cite{anishchenko11_cqw_spreading}.

Motivated by the power-law ditribution exhibited by real world networks showing scale-free characteristics, Ide and Konno have studied the evolution of continuous quantum walks on the threshold network model \cite{ide10_threshold_cqw}, Salimi and Sorouri \cite{salimi10_pseudohermitian} have introduced a model of continuous quantum walks with non-Hermitian Hamiltonians, and Bachman {\it et al} have studied how perfect state transfer can be achieved on quotient graphs \cite{bachman11_transfer}. Finally, we report the works of Konno on continuous time quantum walks on ultrametric spaces \cite{konno06a} and continuous quantum walks  on trees in quantum probability theory \cite{konno06b}, de Falco {\it et al} on speed and entropy of continuous quantum walks \cite{defalco06}, M\"ulken {\it et al} on quantum transport on small-world networks \cite{mulken07}, and Jafarizadeh {\it et al} on studying continuous time quantum walks by using the Krylov subspace-Lanczos algorithm \cite{jafarizadeh07}.

\subsection{Whether discrete or continuous: is it quantum random walks or just quantum walks?}

Randomness is an inherent component of every single step of a classical random walk. In other words, there is no way to predict step $s_i$ of a classical random walk, no matter how much information we have about previous steps $s_{i-1}, s_{i-2}, \ldots, s_1, s_0$. We can only tell the probability associated to each possible step $s_{i+1}^j$.

On the other hand, if we carefully analyze quantum evolution in discrete (unitary operator) and continuous (Schr\"odinger equation) versions, we shall convince ourselves of the fact that quantum evolution is deterministic, i.e. for each computational step denoted by $|\psi\rangle_{i}$ we can  always tell the exact description of step $|\psi\rangle_{i+1}$, as $|\psi\rangle_{i+1}= {\hat U}|\psi\rangle_{i}$.

So, what is random about a quantum walk? Why are quantum walks candidates for developing quantum counterparts of stochastic algorithms? The answer is: randomness comes as a result of either decoherence or measurement processes on either quantum walker(s) and/or quantum coin(s). So, decoherence and quantum measurement allow us to introduce randomness into a quantum walk-based algorithm. Moreover, we are not restricted to introducing chance only at the end of the quantum algorithm execution as we can also exploit several measurement strategies in order to manipulate quantum systems and produce probability distributions suitable for their use in advantageous algorithms; for example, see the \lq top hat' probability distribution \cite{kendon07}, a quasi-uniform distribution created by running a discrete quantum walk and performing measurements on its constituent elements (or, alternatively, allowing such constituent particles to have some interaction  with the environment.)

\subsection{\label{connection_discrete_continuous}How are continuous and discrete quantum walks connected?}

The mathematical models of discrete and continuous quantum walks studied in the previous sections present a serious problem: it is not clear how to transform discrete quantum walks into continuous quantum walks and vice versa. This is an important issue for two reasons: {\bf 1)} in the classical case, discrete and continuous random walks are connected  via a limit process, and {\bf 2)} it is not natural/elegant to have two different kinds of quantum diffusion, one of them with an extra particle (the quantum coin) with no clear connection between them. 

\begin{enumerate}

\item {\bf Strauch's contribution}

In \cite{strauch06}, F.W. Strauch presents a connection between discrete and continuous quantum walks. He starts by using a simplification \cite{childs02} of the continuous quantum walk defined by Eq. (\ref{schrodinger_equation_ctqw_chapter5}), namely

\begin{equation}
{\bf {\hat H}}|j\rangle = -\gamma (|j-1\rangle -2|j\rangle+|j+1\rangle)
\label{continuous_strauch}
\end{equation}
which in \cite{strauch06} is rewritten as

\begin{equation}
i \partial_t \psi(n,t)= -\gamma (\psi(n+1,t) -2 \psi(n,t) + \psi(n-1,t) )
\end{equation}
where $\psi(n,t)$ is a complex amplitude at the continuous time $t$ and the discrete lattice position $n$.

Then, \cite{strauch06} uses results from \cite{aharonov93} and \cite{meyer95} to build a discrete quantum walk represented by the following unitary mapping 

\begin{subequations}
\begin{equation}
\psi_R(n,\tau+1) = ({\text cos} \theta)  \psi_R(n-1,\tau) - (i {\text sin}\theta) \psi_L(n-1,\tau)
\label{discrete_strauch_01}
\end{equation}

\begin{equation}
\psi_L(n,\tau+1) = ({\text cos} \theta)  \psi_L(n+1,\tau) - (i {\text sin}\theta) \psi_R(n+1,\tau)
\label{discrete_strauch_02}
\end{equation}
\end{subequations}
where $\psi_R(n,\tau)$ and $\psi_L(n,\tau)$ are complex amplitudes at the discrete time $\tau$ and discrete lattice position $n$.

Strauch's result  focuses on building a unitary transformation ${\hat U}=exp(-i{\bf {\hat H}}t)$  that allows us to transform Eqs.(\ref{discrete_strauch_01}) and (\ref{discrete_strauch_02}) into Eq. (\ref{continuous_strauch}). There are several important conclusions from the developments shown in \cite{strauch06}:
\\
1. It is indeed possible to transform a discrete quantum walk into a continuous one by means of a limit process (although this is not a straightforward derivation.)
\\
2. Strauch's derivation does not use any coin degree. Thus \cite{strauch06} agrees, from an new perspective, with Patel {\it et al} \cite{patel05} with respect to the irrelevance of the coin degree of freedom in order to obtain the statistical enhancements ($\sigma^2 = O(n))$ that discrete quantum walks show.
\\

\item {\bf Child's contribution}

In \cite{childs10_equivalence}, Childs presents the following mathematical framework for simulating a continuous quantum walk as a limit ($\epsilon-approximation$) of discrete quantum walks (for the sake of clarity and readability of the original paper, we closely follow the notation used in \cite{childs10_equivalence}):

\begin{enumerate}
\item
Let $H$ be a general $N \times N$  Hermitian matrix. We now define a set of $N$ quantum states $|\psi_1\rangle,\ldots,|\psi_N\rangle \in \mathbb{C}^N \otimes \mathbb{C}^N$ as
\begin{align}
 |\psi_j\rangle &:= \frac{1}{\sqrt{||\text{abs}(H)||}} \sum_{k=1}^N \sqrt{H^*_{jk} \frac{d_k}{d_j}} \, |j,k\rangle. 
\label{jstates}
\end{align}

where $\text{abs}(H):= \sum_{j,k=1}^N |H_{jk}| \, |j\rangle\langle k|$ denotes the elementwise absolute value of $H$ in an orthonormal basis $\{|j\rangle: j=1,\ldots,N\}$ of $\mathbb{C}^N$

\item
Define the isometry
\begin{equation}
  T:=\sum_{j=1}^N |\psi_j\rangle\langle j|
\label{isometry}
\end{equation}
 mapping $|j\rangle \in \mathbb{C}^N$ to $|\psi_j\rangle \in \mathbb{C}^N \otimes \mathbb{C}^N$

\item
Enlarge the Hilbert space by building a new set of quantum states from Eq. (\ref{jstates})  to
\begin{equation}
  |\psi_j^\epsilon\rangle := \sqrt{\epsilon} |\psi_j\rangle + \sqrt{1-\epsilon} |{\perp_j}\rangle
\end{equation}
for some $\epsilon \in (0,1]$ and $|{\perp_j}\rangle$ as defined in Eq. (25) of \cite{childs10_equivalence}

\item
From Eq. (\ref{isometry}), build a modified isometry
\begin{equation}
T_\epsilon := \sum_j|\psi_j^\epsilon\rangle \langle j|
\label{modified_isometry}
\end{equation}

\item
Now, given an initial state $|\Psi_0\rangle \in \text{span}\{|j\rangle| j \in \{1,2, \ldots, n \}  \}$ apply the modified isometry given in Eq. (\ref{modified_isometry}) and the operation 
$\frac{1+iS}{\sqrt{2}}$, where $S$ is the swap operator.

\item
Apply $n$ steps of the discrete quantum walk $U = i S(2T_\epsilon T_\epsilon^\dag-1)$ and, finally,

\item Project onto the basis of states $\{\frac{1+ i S}{\sqrt 2}T_\epsilon|j\>: j=1,\ldots,N\}$.

\end{enumerate}

In addition to this protocol, Childs also presents in \cite{childs10_equivalence} a notion of query complexity for continuous-time quantum walk algorithms as well as a continuous-time quantum walk algorithm for solving the distinctness problem \cite{ambainis03}, a problem that was originally solved using a discrete quantum walk-based algorithm by Ambainis \cite{ambainis03}.
\\

\item
As a third contribution to state and clarify the relationships between different models of quantum walks, there are two formulations for {\it discrete} quantum walks: coined \cite{nayak00,ambainis01} and scattering \cite{hillery07,feldman07}. In \cite{andrade09_dqwequivalence}, Andrade and da Luz present a general framework for unitary equivalence of both discrete quantum walk models.

\end{enumerate}

\subsection{\label{quantumness}Are quantum walks really quantum? }

The results presented so far in this review show that superposition and, consequently, interference play an important role in the structure and properties of discrete quantum walks. However, interference is also a characteristic of classical physical systems, like electromagnetic waves. Thus, it makes sense to scrutinize whether the statistical and computational properties of quantum walks are really due to their quantum nature or not. 

Arguments in favor of the plausibility of using classical physics for building experiments which replicate some interference and statistical properties of  quantum walks on a line are given in \cite{jeong04,knight03,knight03b,knight04}, where it was shown that it is possible to develop implementations of a quantum walk on a line  purely described by classical physics (wave interference of  electromagnetic fields) and still be able to reproduce  the variance enhancement that characterizes a discrete quantum walk. For example, the implementation proposed in \cite{knight03b}  utilizes the frequency of a light field as walker and the spatial path or the polarization state of the same light field as the coin.

Arguments in favor of the quantum mechanical nature of quantum walks have been provided by, among others, Kendon and Sanders \cite{kendon05} who showed it would still be necessary to have a quantum mechanical description of such an implementation in order to account for two properties of a quantum walk with one walker: i) the indivisibility of the quantum walker,  and ii) complementarity, which in quantum computation jargon may  be stated as follows: {\it the trade-off between interference  and information about the path followed by the walker (knowing the path followed by a quantum particle decreases the sharpness of the interference pattern \cite{wootters79,kendon07}.)} Furthermore, Romanelli {\it et al} showed in \cite{romanelli03,romanelli04} that the evolution equation of a  quantum walk on a line can be separated into two parts: Markovian and interference terms, and that the quadratic increase in the variance of the quantum walker is a consequence of quantum evolution.

Thus it seems that if we are only interested in some statistical properties of one-walker quantum walks on a line, like its variance enhancement with respect to classical random walks, we could do with either classical or quantum experimental setups. However, the quantum  mechanical nature of walkers and/or coins play an important role in the following cases:

\begin{enumerate}

\item From a purely physical point of view, if one is interested in using quantum walks for testing the quantumness of a quantum computer realization, complementarity would be a very helpful resource as it is a property of quantum mechanical systems that cannot be exactly reproduced in a classical experiment. A similar argument would be applied in the case of  using complementarity as a computational resource.

\item Including more walkers (e.g. \cite{paunkovic04,sva09,mayer11,stefanak11} and/or coins (e.g. \cite{sva05,liu09}) opens up the possibility of detecting, quantifying and harnessing quantum-mechanical properties for information processing purposes. In particular, quantum entanglement has been incoporated into quantum walks research either as a result of performing a quantum walk or as a resource to build new kinds of quantum walks. Since entanglement  is a key component in quantum computation, it is worth keeping in mind that quantum walks can be used either as entanglement generators or as computational processes taking advantage of this quantum mechanical property.  A brief summary of results on quantum walks and entanglement is delivered in subsection \ref{entanglementdiscretewalks}.

\item Genuine quantum computers will be an excellent (and most likely, indispensable) tool to execute exact and efficient simulations of quantum systems (e.g. \cite{feynman82,feynman86,aspuru05,kassal08,kassal11,mohseni08}.)

\end{enumerate}

\subsubsection{\label{entanglementdiscretewalks}Entanglement in quantum walks} 

Carnerio {\it et al} have numerically investigated the variation in entanglement between coin(s) and walker on unrestricted line, trees, and cycles \cite{carneiro05}, conjecturing that for all coin initial states of a Hadamard walk, the entanglement has 0.872 as its limiting value. In \cite{abal05}, Abal {\it et al} have analytically proved this last result. In fact, studying asymptotical behavior of entanglement in various settings is a fruitful research topic: In  \cite{abal07}, Abal {\it et al} have studied the long-term behavior of entanglement for two walkers using non-local coin operators, Venegas-Andraca {\it et al} numerically showed asymptotical properties (particularly the \lq three peak localization phenomenon')  of quantum walks with  entangled coins \cite{sva05} that later on were analytically proved by Liu and Petulante ( the \lq three peak localization phenomenon'  reflects the degeneracy of some eigenvalue of the quantum walk evolution operator) \cite{liu09}. Furthermore, Liu \cite{liu11b} has derived analytical expression for position limit distributions on quantum walks with generalized entangled coins, Annabestani {\it et al}  gave an exact characterization of asymptotic entanglement in $\mathbb{Z}^2$ \cite{annabestani10}, and Ide {\it et al} have produced analytical expressions for limit distributions of Shannon and von Neumann entropies on a one-dimensional quantum walk \cite{ide11}.

Also, Omar {\it et al} have produced several position probability distributions of quantum walks with entangled walkers (fermions and bosons)  \cite{paunkovic04}, Endrejat and B{\"{u}}ttner have presented a multi-coin scheme in order to analyze the effect of entanglement in the initial coin state \cite{endrejat05_dqw_coin}, Pathak and Agarwal \cite{pathak08_qrk_photon} have argued that entanglement generation in discrete-time quantum walks is a physical resource that cannot be exactly reproduced by classical systems, Goyal and Chandrashekar \cite{goyal10} have numerically studied spatial entanglement in $M$-particle quantum walks using the Meyer-Wallach multipartite entanglement measure  \cite{meyer02}, $\check{\text{S}}$tefa$\check{\text{n}}${\'a}k {\it et al} have investigated non-classical effects (directional correlations) in quantum walks with two walkers with $\delta$ interaction \cite{stefanak11}, Ampadu has studied directional correlations among $M$ particles with $\delta$ interaction on a quantum walk on a line \cite{ampadu11e}, and Peruzzo {\it et al} have provided experimental demonstrations of quantum correlations that violate a classical limit by $76$ standard deviations \cite{peruzzo10}. Furthermore, Chandrashekhar has introduced the idea of generating entanglement between two spatially-separated systems using the entanglement generated while performing a discrete quantum walk as a resource \cite{chandrashekar10e} ,  All\'es {\it et al} \cite{alles11} have introduced a shift operator for discrete quantum walks with two walkers which provides conditions for (not highly probable) maximal entanglement generation,  Salimi and Yosefjani \cite{salimi10} have studied the asymptotical behavior of coin-position entanglement under a time-dependent coin regime, and Ampadu \cite{ampadu11c} has proposed limit theorems for the von Neumann and Shannon entropies of discrete quantum walks on $\mathbb{Z}^2$.

Finally, Maloyer and Kendon have numerically calculated the impact of decoherence in the entanglement between walker and coin for quantum walks on a line and on a cycle \cite{maloyer07}, Chandrashekar \cite{chandrashekar06_retain} has proposed a modified discrete-time quantum walk in which the coin toss is no longer needed, Ampadu \cite{ampadu11d} has analyzed the impact of decoherence on the quantification of mutual information in a square lattice, Rohde {\it et al} have studied the dynamical behavior of entanglement on quantum walks running on bounded linear graphs with reflecting boundaries, together with a scheme for realizing their proposal on a linear optics setting \cite{rohde11_quasiperiodicity}, and Romanelli \cite{romanelli10} has defined a global chirality probability distribution (GCD) independent of the walker's position and has proved that GCD converges to a stationary solution.

\subsection{\label{experimental_realizations}Experimental proposals and realizations of quantum walks}

In \cite{roldan05}, Rold\'an {\it et al} have proposed an experimental set-up based on classical optical devices to implement a discrete quantum walk. This is a remarkable result that provide  grounds, together with  \cite{jeong04,knight03,knight03b,knight04}, to reflect on what exactly is quantum when working on the physical and computational properties of quantum walks (more on this on subsection \ref{quantumness}.) Moreover, Rai {\it et al} study the quantum walk of nonclassical light in an array of coupled wave guides \cite{rai08}, Schreiber {\it et al} present a realization of a 5-step quantum walk on passive optical elements \cite{schreiber10} and Zhang {\it et al} have put forward a scheme for implementing quantum walks on spin-orbital angular momentum space of photons \cite{zhang10}.  Also, Rohde {\it et al} have introduced a formal framework for distinguishable and indistinguishable multi-walker quantum walks on several lattices, together with a proposal for implementing such framework on quantum optical settings \cite{rohde11c}, Solntsev {\it et al} have analyzed links between parametric down conversion and quantum walk implementations \cite{solntsev11_spdcqw}, Broome {\it et al} have implemented a discrete quantum walk using single photons in space \cite{broome10_tunable}, Witthaut has explored how the dynamics of spinor atoms in optical lattices can be used for implementing a quantum walker \cite{witthaut10}, van Hoogdalem and Blaauboer introduced the idea of implementing quantum walk step operator in a one-dimensional chain of quantum dots \cite{vanHoogdalem09}, and Souto Ribeiro {\it et al} have presented an implementation of a quantum walk step at single-photon level produced by parametric down-conversion \cite{ribeiro08_photon}. 

Skyrmions are solitons in nonlinear field theory that, as the magnetic field increases, the Skyrmion radius decreases and suddenly shrinks to zero by emitting spin waves. This last phenomenon is known as the Skyrmion burst. In \cite{ezawa11}, Ezawa has proposed to use the remnants of a Skyrmion burst to implement several continuous-time quantum walkers.
In \cite{owens11}, Owens {\it et al} present the architecture of an optical chip with an array of waveguides in which they have implemented a two-photon continuous quantum walk. In \cite{oka05}, Oka {\it et al} show that the Landau-Zener transitions induced in electron systems due to strong electric fields can be mapped to a quantum walk on a lattice,
Hamilton {\it et al} have proposed an experimental setup of a four-dimensional quantum walk using the polarization and orbital angular momentum of a photon \cite{hamilton11}, and K\'alm\'an {\it et al}  have presented a scheme for implementing a coined quantum walk using the ballistic transport of an electron through a series of quantum rings \cite{kalman09}. Indeed, the abundance of experimental proposal and realizations of quantum walks based on optical devices may be a glimpse to future implementations of universal quantum computers \cite{rohde11}. 

Based on the results presented by Xue and Sanders in \cite{xue08a} about the behavior of quantum walks in circle in phase space, Xue {\it et al} have suggested an implementation of quantum walks on circles using superconducting circuit quantum electrodynamics \cite{xue08b}, Manouchehri and Wang proposed implementations of quantum walks on Bose-Einstein condensates \cite{manouchehri08} and quantum dots \cite{manouchehri09},  Xue {\it et al} suggest that a multi-step quantum walk using generalized Hadamard coins may be realized using an ion trap \cite{xue09} while Schmitz {\it et al} have indeed implemented a proof of principle of a quantum walk in a linear ion trap \cite{schmitz09} and Matjeschk {\it et al} have presented an experimental proposal for quantum walks in trapped ions \cite{matjeschk11_ions}. Karski  {\it et al}  have implemented a quantum walk on the line with single neutral atoms by delocalizing them over the sites of a one-dimensional spin-dependent optical lattice \cite{karski09}, Lavi${\check{\text{c}}}$ka {\it et al} have proposed a quantum walk implementation using non-ideal optical multiports \cite{lavicka11_jumps},   and Z{\"{a}}hringer {\it et al} have experimentally demonstrated a 23-step quantum walk on a line in phase space using one and two trapped ions \cite{zahringer10}.

Lahini {\it et al} have studied the dynamics of a two-boson quantum walk on a lattice \cite{lahini11_bosons}, Sansoni {\it et al} have experimentally studied the effect of particle statistics in two-particle coined quantum walks \cite{sansoni11_bosonic}, Mayer {\it et al} have studied the correlations that can be found in a quantum walk built upon interacting and non-interacting particles \cite{mayer11}, and Peruzzo {\it et al} have observed quantum correlations on photons generated using parametric-down conversion techniques and have experimentally found that such correlations critically depend on the actual quantum walk input state \cite{peruzzo10}. Finally, Ahlbrecht {\it et al} have investigated how to use a two-atoms system for executing a quantum walk \cite{ahlbrecht11_boundmolecules}, Regensburger {\it et al} have experimentally shown how a coupled fiber system could be used to implement a quantum walk \cite{regensburger11_zitter}, and Matsuoka {\it et al} have proposed a scheme to implement a continuous-time quantum walk on a diatomic molecule \cite{matsuoka11_cascade}.

\section{\label{qw_based_algorithms} Algorithms based on quantum walks and classical simulation of quantum algorithms-quantum walks}

Let us start with a catchy sentence: efficient search is a Holy Grail in computer science. Indeed, in addition to being searching a core topic in undergraduate and graduate computer science education, many open problems and challenges in both theoretical and applied computer science can be formulated as search problems (e.g. optimization problems, typically within the sphere of NP-hard problems \cite{sipser05,papadimitriou95}, can be seen as \lq detect and/or identify object(s)' problems whose solutions ask for search algorithms.)  Thus, a great deal of efforts and resources have been devoted to build both classical and quantum algorithms for solving a variety of search problems. In particular, due to the central role played by classical random walks in the development of successful stochastic algorithms,  there has been a huge interest in understanding the computational properties of quantum walks over the last few years. Moreover, the development of sucessful quantum-walk based algorithms and the recent proofs of computational universality of quantum walks \cite{childs09,lovett10,underwood10} have boosted this area.

A general strategy for building an algorithm based on quantum walks includes choosing:  a) the unitary operators  for discrete quantum walks or the Hamiltonians for continuous quantum walks, that will be employed to determine the time evolution of the quantum hardware, b) the measurement operators that will be employed to find out the position of the walker and, possibly c) decoherence effects if required for controlling the quantum walk algorithmic effects (e.g. manipulating probability distributions) or mimicking natural phenomena (e.g. \cite{mohseni08}.)

The quantum programmer must bear in mind that the choice of evolution and measurement operators, as well as initial quantum states and (possibly) decoherence models, will determine the shape and other properties of the resulting probability distribution for the quantum walker(s). Moreover, a computer scientist interested in algorithms based on quantum walks must keep in mind that, due to the no-cloning theorem \cite{dieks82_nocloning,wootters82_nocloning}, making copies of arbitrary quantum states is not possible in general thus copying variable content is not allowed in principle. Indeed, it is possible to use cloning machines for imperfect quantum state copying, but it would frequently translate into computational and estimation errors. Since any non-reversible gate can be converted into a reversible gate \cite{bennett73,nielsen00,abramsky05}, errors due to imperfect quantum state cloning are unneccessary and consequently must be avoided. Employing classical computer simulators of quantum walks \cite{our_simulator_08,others_simulators_08} can be a fruitful exercise in order to figure out the operators and initial states required for algorithmic applications of quantum walks (more on classical simulation of quantum algorithms in subsection \ref{classical_simulation}.)
   
Quantum algorithms based on either discrete or continuous quantum walks are built upon detailed and complex mathematical structures and it is not possible to cover all details in a single review paper. Therefore, we shall devote this section to review the fundamental links between quantum walks and computer science (mainly algorithms) and we strongly recommend the reader to go to both the references provided in this section, as well as to the introductions and reviews of quantum walk-based algorithms that can be found in \cite{kempe04,ambainis04a,kendon07,ambainis08_sofsem,santha08,ambainis10_mfcs,konno08,sva08}.

\subsection{Algorithms based on discrete quantum walks}

Let us start by defining an abstract object frequently used in quantum algorithms: an oracle.

\begin{definition}{\bf Oracle}. An oracle is an abstract machine used to study decision problems. It can be thought of as a black box which is able to decide certain decision problems in a single step, i.e. an oracle has the ability to {\it recognize} solutions to certain problems.
\label{oracle_computation}
\end{definition}

An oracle is a mathematical device built to simplify the actual process of algorithm development. Unfortunately, the name \lq oracle' does not help much as it seems to invoke metaphysical entities and powers. However, the nature of an oracle is just that of any other function or procedure: it is defined in terms of what mathematical operations are performed both in terms of computability and complexity \cite{lanzagorta09}. 

Oracles are widely used in classical algorithm design. In the context of quantum computation, we also use oracles to {\it recognize} solutions for the search problem. Additionally, we assume that if an oracle recognizes a solution $|\phi\rangle$ then that oracle is also capable of computing a function with $|\phi\rangle$ as argument \cite{nielsen00,gruska99,lanzagorta09}.

We are interested in searching for $M$ elements in a space of $N$ elements. To do so, we use an index $x \in S$, where $S =\{0,1, \ldots, N-1 \}$, to enumerate those elements. We also suppose we have a function $f:S \rightarrow \{ 0,1\}$ such that $f(x) = 1$ if and only if $x$ is one of the elements we are looking for. Otherwise, $f(x) = 0$. An oracle can be written as a unitary operator ${\hat O}$ defined by

\begin{equation}
{\hat O}(|x\rangle |q\rangle) = |x\rangle |q \oplus f(x)\rangle
\end{equation}
\\
where $|x\rangle$ is the index register, $\oplus$ is addition modulo $2$ (the XOR operation in computer science parlance) and the oracle qubit $|q\rangle$ is a single qubit which is flipped if $f(x) = 1$ and is left unchanged otherwise. As shown in \cite{nielsen00}, we can check whether $x$ is a solution to our search problem by preparing $|x\rangle$, applying the oracle, and checking whether the oracle qubit has been flipped to $|1\rangle$. Grover's algorithm \cite{grover96}, as well as several algorithms we shall review in this section, make use of an oracle. A comparison of quantum oracles can be found in \cite{kashefi02}.

We now proceed to review quantum algorithms based on discrete quantum walks. Let us introduce the following problem:

\begin{definition}{\bf Searching in an unordered list}. 
Suppose we have an unordered list of $N$ items labeled $x_1, x_2, \ldots, x_N$. We want to find one of those elements, say $x_i$.
\label{search_problem}
\end{definition}

Any classical algorithm would take $O(N)$ steps at least to solve the problem given in Def. (\ref{search_problem}). However, one of the jewels of quantum computation, Grover's search algorithm \cite{grover96}, would do much better.  By using an oracle and a technique called {\bf Amplitude Amplification}, the search algorithm proposed in \cite{grover96} would only take $O(\sqrt{N})$ time steps to solve the same search problem. In addition to its intrinsic value for outperforming classical algorithms, Grover's algorithm has relevant applications in computer science, including solutions to the 3-SAT problem \cite{ambainis04a}.

In \cite{shenvi02}, Shenvi {\it et al} proposed an algorithm based on a discrete quantum walk to solve the search problem given in Def. (\ref{search_problem}). For the sake of completeness and in order to present the results contained in \cite{shenvi02}, let us remember the definition of a hypercube (Def. \ref{hypercube_1}).

\begin{definition}{\bf The hypercube}.  The hypercube is an undirected graph with $2^n$ nodes, each of which is labeled by a binary string of $n$ bits. Two nodes $\vec{x},\vec{y}$ in the hypercube are connected by an edge if $\vec{x},\vec{y}$ differ only by a single bit flip, i.e. if $|\vec{x} - \vec{y}|=1$, where $|\vec{x} - \vec{y}|$ is the Hamming distance between $\vec{x}$ and $\vec{y}$. As an example, the 3-dimensional hypercube is shown in Fig. \label{hypercube_definition}
\end{definition}

\begin{figure}
\begin{center}
\scalebox{0.5}{\includegraphics{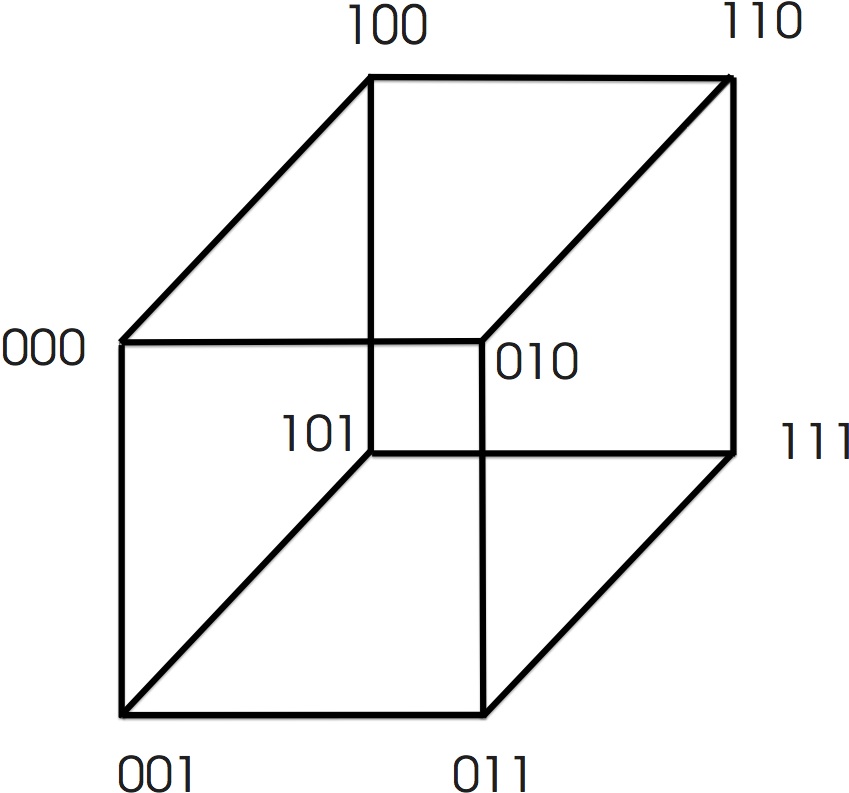}}
\end{center}
\caption{{\small A 3-dimensional hypercube. Nodes are labeled following the formula $d \oplus e_d$ where $d \in \{000,001, 010, 011, 100, 101, 110, 111  \}$ and $e_d \in \{001, 010, 100 \}$.}}
\label{hypercube_3d}
\end{figure}

An example of a 3-dimensional hypercube can be seen in Fig. (\ref{hypercube_3d}). Since each node of the hypercube has degree $n$ and there are $2^n$ distinct nodes then the Hilbert space upon which the discrete quantum walk is defined is  ${\cal H} = {\cal H}^n \otimes {\cal H}^{2^n}$, and each state $|\psi\rangle \in {\cal H}$ is described  by a bit string ${\vec x}$ and a direction $d$. We now define the following coin and shift operators

\begin{equation}
{\hat C}= {\hat C}_0 \otimes {\hat I} = (- {\hat I} + 2|s^c \rangle \langle s^c|) \otimes {\hat I}
\end{equation}
where $|s^c\rangle $ is the equal superposition over all $n$ directions, i.e. $|s^c\rangle = {1 \over \sqrt{n}}\sum_{d=1}^{n} |d\rangle$, and

\begin{equation}
{\hat S} = \sum_{d=0}^{n-1} \sum_{\vec x} |d,{\vec x} \otimes {\vec e}_d\rangle \langle d,{\vec x}|
\end{equation}

where $|{\vec e}_d\rangle$ is the $d^{th}$ basis vector of the hypercube. Using the eigenvalues and eigenvectors of the evolution operator $\hat {U}={\hat S}{\hat C}$ of the quantum walk on the hypercube \cite{moore02} in order to build a slightly modified  coin operator $\hat{C}'$ (which works within the algorithm structure as an oracle (Def.(\ref{oracle_computation})))  and an evolution operator $\hat {U}'$, and by collapsing the hypercube into a line, the quantum walk designed by evolution operator $\hat{U}'$ is used to search for element $x_\text{target} \in \{0,1 \}^n$.

It is claimed in \cite{shenvi02} that, after applying $\hat{U}'$ a number of  $t_f = {\pi \over 2} \sqrt{2^n} = O(\sqrt{N})$ times, the outcome of their  algorithm is $x_\text{target}$ with probability ${1 \over 2} - O({1 \over n})$.  A summary of similarities and differences between this quantum walk algorithm and  Grover's algorithm can be found in the last pages of \cite{shenvi02}, G\'{a}bris {\it et al} \cite{gabris07} studied the impact of noise on the algorithmic  performance given in \cite{shenvi02} using a scattering quantum walk \cite{hillery07}, Lovett {\it et al} \cite{lovett10_portugal} have numerically studied the behavior of the algorithm presented in \cite{shenvi02} on different two-dimensional lattices (e.g. honeycomb lattice), and Poto\ifmmode \check{c}\else \v{c}\fi{}ek {\it et al} \cite{potocek09_qwsearch} have introduced strategies for improving  both success probability and query complexity computed in \cite{shenvi02}.

Now, let us think of the following problem: we have a hypercube as defined in Def. (\ref{hypercube_definition}) and we are interested in measuring the time (or, equivalently, the number of steps) an algorithm would take to go from node $i$ to node $j$, i.e. its {\it hitting time} (Def. (\ref{hitting_time_classical})). Since defining the notion of hitting time for a quantum walk is not straightforward, Kempe \cite{kempe03} has proposed the following definitions

\begin{definition}{\bf One-shot hitting time}.
A quantum walk $U$ has a $(T,p)$ one-shot $(|\phi_0\rangle,|x\rangle)$ hitting time if the probability to measure state $|x\rangle$ at time $T$ starting in $|\phi\rangle_0$ is larger than $p$, i.e. $|| \langle x | U^T | \phi_0 \rangle ||^2 \geq p$.
\label{one_shot_ht}
\end{definition}

\begin{definition}{\bf $|x\rangle$- stopped walk}.
A $|x\rangle$-stopped walk from $U$ starting in state $|\phi_0\rangle$ is the process defined as the iteration of a measurement with the two projectors $\hat {\Pi}_0 = \hat {\Pi}_x = | x \rangle \langle x |$ and $\hat {\Pi}_1 = \hat {I} - \hat {\Pi}_0$. If $\hat {\Pi}_1$ is measured, an application of $U$ follows. If $\hat {\Pi}_0$ is measured the process is stopped.
\end{definition}

\begin{definition}{\bf Concurrent hitting time}.
A quantum walk $U$ has a $(T,p)$ concurrent $(|\phi_0\rangle,|x\rangle)$ hitting time if the $|x\rangle$-stopped walk from $U$ and initial state $|\phi_0\rangle$ has a probability $\geq p$ of stopping at a time $t \leq T$.
\label{concurrent_ht}
\end{definition}

In both cases (Defs. (\ref{one_shot_ht}) and (\ref{concurrent_ht})), it has been shown by Kempe \cite{kempe03} that the hitting time from one corner to its opposite is polynomial. However, although it was thought that this polynomial hitting time would imply an exponential speedup over corresponding classical algorithms, that is not the case as it is possible to build a polynomial time classical algorithm to traverse the hypercube from one corner to its opposite, as shown by Childs {\it et al} in \cite{childs03}. Further studies on hitting times of quantum walks on graphs have been produced by  Ko\v{s}\'{\i}k and Bu\v{z}ek \cite{kosik04} as well as Krovi and Brun \cite{krovi05,krovi06}.

A natural step further along employing discrete quantum walks for solving search problems is to use quantum computation techniques to find items stored in spaces of 2 or more dimensions.  In \cite{benioff02}, Benioff proposed the use of Grover's algorithm for searching items in a grid  of $\sqrt{N} \times \sqrt{N}$ elements, and showed that a direct application of such algorithm would take $O(N)$ times steps to find one item, i.e. there would be no more quantum speedup. Later on, in \cite{aaronson03} Aaronson and Ambainis used Grover's algorithm and multilevel recursion to build algorithms capable of searching in a 2-dimensional grid in $O(\sqrt{N}\log^2N)$ steps and a 3-dimensional grid in $O(\sqrt{N})$ steps,  and Ambainis {\it et al}  \cite{ambainiskempe04} proposed  algorithms based on discrete quantum walks (evolution operators used in \cite{ambainiskempe04} are those \lq perturbed' operators defined in \cite{shenvi02}) that would take $O(\sqrt{N}\log N)$ steps to search in a 2-dimensional grid and would reach an optimal performance of $O(\sqrt{N})$ for 3 and higher dimensional grids (an important contribution of \cite{ambainiskempe04} was to show that the performance of search algorithms based on quantum walks is sensitive to the selection of coin operators, i.e. the performance of a search algorithm may be optimal or not depending on the coin operator choice),  Aaronson and Ambainis \cite{aaronson05} have shown  how to build algorithms based on discrete quantum walks to search on a 2-dimensional grid using a total number of $O(\sqrt{N} \log^{5/2} N)$ steps, and a 3-dimensional grid with $O(\sqrt{N})$ number of steps, Tulsi \cite{tulsi08_qwsearch} has presented a $O(\sqrt{N\log N})$ modified version of Ambainis {\it et al}'s quantum walk search algorithm \cite{ambainiskempe04}, and Ambainis {\it et al} \cite{ambainis11_modified_classical} have proved that executing the  algorithm presented in \cite{ambainiskempe04} $O(\sqrt{N\log N})$ times would leave the walker within a neighbourhood $O(\sqrt{N})$ with probability $\Theta(1)$, thus classical  algorithm for local search could be used instead of performing the amplitude amplification technique designed in \cite{ambainiskempe04}. Numerical studies on how dimensionality, tunneling and connectivity affect a discrete quantum-walk based search algorithm are presented  by Lovett {\it et al} in \cite{lovett11_modified_search_factors}, and more numerical studies on potential improvements on algorithmic complexity on hypercubic lattices using the Dirac operator have been presented by Patel {\it et al} in \cite{patel10_search_01,patel10_search_02}. Finally, Childs and Goldstone \cite{childs04} developed a continuous quantum walk algorithm to solve the search problem in a grid and discovered algorithms that would have an optimal performance of $O(\sqrt{N})$ in grids of 5 or more dimensions.

A variant of Def. (\ref{search_problem}), the {\bf element distinctness problem}, was analyzed by Ambainis in \cite{ambainis04b}:

\begin{definition}{\bf Element distinctness problem} \cite{sipser05}.
Given a list of strings over $\{ 0,1\}$ separated by \#, determine
if all the strings are different.
\label{element_distinctness}
\end{definition}

A quantum algorithm for solving the element distinctness problem is given in \cite{ambainis04b}. This algorithm combines the quantum search of spatial regions proposed in  \cite{aaronson05} with a quantum walk. 

The first part of \cite{ambainis04b} transforms the string list from Def. (\ref{element_distinctness}) into a graph $G$ with marked and non-marked vertices; in this process, \cite{ambainis04b} uses an oracle (Def. (\ref{oracle_computation}).) The second part of the algorithm employs a discrete quantum walk to search graph $G$. As a result, the algorithm solves the distinctness problem in a total number of $O(N^{2/3})$ steps and  $O(N^{{k \over {k+1}}})$ steps for $k$ identical strings, among $N$ items. Upon the work presented in \cite{ambainis04b}, Magniez {\it et al} proposed in \cite{magniez07b} a new quantum algorithm for solving the {\it triangle problem}, which can be stated as

\begin{definition}
Let $G$ be a graph. Any complete subgraph of $G$ on three vertices is called a triangle. The triangle problem (in oracle version) can be posed as follows:
\\
Oracle input: the adjacency matrix $f$ of a graph $G$ on $n$ nodes.
\\
Oracle output: a triangle if there is any, otherwise reject.
\end{definition}

Additionally, another quantum algorithm, based on Grover's search quantum algorithm  \cite{grover96}, is presented in \cite{magniez07b} for solving the same triangle problem.

One more application of \cite{ambainis04b} has been proposed by Childs and Eisenberg in \cite{childs05}, where it has been proposed to employ the quantum algorithm developed for the distinctness  problem (Def. (\ref{element_distinctness})) to solve the L-subset finding (oracle) problem, which can be stated as

\begin{definition}{\bf The triangle problem (oracle version)}.
\\
Oracle input: 
1) A black box function $f:D \rightarrow R$, where $D,R$ are finite sets and $|D|=n$ is the problem size.
2) Property $P \subset (D \times R)^L$.\\
Oracle output:
Some subset $L= \{x_1, \ldots, x_L \} \subset D$ such that $((x_1,f(x_1), \ldots, (x_L,f(x_L)) \in P$, or reject
if none exists. 
\end{definition}

An alternative, refreshing and highly influential approach to discrete quantum walks has been presented  by M. Szegedy in \cite{szegedy04}, where a new definition of a discrete quantum walk in presented via the quantization of a stochastic matrix, as well as an alternative definition of hitting time for discrete quantum walks. \cite{szegedy04} begins by defining the search problem as follows:

\begin{definition}{\bf Search problem via stochastic processes}
Given a Markov chain with transition probability matrix $P = (p_{x,y})$  on a discrete state space $X$, with $|X|=n$, $u$ a given probability distribution on $X$, and a subset of marked elements $M \subseteq X$, compute an estimate for the number $t$ of iterations required to find an element of $M$, assuming that the Markov chain  is started from a u-distributed element of $X$. 
\label{search_via_stochastic}
\end{definition}

\cite{szegedy04} continues by defining the following concepts:

\begin{definition}
$P_M$ is the matrix obtained from $P$ by deleting its rows and columns indexed from $M$.
\end{definition}

Since there is no \lq natural' (i.e. straightforward) method for quantizing a discrete Markov chain, \cite{szegedy04} proposes a quantization method of $P$ which uses bipartite random walks.

\begin{definition}
Let $X$ and $Y$ be two finite sets and $P=(p_{x,y})$ and $Q=(q_{y,x})$ be matrices describing probabilistic maps  $X \rightarrow Y$ and $Y \rightarrow X$, respectively. If we have a single probabilistic function $P$ from $X$ to $X$, i.e. a Markov chain, in order to create a bipartite walk we can set $q_{y,x}=p_{x,y}$ for every $x,y \in X$ (that is, we set $Q=P$.)
\end{definition}

The quantization method for $(P,Q)$ proposed by Szegedy is as follows. We start by creating two operators on the Hilbert space with basis states $|x\rangle, |y\rangle \text{, where } x \in X \text{ and } y \in Y$. Let us define the states

\begin{subequations}
\begin{equation}
\phi_x = \sum_{y \in Y} \sqrt{p_{x,y}}|x\rangle |y\rangle
\end{equation}
\begin{equation}
\psi_y = \sum_{x \in X} \sqrt{q_{y,x}}|x\rangle |y\rangle
\end{equation}
\end{subequations}
for every $x \in X$, $y \in Y$. Finally, let us define $A=(\phi_x)$ as the matrix composed of columns vectors $\phi_x$ ($x \in X$), and $B=(\psi_y)$  as the matrix composed of columns vectors $\psi_y$ ($y \in Y$). Then,  \cite{szegedy04} defines the unitary operator $W$, the quantization of  the bipartite walk $(P,Q)$, as

\begin{definition}
$W = (2AA^* - I)(2BB^* - I)$
\end{definition}

\cite{szegedy04} proceeds to build definitions and theorems for new quantum hitting time and upper bounds for finding a marked element as in Def. (\ref{search_via_stochastic}). A relevant result presented in this paper is: for every ergodic Markov chain whose transition probability matrix is equal to its transpose, the quantum walk hitting time as defined in \cite{szegedy04} is at most the square root of the classical one. Furthermore, a remarkable feature of \cite{szegedy04} is a proposal for a new link between classical and quantum walks, namely the development of a quantum walk evolution operator $W$  via a classical stochastic matrix $P$. Inspired in the quantum walk model presented in \cite{szegedy04}, Ide {\it et al} have investigated the time averaged distribution of discrete quantum walks \cite{ide11_average_dtqw} and Segawa has studied the relation between recurrent properties of random walks and localization phenomena in quantum walks \cite{ide11_localization_dtqw}. Also, Chiang \cite{chiang10} and Chiang and Gomez \cite{chiang11} have proposed a model of noise based on system precision limitations and noisy environments in order to introduce a model of evolution perturbation for quantum walks and, based on the results presented in \cite{szegedy04} and Weyl's perturbation theorem on classical matrices, Chiang and Gomez \cite{chiang11} have studied how perturbation affects quantum hitting time as originally defined in \cite{szegedy04}.

Upon the quantum walk definition given in \cite{szegedy04}, Magniez {\it et al} \cite{magniez07} proposed a quantum walk-based algorithm for solving the following problem:

\begin{theorem} \cite{magniez07}
Let $\delta > 0$ be the eigenvalue gap of a reversible, ergodic Markov chain $P$,  and let $\epsilon > 0$ be a lower bound on the probability that an element chosen from the stationary distribution of $P$ is marked whenever $M$ is non-empty. Then, there is a quantum algorithm that with high probability determines if $M$ is empty or finds an element of $M$, with cost of order $S + {1 \over \sqrt{\epsilon}} ({1 \over \sqrt{\delta}} U + C)$, where $S$ is the computational cost of constructing superposition states, and $U,C$ are costs of constructing unitary transformations as defined on page 2 of \cite{magniez07}.
\end{theorem}

Furthermore, in \cite{magniez10} Magniez {\it et al}  have presented an algorithm for detecting marked elements that improves the complexity of the detection algorithm presented in \cite{szegedy04} and Ide {\it et al} \cite{ide11_average_dtqw} have derived a time average distribution for a quantum walk following \cite{szegedy04}. In addition, Krovi {\it et al} have constructed quantum walk-based algorithms that both detect and find marked vertices on a graph \cite{krovietal10}, Buhrman and \v{S}palek \cite{buhrmanspalek06} have presented a bounded error quantum algorithm with complexity $O(n^{5/3})$ for veryfying whether the product of two matrices of order $n \times n$ equals a third (i.e. the matrix multiplication verification problem), and Magniez and Nayak \cite{magnieznayak07_groupcomplexity} have presented a quantum algorithm for testing the commutativity of a black-box group, all three algorithms based on the formalisms introduced by Szegedy \cite{szegedy04}.

A novel application of discrete quantum walks is shown by Somma {\it et al} in \cite{somma07}, where a quantum algorithm for combinatorial optimization problems is proposed: this quantum algorithm combines techniques from discrete quantum walks, quantum phase estimation, and quantum Zeno effect, and can be seen as a quantum counterpart of classical  simulated annealing based on Markov chains (also, the Zeno effect in quantum-walk dynamics  under the influence of periodic measurements in position space is studied by Chandrashekar in \cite{chandrashekar10_zeno}), and Hillery {\it et al} have presented in \cite{hillery10_searching_scattered} a discrete quantum walk algorithm for detecting a marked edge or a marked complete subgraph within a graph. 

Finally, Paparo and Martin-Delgado  present a  novel and refreshing proposal developed  upon the notion of Szegedy's quantum walk \cite{szegedy04}: a quantum-mechanical version of Google's PageRank algorithm \cite{paparo12}.

\subsection{Algorithms based on continuous quantum walks}

The operation and mathematical formulation of discrete quantum walks  fits very well into the mindset of a computer scientist, as time evolves in discrete steps (as a typical classical algorithm would) and the model employs walkers and coins, usual elements of stochastic processes when employed in algorithm development. However, the most successful applications of quantum walks are found within the realm of continuous quantum walks. Given the seminal result derived by F. Strauch in \cite{strauch06} about the connection between discrete and continuous quantum walks,  we now know that results from continuous quantum walks should be translatable, at least in principle, to discrete quantum walks and vice versa.

Nonetheless, the mathematical structure of continuous quantum walks and the physical meaning of corresponding equations provide an accurate picture of several physical systems upon which we may implement quantum walks and quantum computers. Although many physical implementations in this field have been based on the discrete quantum walk model (please see subsection \ref{experimental_realizations}), the additional stimulus provided by \cite{strauch06} as well as the computational universality of quantum walks \cite{childs09,lovett10,underwood10} and recent connections found between quantum walks and adiabatic quantum computation \cite{chase08}, another model of continuous quantum computation, it is reasonable to expect new implementations based on continuous quantum walks.


Readers interested in acquiring a deeper understanding of the physics and mathematics of continuous quantum systems (particularly continuous quantum walks) may find the following references useful: \cite{feynman_lecturesIII,cohen77,tannor07}.

\subsubsection{Exponential algorithmic speedup by a quantum walk}

\begin{figure}
\begin{center}
(a)\scalebox{0.4}{\includegraphics{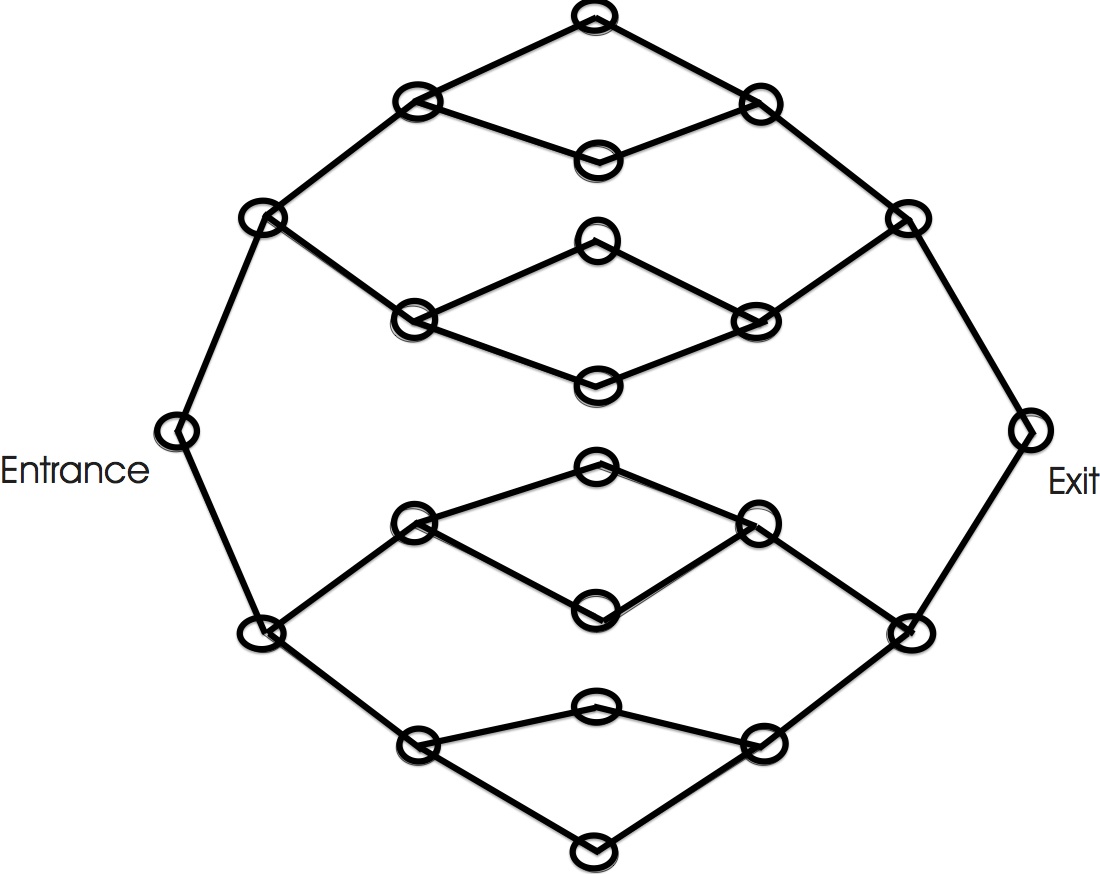}}
(b)\scalebox{0.4}{\includegraphics{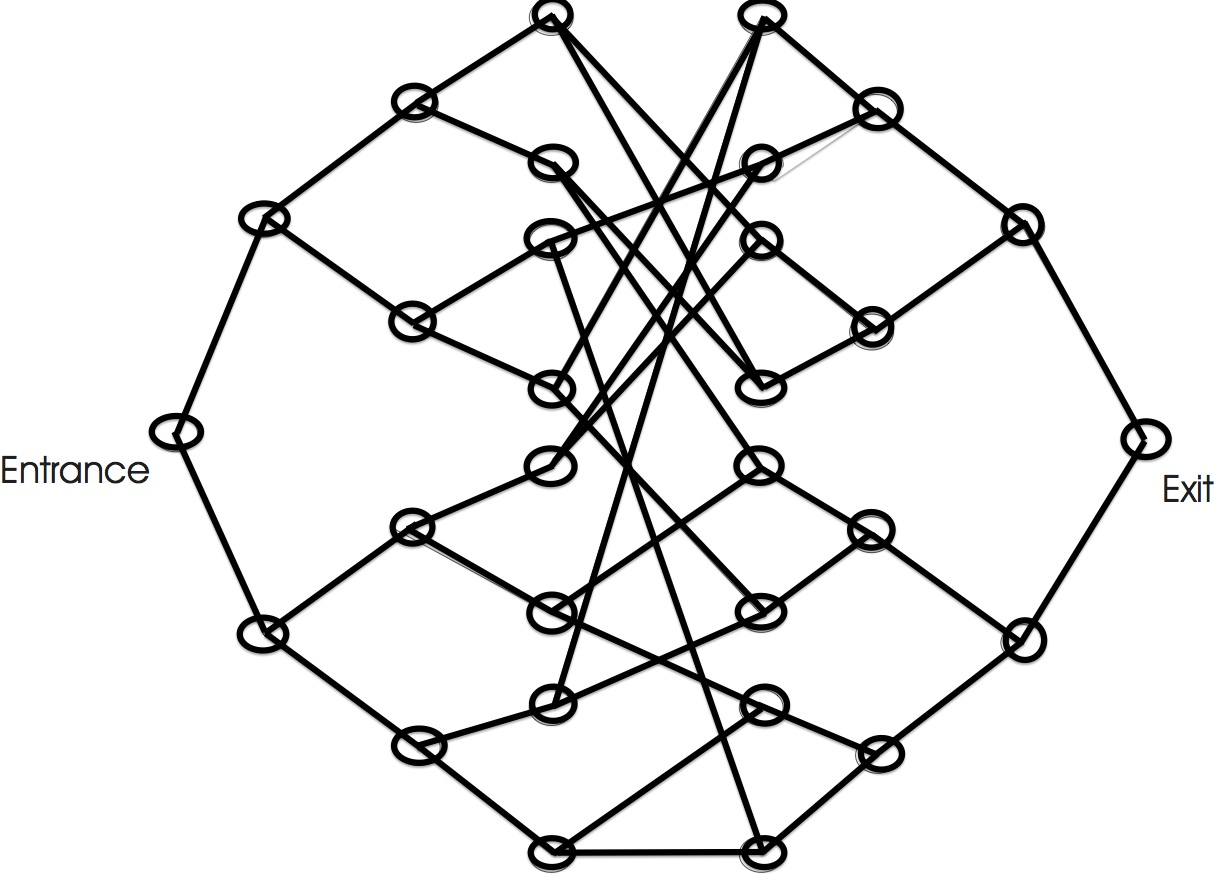}}
\end{center}
\caption{{\small Balanced and unbalanced trees.}}
\label{trees}
\end{figure}

In \cite{farhi98},
E. Farhi and S. Gutmann introduced an algorithm based on a continuous quantum walk that solves the following problem: Given a graph $G_s$ consisting of two balanced binary trees of height $n$ with the $2^n$ leaves of the left tree identified with the $2^n$ leaves of the right tree according to the way shown in Fig. (\ref{trees}(a)), and with two marked nodes {\it ENTRANCE} and {\it EXIT}, find an algorithm to go from {\it ENTRANCE} to {\it EXIT}.

It was shown in \cite{farhi98} that it is possible to build a quantum walk that traverses graph $G_s$ from {\it ENTRANCE} to {\it EXIT} which is exponentially faster than its corresponding classical random walk \cite{childs02}. In other words, the {\it hitting time} of the continuous quantum walk proposed in \cite{farhi98} is of polynomial order,  while the hitting time of the corresponding classical random walk is of exponential order. However, this advantage does not lead to an exponential speedup due to the fact that it is possible to build a deterministic algorithm that traverses the same graph in polynomial time \cite{childs03}.

Ideas from \cite{farhi98} were taken one step further by A. Childs {\it et al} in \cite{childs03}, where the authors introduced a more general type of graphs $G_r$ to be crossed, proved that those graphs could not be passed across efficiently with any classical algorithm, and delivered an algorithm based on a continuous quantum walk that traverses the graph in polynomial time.

Graphs $G_r$ are built as follows.
Begin by constructing two balanced binary trees of height $n$ (i.e. with $2^n$ leaves), but instead of identifying the leaves, they are connected by a random cycle that alternates between the leaves of the two trees, that is, we choose a leaf on the left at random and connect it to a leaf on the right chosen at random too. Then, we connect the latter to a leaf on the left chosen randomly among the remaining ones. The process is continued, always alternating sides, until every leaf on the left is connected to two leaves on the right, and vice versa. See  Fig. (\ref{trees}(b)) for an example of graphs $G_r$.

In order to build the quantum walk that will be used to traverse a graph $G_r$, the authors of \cite{childs03} defined a Hamiltonian $\hat{H}$ based on $G$'s adjacency matrix $A$.  $\hat{H}$ has matrix elements given by

\begin{equation}
\langle a | \hat{H} |a\rangle =
\begin{cases}
\gamma, & a \neq a', aa' \in G_r  \\
0,& \text{otherwise}
\end{cases}
\label{continuous_quantum_walk_childs}
\end{equation}

In the continuous quantum walk algorithm proposed in \cite{childs03}, the authors used an oracle to learn about the structure of the graph $G_r$, i.e. information about the Hamiltonian given by Eq. (\ref{continuous_quantum_walk_childs}) is extracted using an oracle. By doing so, it is proved in \cite{childs03} that  it is possible to construct a continuous quantum walk that would efficiently traverse any graph $G_r$. An improved lower bound for any classical algorithm traversing $G_r$ has been proposed in \cite{fenner03}, but the performance difference between quantum and classical algorithms in \cite{childs03} remains as previously stated. 

I now provide a succinct review of more continuous-time quantum walk algorithms. Focusing on finding hidden nonlinear structures over finite fields, Childs {\it et al} \cite{childsschulman07} have developed efficient quantum algorithms to solve the hidden radius problem and the hidden flat of centers problems. Moreover, Farhi {\it et al} \cite{fgg08_nandtrees} have produced a $O(\sqrt{N})$ quantum algorithm for solving the NAND tree problem (which consists of evaluating the root node of a perfectly bifurcating tree whose $N$ leaves are either \lq 0' or \lq 1' and the value of any other node is the NAND of corresponding children leaves) and Cleve {\it et al} have built quantum algorithms for evaluating MIN-MAX trees \cite{cleveetal08_minmax}. Finally, Agliari {\it et al} \cite{agliari10_fractal} have proposed a quantum walk-based search algorithm on fractal structures.

Let us present a final reflection with respect to algorithms purely based on quantum walks. As stated in the beginning of this section and rightly argued by Ritcher \cite{ritcher07}, the quantum algorithms reviewed in this section are instances of an abstract search problem: given a state space which can be translated into a graph structure, find a marked state (or set of states) by performing a quantum walk on the graph. With this abstraction in mind as well as with the purpose of combining the power of quantum walks with classical sampling algorihtms, Ritcher  \cite{ritcher07} has proposed a method for almost-uniform sampling based on repeated measurements of a continuous quantum walk.

\subsection{Simulation of quantum systems using quantum walks}

One of the main goals of quantum computation is the simulation of quantum systems, i.e.
the realization of programmable quantum systems whose physical properties 
allow us to model the behavior of other quantum systems \cite{qipc07,aspuru05,kassal08}.

A novel use of continuous quantum walks for simulation of quantum processes
has been presented by Mohseni {\it et al} in \cite{mohseni08}. In this contribution, the authors
have developed a theoretical framework for studying quantum interference effects in energy transfer 
phenomena, with the purpose of modeling photosynthetic processes. The main contribution of \cite{mohseni08} is to analyze the action of the environment in the coherent dynamics of quantum systems
related to photosynthesis. The framework developed in  \cite{mohseni08} includes a generalization 
of a non-unitary continuous quantum walk in a directed graph (as opposed to a previous 
definition of a unitary continuous quantum walk on undirected graphs \cite{childs03}.)

\subsection{\label{classical_simulation}Classical computer simulation of quantum algorithms and quantum walks}

Exact simulation of quantum systems using the mathematical model of the Universal Turing Machine (or any other universal automaton equally or less powerful than the Universal Turing Machine) is either an impossible task (for example, if we try to exactly simulate uniquely quantum mechanical behavior for which no classical counterpart is known  \cite{feynman82,feynman_lectures_computation}) or a very difficult one (for example, when trying to replicate physical phenomena in which the number of possible combinations or outcomes increases exponentially or factorially with respect to the number of physical systems involved in the experiment.) Still, as long as quantum computers are not available in the market in order to run quantum algorithms on them, physicists and computer scientists need an alternative tool to explore ideas and emergent properties of quantum systems and sophisticated quantum algorithms. 

Classical computer simulation of quantum algorithms is crucial for understanding and developing intuition about the behavior of quantum systems used for computational purposes, as well as to realize the approximate behavior of practical implementations of quantum algorithms.  Moreover, we may use classical simulation of quantum systems in order to learn which properties and operations of quantum systems cannot be efficiently simulated by classical systems (see \cite{nielsenknillgotesman00} and \cite{browne07} for most interesting results), as well as to find out how exclusive quantum-mechanical systems and operations can be employed for algorithm speed-up. Given the relevance of quantum walks in quantum computing both as a universal model of quantum computation and as an advanced tool for building quantum algorithms, as well as the daunting complexity of designing and coding classical algorithms for running on stand-alone, distributed or parallel hardware platforms, simulating quantum algorithms and quantum walks on classical computers has become a field on its own merit.

In the following lines, we summarize several theoretical developments and practical software implementations of  classical simulators of quantum algorithms, being all these developments suitable for (approximately) simulating both discrete and continuous quantum walks.

\"Omer \cite{omer00}, Bettelli {\it et al}  \cite{bettelli03}, Viamontes \emph{et al}  \cite{viamontes03}, Selinger \cite{selinger04a}, and Ba\~nuls {\it et al} \cite{banuls06}, among others, have introduced mathematical frameworks for implementing quantum algorithms simulators using classical computer languages. Later and among many other relevant contributions, Nyman proposed using symbolic classical computer languages for simulating quantum algorithms \cite{nyman09},  \"Omer introduced abstract semantic structures for modelling quantum algorithms in classical environments \cite{omer05}, and Altenkirch \emph{et al} proposed a quantum programming language based on classical functional programming \cite{altenkirch05}. Selinger \cite{selinger04b} and Gay \cite{gay06} provided an early description of quantum programming languages and Miszczak \cite{miszczak11} presented a summary of models of quantum computation and current quantum programming languages.

Among several software packages and platforms that have been developed for quantum algorithm simulation, I would like to mention the contributions of Marquezino and Portugal \cite{marquezino08} (quantum walk simulator for one- and two-dimensional lattices),  G\'{o}mez-Mu{\~{n}}oz \cite{our_simulator_08} (Mathematica add-on for quantum algorithm simulation), De Raedt {\it et al} \cite{deraedt07} (quantum algorithm simulation on parallel computers), Caraiman and Manta \cite{caraiman10} (quantum algorithm simulation on grids), D\'{\i}az-Pier {\it et al} \cite{diaz11} (this is an extension of \cite{our_simulator_08} built for simulating adiabatic quantum algorithms on GPUs), and Machnes {\it et al} \cite{machnes11} (a Matlab toolset for simulating quantum control algorithms.) The interested reader will find a comprehensive list of currently available classical simulators of quantum algorithms in \cite{others_simulators_08}.

An example of the importance of realizing whether truly quantum properties can be used for algorithm speed-up was provided in the field of quantum walks a few years ago. As already explained in this review paper (subsection \ref{quantumness}), since the publication of \cite{nayak00} it had been believed that the enhanced variance of position distribution in quantum walks was  responsible (partially at least) for quadratic speed-up of quantum walk-based  algorithms. However,  it has been shown \cite{knight03,knight03b,knight04,jeong04} that it is possible to develop implementations of a quantum walk on a line purely described by classical physics and still be able to reproduce the variance enhancement that characterizes a discrete quantum walk. Thus, {\it it remains as an open question what exclusive quantum-mechanical properties and operations are relevant for enhancing our computing capabilities}.

\section{\label{qw_computational_universality}Universality of quantum walks}

Universality is a highly desirable property for a model of computation because it shows that such a model is capable of simulating any other model of computation. Basically, models of computation that are labeled as universal are capable of solving the same problems, although it could happen in different time regimes. The history of quantum computing includes the recollection of significant efforts  to prove the universality of several models of quantum computers, i.e. that any  algorithm that can be computed by a general-purpose quantum computer \cite{deutsch85} can also be executed by quantum gates \cite{nielsen00,kitaev99}, and computers based on the quantum adiabatic theorem \cite{messiah99,farhi00,aharonov07}, for example.

In the field of quantum walks, Hines and Stamp have shown in \cite{hines07} how to map quantum walk Hamiltonians and Hamiltonians for other quantum systems on hypercubes and hyperlattices.  Later on, formal proofs of computational universality of quantum walks have been presented by Childs (2009) \cite{childs09}, Lovett {\it et al} (2010) \cite{lovett10}, and Underwood and Feder (2010) \cite{underwood10}. Let us now dwell on the properties and details of \cite{childs09,lovett10} and \cite{underwood10}.
\\\\

a) {\bf Universal computation by continuous-time quantum walk \cite{childs09}}
\\

In his seminal work \cite{childs09}, Childs proved that the model known as continuous-time quantum walk is universal for quantum computation. This means that, for an arbitrary problem $A$ that is computable in a general-purpose quantum computer, it is possible to employ the continuous-time quantum walk model to build computational processes that would also solve $A$. Since it has already been proved by Childs {\it et al} \cite{childs03} and Aharonov and Ta-Shma \cite{aharonovtashama03} that it is possible to simulate a continuous quantum walk using poly(log$N$) gates, we then conclude that quantum walks and quantum circuits have essentially the same computational power.

\begin{figure}
\begin{center}
(a) \scalebox{0.5}{\includegraphics{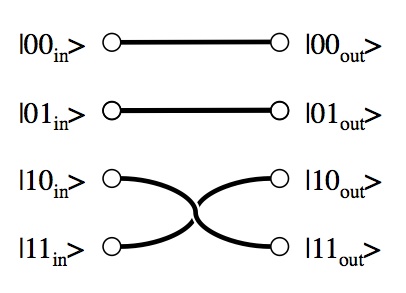}}
(b) \scalebox{0.5}{\includegraphics{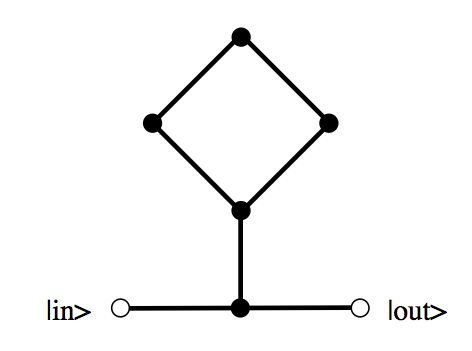}}\\
(c) \scalebox{0.5}{\includegraphics{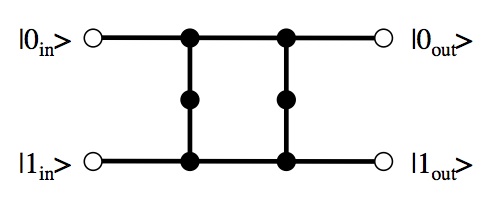}}
\end{center}
\caption{{\small (a) Widget for $\text{C}_{\text{not}}$ gate, (b) widget for phase gate, and (c) widget for the basis-changing gate.}}
\label{graphs_for_quantum_gates}
\end{figure}

The proof of universal computation delivered in \cite{childs09} is based on the following ideas:

\begin{enumerate}

\item
Executing a continuous-time quantum walk-based algorithm is equivalent to propagating a continuous-time quantum walk on a graph $G$. Propagation occurs via scattering theory.

\item
The particular structure of graph $G$ depends on the problem to solve (i.e. on the algorithm that one would like to implement.) Nevertheless and in all cases, graph $G$ consists of sub-graphs (with maximum degree equal to three) representing quantum-mechanical operators connected by quantum wires.  Moreover, graph $G$ is finite in terms of both the number of quantum gates as well as the number and length of quantum wires.

\item
Quantum wires do not represent qubits: they represented quantum states, instead. Consequently, the number of quantum wires in a graph $G$ will grow exponentially with respect to the number of qubits to be employed. Indeed, if we meant to simulate the propagation of a continuous-time quantum walk in $G$ on a classical computer we would certainly need an exponential amount of computational resources for representing quantum wires; however, both $G$ and the propagation of a continuous-time quantum walk on it are to be simulated by a general purpose quantum computer which, as previously stated in the beginning of this section, can simulate a continuous-time quantum walk in poly(log$N$) \cite{childs03,aharonovtashama03}.

\item
A set of gates is labelled as {\it universal for quantum computation} if any unitary operation may be approximated to arbitrary accuracy by a quantum circuit involving those gates \cite{nielsen00}. The core of \cite{childs09} is to simulate a universal gate set for quantum computation by propagating a continuous-time quantum walk on different graph shapes. The universal set chosen by Childs in \cite{childs09} is composed by the controlled-not, phase and and basis-changing gates with matrix representations given in Eqs. (\ref{cnot_childs},\ref{phase_childs},\ref{basis_changing}), which together constitute a dense subset of $SU(2)$. Graphs employed to represent these three quantum gates are shown in Fig. (\ref{graphs_for_quantum_gates}).

\begin{subequations}
\begin{equation}\label{cnot_childs}
C_{\text{not}} =\begin{pmatrix} 1 & 0 & 0 & 0 \\ 0 & 1 & 0 & 0 \\ 0 & 0 & 0 & 1 \\ 0 & 0 & 1 & 0 \\ \end{pmatrix}
\end{equation}
\begin{equation}\label{phase_childs}
U_b =\begin{pmatrix} 1 & 0\\ 0 &  e^{\frac{i\pi}{4}}\end{pmatrix}
\end{equation}
\begin{equation}\label{basis_changing}
U_c =\frac{1}{\sqrt{2}}\begin{pmatrix} 1 & i\\ i & 1\end{pmatrix}
\end{equation}
\end{subequations}

\item
The eigenvalues and eigenvectors of those graphs employed to simulate a universal gate set for quantum computation play a central role in this discussion.

\item
\cite{childs09} constitutes a theoretical proposal for proving and exhibiting the computational power of continuous quantum walks. In particular,  \cite{childs09} does {\it not} constitute a hardware-oriented proposal for implementing a general-purpose quantum computer based on continuous quantum walk. 

\end{enumerate}

To put it in a few words of my own, \cite{childs09} proposes quantum computation as the flow of quantum information, via the dynamics of a continuous-time quantum walk, on graphs. Let us now work out the details of \cite{childs09} (for the sake of clarity  and readability of the original paper, hereinafter I will closely follow the notation used in \cite{childs09}.)
\\

-{\it Scattering on an infinite line}. 
\cite{childs09} starts by reviewing some properties of scattering theory on infinite lines. Let $L$ be an infinite line of vertices. Each vertex $x$ corresponds to a basis state $|x\rangle \in \mathbb{Z}$ and is, of course, connected only to vertices $x \pm 1$. Then, the eigenstates of the adjacency matrix of this graph are the momentum states $|k\rangle$, $k \in [-\pi,\pi)$ with corresponding eigenvalues $2\cos k$. The eigenstates $|k\rangle$ fulfill the following condition:

\begin{equation}
\langle x | \tilde k \rangle = e^{i k x} 
\label{uno_childs}
\end{equation}

\begin{figure}
\begin{center}
\scalebox{0.5}{\includegraphics{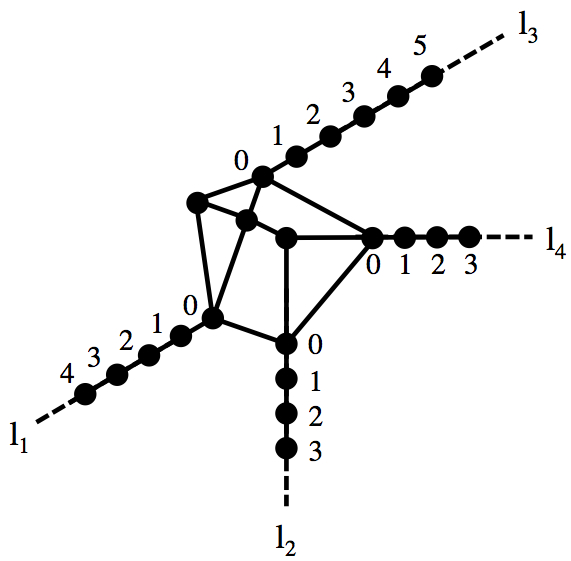}}
\end{center}
\caption{{\small An example of a graph with three semi-infinite lines.}}
\label{semi_infinite_line_childs} 
\end{figure}

-{\it Scattering on a semi-infinite line}. The next step toward calculating expressions for scattering on finite graphs is to study semi-infinite lines. Let us consider a graph $G$ and construct an infinite graph with adjacency matrix $H$ by attaching a semi-infinite line to each of $N$ of its vertices (i.e. it is not compulsory to attach infinite lines to all vertices in $G$, just some vertices would suffice.) We shall enumerate the vertices of  each infinite line attached to $G$ by labelling the vertex in the original graph with $x=0$ and assigning the values $x=1, 2, \ldots, n, \ldots$ to the vertices we find as we move out along the line (see Fig. (\ref{semi_infinite_line_childs}) for an example of a graph with semi-infinite lines.)

\begin{figure}
\begin{center}
\scalebox{0.5}{\includegraphics{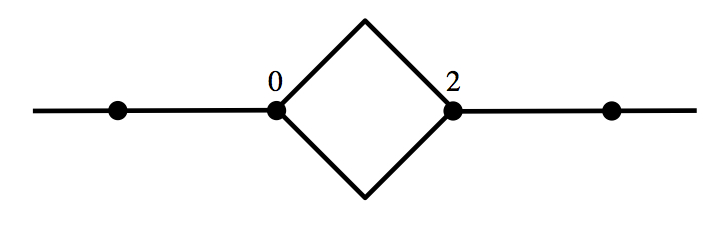}}
\end{center}
\caption{{\small An example of a semi-infinite graph with a diamond-shape.}}
\label{semi_infinite_line_feldman}
\end{figure}

A nice example of this kind of semi-infinite graphs on discrete-time quantum walks is provided by Feldman and Hillery in \cite{feldman04} which we reproduce here. Let $G_{d}$ be the graph  given in Fig. (\ref{semi_infinite_line_feldman}). The graph goes to $-\infty$ on the left and to $+\infty$ on the right. One set of unnormalized eigenstates of this graph can be described as having an incoming wave from the left, an outgoing transmitted wave going to the right, and a reflected wave going to the left. The eigenstates with a wave incident from the left take the form

\begin{eqnarray}
|\Psi\rangle & = &  \sum_{j=-\infty}^{-1}(e^{ij\theta}|j,j+1\rangle + r(\theta )e^{-i(j+1)\theta}|j+1,j\rangle )+|\Psi_{02}\rangle \nonumber \\  & &+\sum_{j=2}^{\infty}t(\theta )e^{i(j-2)\theta}|j,j+1\rangle ,
\end{eqnarray}

where $|\Psi_{02}\rangle$ is the part of the eigenfunction between  vertices $0$ and $2$, and $e^{-i\theta}$ is the eigenvalue of the operator $U$ that advances the walk one step.  The first term can be thought
of as the incoming wave (from $-\infty$ to zero), the term proportional to $r(\theta )$ is the reflected wave (from zero to $-\infty$), and the term proportional to $t(\theta )$ is the transmitted wave (from 2 to $+\infty$). Please notice the crucial role that eigenvalue $e^{-i\theta}$ plays in the quantification of phases.

Let us now go back to \cite{childs09}. For each $j \in \{1, 2, \ldots, N \}$ (i.e. for each infinite line attached to $G$)  there is an incoming scattering state of momentum $k$ denoted 
$|\tilde k, \text{sc}^\rightarrow_j\rangle$ given by

\begin{eqnarray}
  \langle x,j |\tilde k, \text{sc}^\rightarrow_j\rangle &= e^{-i k x} + R_j(k) \, e^{i k x} \\
  \langle x,j'|\tilde k, \text{sc}^\rightarrow_j\rangle &= T_{j,j'}(k) \, e^{i k x}
  ,\quad
  j' \ne j
\label{dos_childs}
\end{eqnarray}

The reflection coefficient $R_j(k)$, the transmission coefficients $T_{j,j'}(k)$ and the form of $|\tilde k, \text{sc}^\rightarrow_j\rangle$ are determined by the eigenequation 
$H|\tilde k, \text{sc}^\rightarrow_j\rangle = 2 \cos k |\tilde k, \text{sc}^\rightarrow_j\rangle$.  Eigenstates $|\tilde k, \text{sc}^\rightarrow_j\rangle$ together with the bound states defined in section II of \cite{childs09} form a complete a orthogonal set of eigenfunctions of $H$ that are employed to calculate the propagator for scattering through $G$ (we do not write the mathematical expressions for bound states as it is proved in \cite{childs09} that the role of those states on the scattering process through $G$ can be neglected):

\begin{eqnarray}
\langle y,j'|e^{-i Ht}|x,j\rangle & = & {\int_{-\pi}^0 e^{-2i t\cos k} \big( T_{j,j'} e^{ik(x+y)} + T_{j',j}^* e^{-ik(x+y)} \big)} \, \frac{dk}{2\pi} \nonumber \\ & & +\sum_{\kappa,\pm} e^{\mp 2i t\cosh\kappa}
    B^\pm_{j'}(\kappa) B^\pm_j(\kappa)^* (\pm e^{-\kappa})^{x+y}
\label{tres_childs}
\end{eqnarray}

where $\kappa$ is a parameter of bound states. The whole purpose of this exercise is to have the mathematical tools needed to compute the propagation of the continuous-time quantum walk on the graphs that act as quantum gates (Fig. (\ref{graphs_for_quantum_gates}).) Finally with respect to this introductory mathematical treatment, it is stated in \cite{childs09} that finite graphs can be modelled with Eqs. (\ref{uno_childs},\ref{dos_childs},\ref{tres_childs}) without significant changes.
\\

-{\it Universal gate set}. As previously stated in this review,  the universal gate set chosen by Childs is composed of the controlled-not, phase and and basis-changing gates.

The implementation of the controlled-not gate is straightforward  as it suffices just to exchange the quantum wires corresponding to the basis states $|10\rangle$ and $|11\rangle$ as shown in Fig. (\ref{graphs_for_quantum_gates}.a). This wire-exchange may sound unfeasible, but it is not:  \cite{childs09} is a theoretical proposal that describes the logical/mathematical processes that must be performed in order to achieve universal quantum computation, not the implementation of quantum walk-based universal computation on actual quantum hardware.

As for the phase gate, the process to be performed is to apply a nontrivial phase to the $|1\rangle$, leaving the $|0\rangle$ unchanged. To do so, Childs has proposed to propagate the quantum walk through the widget shown in Fig. (\ref{graphs_for_quantum_gates}.b). The process is as follows: attach  semi-infinite lines to the ends (open circles) of Fig. (\ref{graphs_for_quantum_gates}.b) and compute the transmission coefficient for a wave of momentum $k$ incident on the input terminal (LHS open circle.) The value for $T^{(b)}_{\text{in,out}}$ reported in \cite{childs09} is 

\begin{equation}
T^{(b)}_{\text{in,out}} = \frac{8}{8+i \cos 2k \csc^3 k \sec k}
\label{childs_cuatro}
\end{equation}

As direct substitution in Eq. (\ref{childs_cuatro}) shows, at $k = \frac{-\pi}{4}$ the widget has perfect transmission (i.e.  $T^{(b)}_{\text{in,out}}=1$.) Furthermore, also at  $k=\frac{-\pi}{4}$, the widget shown in Fig. (\ref{graphs_for_quantum_gates}.b.) introduces a phase of $e^{\frac{i\pi}{4}}$ to the quantum information that is being propagated through it. This last result is not explicitly derived in \cite{childs09} but it can be calculated from the eigenvalues of the corresponding adjacency matrix and the mathematical model for propagation for scattering through graphs (Eq. (\ref{tres_childs}).) The same rationale applies to the construction of the basis-changing single-qubit gate proposed by Childs: propagating a continuous-time quantum walk at  $k=\frac{-\pi}{4}$ through the graph shown in Fig. (\ref{graphs_for_quantum_gates}.c) would be equivalent to applying the unitary transformation given in Eq. (\ref{basis_changing}.) 

Now, assuming that $k$ will only take the value $\frac{-\pi}{4}$ could be very difficult to implement. Consequently, \cite{childs09} introduces two more gates: a momentum filter and a momentum separator, which are to be used for appropriately tuning the algorithm input. Finally, it is stated in \cite{childs09} that for the actual implementation of a general quantum gate as well as a continuous-time quantum-walk algorithm, we would only need to connect appropriate widgets using quantum wires.

Let us now review the main ideas and properties of universal computation of discrete-time quantum walks.
\\\\
b) {\bf Universal computation by discrete quantum walk \cite{lovett10}}
\\
In \cite{lovett10}, Lovett {\it et al} have presented a proof of computational universality for discrete-time quantum walks. The arguments delivered in \cite{lovett10} keep a close link with the ideas presented in \cite{childs09}, in terms of the universal gate set upon which the simulation of an arbitrary quantum gate can be achieved as well as on the nature of quantum wires (as in \cite{childs09}, quantum wires represent basis states rather than qubits.) Here a summary of relevant properties:

\begin{enumerate}

\item
Executing a discrete-time quantum walk-based algorithm is equivalent to propagating a discrete-time quantum walk on a graph $G$ via state transfer theory. In contrast to the behavior of continuous-time quantum walks, coined discrete-time quantum walks do exhibit back-propagation, hence the need to look for an efficient way to propagate the discrete-time quantum walk. 

It has been shown \cite{tregenna03,milburn-travaglione} that perfect state transfer can be achieved in graphs (for example, an eight-node cycle gives perfect state transfer from the initial vertex to the opposite vertex in 12 time steps \cite{lovett10}.) Thus, Lovett {\it et al} propose a scheme based on two-edge quantum wires (i.e. a cycle of two nodes) for achieving perfect state transfer. The basic wire used to propagate a discrete-time quantum walk is shown in Fig. (\ref{basic_wire_lovett}). In this setup, the state $|\Psi\rangle = \alpha|0\rangle + \beta|1\rangle$ would be split as initial state $|\psi\rangle = \frac{1}{\sqrt{2}} \big(  \alpha |0\rangle_a + \alpha |0\rangle_b + \alpha |1\rangle_a + \alpha |1\rangle_b   \big)$.  I shall describe the propagation method proposed in \cite{lovett10} in the following lines.

\item
As in \cite{childs09}, the particular structure of graph $G$ depends on the problem to solve (i.e. on the algorithm that one would like to implement.) Nevertheless and in all cases, graph $G$ consists of sub-graphs representing quantum-mechanical operators connected by quantum wires (in contrast with \cite{childs09}, in \cite{lovett10} graphs representing quantum gates have maximum degree equal to eight.)  Furthermore, graph $G$ is finite in terms of both the number of quantum gates as well as the number and length of quantum wires.

\item
Quantum wires do not represent qubits: they represented quantum states, instead. As in \cite{childs09}, the number of quantum wires in a graph $G$ will grow exponentially with respect to the number of qubits to be employed but, as previously stated in the beginning of this section, both $G$ and the propagation of a discrete-time quantum walk on it are to be simulated by a general purpose quantum computer which can simulate a discrete-time quantum walk using poly(log$N$) gates \cite{childs03,aharonovtashama03}.

\item
it is proposed in \cite{lovett10} to simulate a universal gate set for quantum computation by propagating a discrete-time quantum walk on different graph shapes. The universal set chosen by Lovett {\it et al} in \cite{lovett10} is composed by the controlled-not, phase and Hadamard gates with matrix representations given in Eqs. (\ref{cnot_lovett},\ref{phase_lovett},\ref{hadamard_lovett}). Graphs employed to represent these three quantum gates are shown in Fig. (\ref{graphs_for_quantum_gates_lovett}).  Also, as in \cite{childs09}, Lovett {\it et al} have presented a theoretical proposal for proving and exhibiting the computational power of discrete-quantum walks and it does {\it not} constitute a straightforward quantum computer architecture proposal for implementing a general-purpose quantum computer based on  discrete-time quantum walks (pretty much in the same spirit that a classical algorithm is not straightforwardly implemented in classical digital hardware.)

\begin{subequations}
\begin{equation}\label{cnot_lovett}
C_{\text{not}} =\begin{pmatrix} 1 & 0 & 0 & 0 \\ 0 & 1 & 0 & 0 \\ 0 & 0 & 0 & 1 \\ 0 & 0 & 1 & 0 \\ \end{pmatrix}
\end{equation}
\begin{equation}\label{phase_lovett}
P(\pi/8) =\begin{pmatrix} 1 & 0\\ 0 &  e^{\frac{i\pi}{4}}\end{pmatrix}
\end{equation}
\begin{equation}\label{hadamard_lovett}
U_c =\frac{1}{\sqrt{2}}\begin{pmatrix} 1 & 1\\ 1 & -1\end{pmatrix}
\end{equation}
\end{subequations}

\begin{figure}
\begin{center}
\scalebox{0.5}{\includegraphics{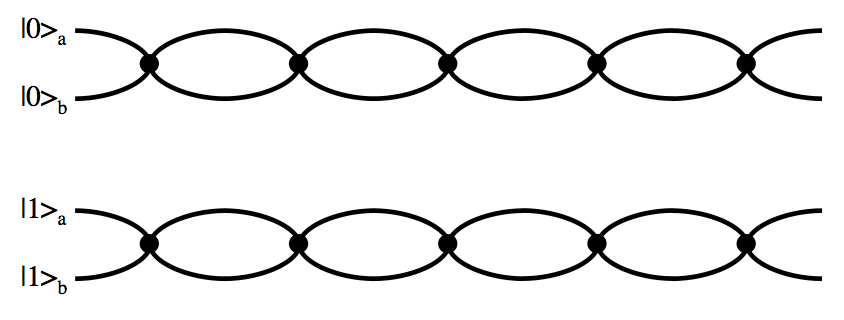}}
\end{center}
\caption{{\small Basic wire used to propagate  quantum information via discrete-time quantum walk. The state $|\Psi\rangle = \alpha|0\rangle + \beta|1\rangle$ would be split as initial state
$|\psi\rangle = \frac{1}{\sqrt{2}} \big(  \alpha |0\rangle_a + \alpha |0\rangle_b + \alpha |1\rangle_a + \alpha |1\rangle_b   \big)$. Also, the physical process used to propagate the quantum walk consists of applying a $4-d$ Grover diffusion coin (note that each node is a vertex of degree 4), together with an implementation-related shift operator (the shift operator described in \cite{lovett10} consists only of its expected behavior and does not deal with particular physical implementations.)}}
\label{basic_wire_lovett}
\end{figure}

\begin{figure}
\begin{center}
(a) \scalebox{0.5}{\includegraphics{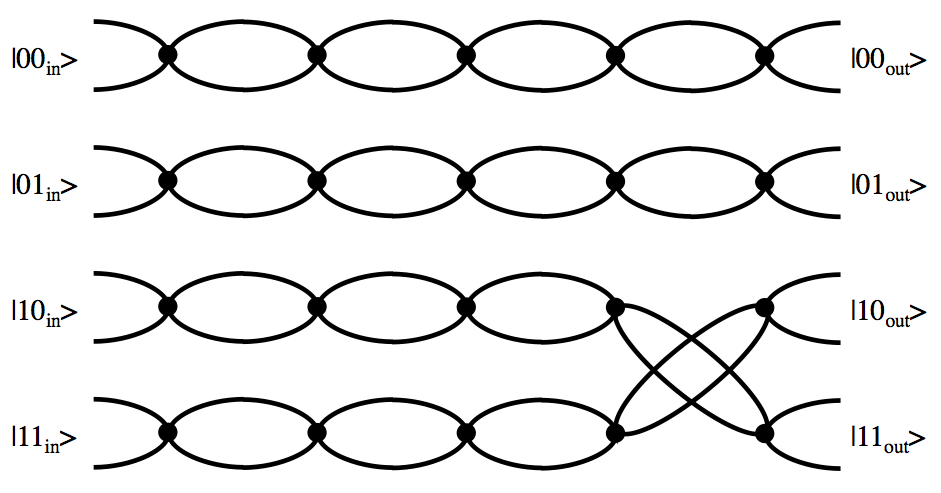}}\\
(b) \scalebox{0.5}{\includegraphics{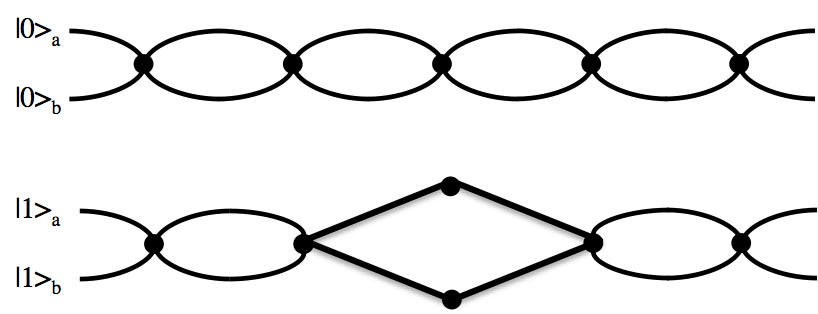}}\\
(c) \scalebox{0.5}{\includegraphics{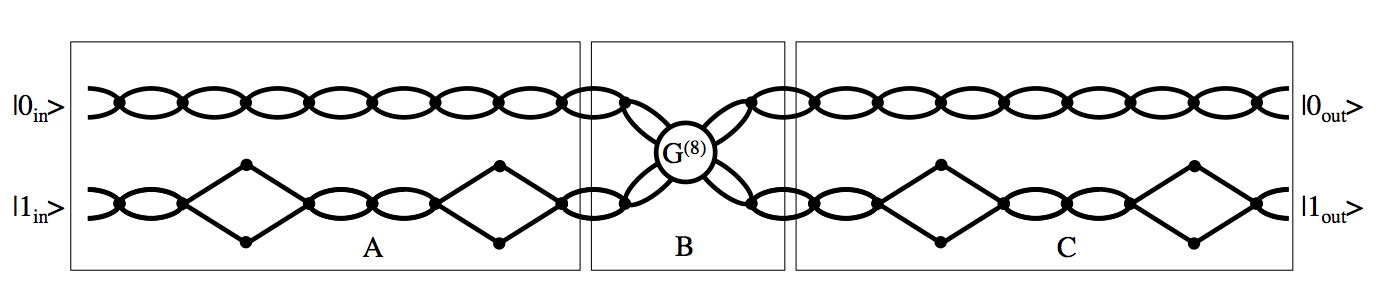}}\\
\end{center}
\caption{{\small Graphs for simulating the effect of  (a) $\text{C}_{\text{not}}$ gate, (b) phase $\pi/8$ gate, and (c) Hadamard gate.}}
\label{graphs_for_quantum_gates_lovett}
\end{figure}

\end{enumerate}

-{\it State transfer on the basic wire using a four-dimensional Grover coin}. 
Let us now describe the propagation method proposed in \cite{lovett10}. Suppose that we need to transmit a qubit that has been initialized as 

\begin{equation}
|\psi\rangle = \alpha|0\rangle + \beta|1\rangle
\label{lovett_initial_state}
\end{equation}

\begin{figure}
\begin{center}
\scalebox{0.5}{\includegraphics{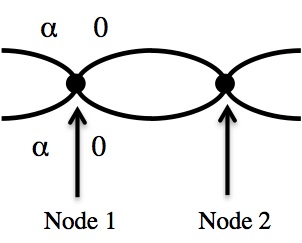}}
\end{center}
\caption{{\small Both LHS arms of $|0\rangle$, $|0\rangle_a$ and $|0\rangle_b$, are initialized to the same quantum information $\alpha$. Also, both arms on the RHS of this graph have been initialized to $0$. We may think of this graph as a dynamical quantum process which consists of quantum information flowing through the graph, from left to right. Furthermore, the {\it same} quantum information flows through both upper and lower arms.}}
\label{propagation_grover_coin_01}
\end{figure}

\begin{figure}
\begin{center}
(a) \scalebox{0.5}{\includegraphics{propagation_grover_coin_01.jpg}}
(b) \scalebox{0.5}{\includegraphics{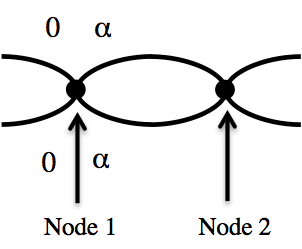}}
\end{center}
\caption{{\small Fig. (a) represents the system immediately before $G^{(4)}$ (Eq. (\ref{grover_4d})) is applied, and (b) represents the system immediately after $G^{(4)}$ (Eq. (\ref{grover_4d})) has been applied.}}
\label{propagation_grover_coin_02}
\end{figure}

\begin{figure}
\begin{center}
(a) \scalebox{0.5}{\includegraphics{propagation_grover_coin_01.jpg}}
(b) \scalebox{0.5}{\includegraphics{propagation_grover_coin_02.jpg}}
(b) \scalebox{0.5}{\includegraphics{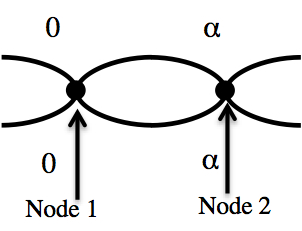}}
\end{center}
\caption{{\small (a) represents the system immediately before $G^{(4)}$ (Eq. (\ref{grover_4d})) is applied and (b) represents the system immediately after $G^{(4)}$ (Eq. (\ref{grover_4d})) has been applied. Please observe that, on step (b), the quantum information represented by $\alpha$ is near Node 1. The third step of this basic operation, consisting of applying a shift operator to (b), would produce graph (c), i.e. would shift amplitude $\alpha$ to the right, near Node 2, so that a new computational step can be performed.}}
\label{propagation_grover_coin_03}
\end{figure}

Then:
\begin{itemize}
\item
The initial state of the basic wire consists of preparing both LHS arms $|0\rangle_a$ and $|0\rangle_b$ with the same quantum information $\alpha$, i.e the actual amplitude assigned to basis state $|0\rangle$ from Eq. (\ref{lovett_initial_state}.) The same rationale applies to LHS arms $|1\rangle_a$ and $|1\rangle_b$: they both are initialized with the same quantum information $\beta$, i.e. the amplitude assigned to basis state $|1\rangle$ from Eq. (\ref{lovett_initial_state}.) This initialization, visually presented in Fig. (\ref{propagation_grover_coin_01}) for $\alpha \in \mathbb{C}$, may be written as shown in Eq. (\ref{initial_state_lovett}.) Note that the RHS of Fig. (\ref{propagation_grover_coin_01}) is initialized to $0$.

\begin{equation}
|\Psi\rangle_{t_1} =  \begin{pmatrix} \alpha \\ \alpha \\ 0 \\0 \\ \end{pmatrix}
\label{initial_state_lovett}
\end{equation}

\item
Now, a crucial point comes into the scene: the application of $4-d$ Grover diffusion operator (Eq. (\ref{grover_4d})) to $|\Psi\rangle_{t_1}$. It is stated in \cite{lovett10} that, for any vertex of even degree, the Grover coin $G^{(4)}$ will transfer the entire state from all input edges to all output edges, {\it provided the inputs are all equal in both amplitude and phase}.

\begin{equation}
G^{(4)} =\frac{1}{2} \begin{pmatrix} {-1} & 1 & 1 & 1\\1 & {-1} & 1 & 1\\1 & 1 & {-1} & 1\\1 & 1 & 1 & {-1}\\ \end{pmatrix}
\label{grover_4d}
\end{equation}

Mathematically speaking, computing $G^{(4)}|\Psi\rangle_{t_1}$ is a straightforward procedure. Physicall speaking, applying $G^{(4)}$ to $|\Psi\rangle_{t_1}$ would be equivalent to applying a unitary operator that does perfect quantum information transfer from the LHS of the graph to the RHS of that same graph, as shown in Fig. (\ref{propagation_grover_coin_02}). In principle and depending on the particular properties of quantum hardware we may try to translate and implement this protocol, we should be able to find such a transfer physical operation as we are modelling it as a quantum-mechanical unitary operator. 

So, $G^{(4)}|\Psi\rangle_{t_1}$ yields Eq. (\ref{first_operation})

\begin{equation}
\frac{1}{2} \begin{pmatrix} {-1} & 1 & 1 & 1\\1 & {-1} & 1 & 1\\1 & 1 & {-1} & 1\\1 & 1 & 1 & {-1}\\ \end{pmatrix} 
\begin{pmatrix} \alpha \\ \alpha \\ 0 \\0 \\ \end{pmatrix} = \begin{pmatrix} 0 \\ 0 \\ \alpha \\ \alpha \\ \end{pmatrix}
\label{first_operation}
\end{equation}

\item
The third and last step of this basic quantum operation consists of shifting quantum information from the zone nearby Node 1 to the sorrounding area of Node 2. This step is equivalent to preparing the input of the next algorithmic operation. The full three-step basic operation is shown in Fig. (\ref{propagation_grover_coin_03}).

\end{itemize}

\paragraph{}
-{\it Construction of the Universal gate set}. Let us now describe how to construct, according to \cite{lovett10}, the controlled-not, phase and and Hadamard gates (Eqs. (\ref{cnot_lovett},\ref{phase_lovett},\ref{hadamard_lovett}).)  

As in \cite{childs09}, the controlled-not gate is trivial to implement: we only need to exchange corresponding basis states wires as shown in Fig. (\ref{graphs_for_quantum_gates_lovett}.(a).) As previously declared in this review, this wire-exchange describes the logical/mathematical processes that must be performed in order to achieve universal quantum computation, not the implementation of quantum walk-based universal computation on actual quantum hardware.

\begin{figure}
\begin{center}
\scalebox{0.7}{\includegraphics{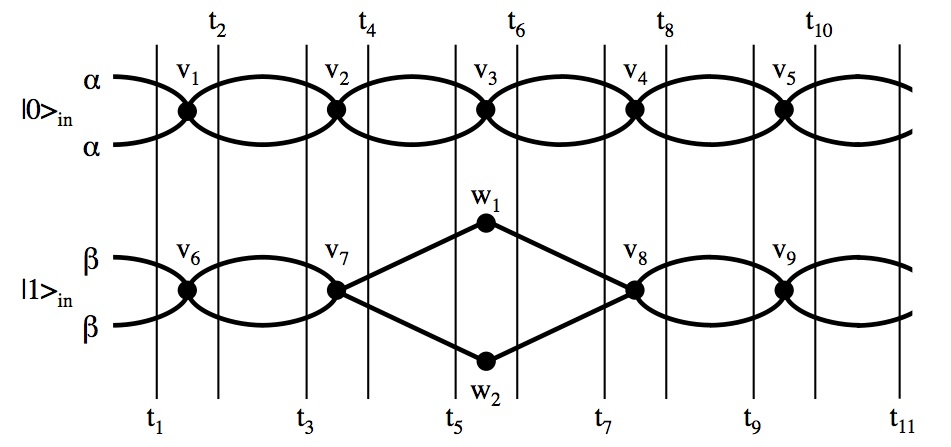}}
\end{center}
\caption{{\small Phase gate proposed in \cite{lovett10}  divided up in 11 steps. We provide a detailed analysis of each step in the main text of this paper.}}
\label{phase_gate}
\end{figure}

As for the phase gate, the  process to be performed is to apply a nontrivial phase to the $|1\rangle$, leaving the $|0\rangle$ unchanged. Fig. (\ref{phase_gate}) shows the detailed graph structure of this gate. The rationale behind  Fig. (\ref{phase_gate})  is as follows:

- For each four-edge vertex, apply to to $|0\rangle_a$, $|0\rangle_b$, $|1\rangle_a$, and $|1\rangle_b$ a $4-d$ Grover diffusion operator, a relative phase gate and the shift operator, i.e. apply the full operator  $S(e^{-i\pi/4}(G^{(4)}))$. $G^{(4)}$ is given in matrix representation in Eq. (\ref{grover_4d}) and we propose the following definitions for the relative phase gate $e^{-i\pi/4}$ (Eq. (\ref{proposed_phase_factor_matrix})) and the shift operator $S$ (Eq. (\ref{proposed_shift_matrix})):

\begin{equation}
PF_{-\pi/4} = \begin{pmatrix} 1 & 0 & 0 & 0\\0 & 1 & 0 & 0\\0 & 0 & e^{{-i\pi}/4} & 0\\0 & 0 & 0 & e^{{-i\pi}/4}\\ \end{pmatrix}
\label{proposed_phase_factor_matrix}
\end{equation}

\begin{equation}
S = \frac{1}{2} \begin{pmatrix} {-1} & 1 & 1 & 1\\1 & {-1} & 1 & 1\\1 & 1 & {-1} & 1\\1 & 1 & 1 & {-1}\\ \end{pmatrix}.
\label{proposed_shift_matrix}
\end{equation}

That is, $S = G^{(4)}$. Now, the diamond-shaped graph that is located in the middle of  $|1\rangle_{a,b}$ applies a shift operation to the quantum information that is propagated along that wire {\it without applying a relative phase gate}. Consequently, at step $t_6$ of  Fig. (\ref{phase_gate}), the quantum information running on $|1\rangle_{a,b}$ has a different phase from the one found on the quantum information running on $|0\rangle_{a,b}$.

Let us now, for each time step $t_i$, take a look at quantum operations and corresponding calculations. 
\\\\
-- {\large{\bf Time step $t_1$}}. 
\\\\
For $|0\rangle_{a,b}$
\begin{equation}
|\Psi\rangle_{t_1} = \begin{pmatrix} \alpha \\ \alpha \\ 0 \\0 \\ \end{pmatrix}
\label{phase_gate_t1_zero} 
\end{equation}

For $|1\rangle_{a,b}$ 
\begin{equation}
|\Phi\rangle_{t_1} = \begin{pmatrix} \beta \\ \beta \\ 0 \\0 \\ \end{pmatrix}
\label{phase_gate_t1_one} 
\end{equation}

-- {\large{\bf Time step $t_2$}}. 
\\\\
For $|0\rangle_{a,b}$: $|\Psi\rangle_{t_2} = PF_{-\pi/4}(G^{(4)}|\Psi\rangle_{t_1})$, i.e. 

\begin{equation}
|\Psi\rangle_{t_2} = \frac{1}{2} \begin{pmatrix} 1 & 0 & 0 & 0\\0 & 1 & 0 & 0\\0 & 0 & e^{{-i\pi}/4} & 0\\0 & 0 & 0 & e^{{-i\pi}/4} \end{pmatrix}
\begin{pmatrix} {-1} & 1 & 1 & 1\\1 & {-1} & 1 & 1\\1 & 1 & {-1} & 1\\1 & 1 & 1 & {-1}\\ \end{pmatrix} 
\begin{pmatrix} \alpha \\ \alpha \\ 0 \\0 \\ \end{pmatrix} = \begin{pmatrix} 0 \\ 0 \\ e^{{-i\pi}/4}\alpha \\ e^{{-i\pi}/4}\alpha \\ \end{pmatrix}
\label{phase_gate_t2_zero}
\end{equation}

For $|1\rangle_{a,b}$ the rationale is identical:
\begin{equation}
|\Phi\rangle_{t_2} = \frac{1}{2} \begin{pmatrix} 1 & 0 & 0 & 0\\0 & 1 & 0 & 0\\0 & 0 & e^{{-i\pi}/4} & 0\\0 & 0 & 0 & e^{{-i\pi}/4} \end{pmatrix}
\begin{pmatrix} {-1} & 1 & 1 & 1\\1 & {-1} & 1 & 1\\1 & 1 & {-1} & 1\\1 & 1 & 1 & {-1}\\ \end{pmatrix} 
\begin{pmatrix} \beta \\ \beta \\ 0 \\0 \\ \end{pmatrix} = \begin{pmatrix} 0 \\ 0 \\ e^{{-i\pi}/4}\beta \\ e^{{-i\pi}/4}\beta \\ \end{pmatrix}
\label{phase_gate_t2_one} 
\end{equation}

-- {\large{\bf Time step $t_3$}}. 
\\\\
For $|0\rangle_{a,b}$: $|\Psi\rangle_{t_3} = S|\Psi\rangle_{t_2}$, i.e. 

\begin{equation}
|\Psi\rangle_{t_3} = \frac{1}{2} \begin{pmatrix} {-1} & 1 & 1 & 1\\1 & {-1} & 1 & 1\\1 & 1 & {-1} & 1\\1 & 1 & 1 & {-1}\\ \end{pmatrix} 
\begin{pmatrix} 0 \\ 0 \\ e^{{-i\pi}/4}\alpha \\ e^{{-i\pi}/4}\alpha \\ \end{pmatrix} = \begin{pmatrix} e^{{-i\pi}/4}\alpha \\ e^{{-i\pi}/4}\alpha \\ 0 \\ 0 \\ \end{pmatrix}
\label{phase_gate_t3_zero}
\end{equation}

For $|1\rangle_{a,b}$ the rationale is identical:
\begin{equation}
|\Phi\rangle_{t_3} = \frac{1}{2} \begin{pmatrix} {-1} & 1 & 1 & 1\\1 & {-1} & 1 & 1\\1 & 1 & {-1} & 1\\1 & 1 & 1 & {-1}\\ \end{pmatrix} 
\begin{pmatrix} 0 \\ 0 \\ e^{{-i\pi}/4}\beta \\ e^{{-i\pi}/4}\beta \\ \end{pmatrix} = \begin{pmatrix} e^{{-i\pi}/4}\beta \\ e^{{-i\pi}/4}\beta \\ 0 \\ 0 \\ \end{pmatrix}
\label{phase_gate_t3_one} 
\end{equation}

-- {\large{\bf Time step $t_4$}}. 
\\\\
For $|0\rangle_{a,b}$: $|\Psi\rangle_{t_4} = PF_{-\pi/4}(G^{(4)}|\Psi\rangle_{t_3})$, i.e. 

\begin{equation}
|\Psi\rangle_{t_4} = \frac{1}{2} \begin{pmatrix} 1 & 0 & 0 & 0\\0 & 1 & 0 & 0\\0 & 0 & e^{{-i\pi}/4} & 0\\0 & 0 & 0 & e^{{-i\pi}/4} \end{pmatrix}
\begin{pmatrix} {-1} & 1 & 1 & 1\\1 & {-1} & 1 & 1\\1 & 1 & {-1} & 1\\1 & 1 & 1 & {-1}\\ \end{pmatrix} 
\begin{pmatrix} e^{{-i\pi}/4}\alpha \\ e^{{-i\pi}/4}\alpha \\ 0 \\0 \\ \end{pmatrix} = \begin{pmatrix} 0 \\ 0 \\ e^{{-2i\pi}/4}\alpha \\ e^{{-2i\pi}/4}\alpha \\ \end{pmatrix}
\label{phase_gate_t4_zero}
\end{equation}

For $|1\rangle_{a,b}$ the rationale is identical:
\begin{equation}
|\Phi\rangle_{t_4} = \frac{1}{2} \begin{pmatrix} 1 & 0 & 0 & 0\\0 & 1 & 0 & 0\\0 & 0 & e^{{-i\pi}/4} & 0\\0 & 0 & 0 & e^{{-i\pi}/4} \end{pmatrix}
\begin{pmatrix} {-1} & 1 & 1 & 1\\1 & {-1} & 1 & 1\\1 & 1 & {-1} & 1\\1 & 1 & 1 & {-1}\\ \end{pmatrix} 
\begin{pmatrix} e^{{-i\pi}/4}\beta \\ e^{{-i\pi}/4}\beta \\ 0 \\0 \\ \end{pmatrix} = \begin{pmatrix} 0 \\ 0 \\ e^{{-2i\pi}/4}\beta \\ e^{{-2i\pi}/4}\beta \\ \end{pmatrix}
\label{phase_gate_t4_one} 
\end{equation}

-- {\large{\bf Time step $t_5$}}. 
\\\\
For $|0\rangle_{a,b}$: $|\Psi\rangle_{t_5} = S|\Psi\rangle_{t_4}$, i.e. 

\begin{equation}
|\Psi\rangle_{t_5} = \frac{1}{2} \begin{pmatrix} {-1} & 1 & 1 & 1\\1 & {-1} & 1 & 1\\1 & 1 & {-1} & 1\\1 & 1 & 1 & {-1}\\ \end{pmatrix} 
\begin{pmatrix} 0 \\ 0 \\ e^{{-2i\pi}/4}\alpha \\ e^{{-2i\pi}/4}\alpha \\ \end{pmatrix} = \begin{pmatrix} e^{{-2i\pi}/4}\alpha \\ e^{{-2i\pi}/4}\alpha \\ 0 \\ 0 \\ \end{pmatrix}
\label{phase_gate_t5_zero}
\end{equation}

For $|1\rangle_{a,b}$ the rationale is identical:
\begin{equation}
|\Phi\rangle_{t_5} = \frac{1}{2} \begin{pmatrix} {-1} & 1 & 1 & 1\\1 & {-1} & 1 & 1\\1 & 1 & {-1} & 1\\1 & 1 & 1 & {-1}\\ \end{pmatrix} 
\begin{pmatrix} 0 \\ 0 \\ e^{{-2i\pi}/4}\beta \\ e^{{-2i\pi}/4}\beta \\ \end{pmatrix} = \begin{pmatrix} e^{{-2i\pi}/4}\beta \\ e^{{-2i\pi}/4}\beta \\ 0 \\ 0 \\ \end{pmatrix}
\label{phase_gate_t5_one} 
\end{equation}

-- {\large{\bf Time step $t_6$}}. 
\\\\
Here we have a most important result. For $|0\rangle_{a,b}$: $|\Psi\rangle_{t_6} = PF_{-\pi/4}(G^{(4)}|\Psi\rangle_{t_5})$, i.e. 

\begin{equation}
|\Psi\rangle_{t_6} = \frac{1}{2} \begin{pmatrix} 1 & 0 & 0 & 0\\0 & 1 & 0 & 0\\0 & 0 & e^{{-i\pi}/4} & 0\\0 & 0 & 0 & e^{{-i\pi}/4} \end{pmatrix}
\begin{pmatrix} {-1} & 1 & 1 & 1\\1 & {-1} & 1 & 1\\1 & 1 & {-1} & 1\\1 & 1 & 1 & {-1}\\ \end{pmatrix} 
\begin{pmatrix} e^{{-2i\pi}/4}\alpha \\ e^{{-2i\pi}/4}\alpha \\ 0 \\0 \\ \end{pmatrix} = \begin{pmatrix} 0 \\ 0 \\ e^{{-3i\pi}/4}\alpha \\ e^{{-3i\pi}/4}\alpha \\ \end{pmatrix}
\label{phase_gate_t6_zero}
\end{equation}

However, for $|1\rangle_{a,b}$, we only apply the  coin operator $G^{(2)}=\begin{pmatrix} 0 & 1 \\1 & 0\\ \end{pmatrix}$, suitable for propagating quantum information through the two-edge vertices $W_1$ and $W_2$ {\it without applying an additional relative phase operator}: 

\begin{subequations}
\begin{equation}\label{phase_gate_t6_one_w1}
|\Phi\rangle^{W_1}_{t_6} = \begin{pmatrix} 0 & 1\\1 & 0\\ \end{pmatrix} \begin{pmatrix} e^{{-2i\pi}/4}\beta \\ 0 \\ \end{pmatrix} = \begin{pmatrix} 0 \\ e^{{-2i\pi}/4}\beta \\ \end{pmatrix}
\end{equation}
\begin{equation}\label{phase_gate_t6_one_w2}
|\Phi\rangle^{W_2}_{t_6} = \begin{pmatrix} 0 & 1\\1 & 0\\ \end{pmatrix} \begin{pmatrix} e^{{-2i\pi}/4}\beta \\ 0 \\ \end{pmatrix} = \begin{pmatrix} 0 \\ e^{{-2i\pi}/4}\beta \\ \end{pmatrix}
\end{equation}
\end{subequations}

Thus, the state of this computation at time $t_6$ is given by

\begin{subequations}
\begin{equation}
|\Psi\rangle_{t_6} = \begin{pmatrix} 0 \\ 0 \\ e^{{-3i\pi}/4}\alpha \\ e^{{-3i\pi}/4}\alpha \\ \end{pmatrix}
\label{total_state_zero}
\end{equation}
\begin{equation}
|\Phi\rangle{t_6} =  \begin{pmatrix} 0 \\ 0 \\ e^{{-2i\pi}/4}\beta \\ e^{{-2i\pi}/4}\beta \\ \end{pmatrix}
\label{total_state_one}
\end{equation}
\end{subequations}

Direct calculations would produce the following states:

-- {\large{\bf Time step $t_7$}}. 
\begin{subequations}
\begin{equation}
|\Psi\rangle_{t_7} = \begin{pmatrix} e^{{-3i\pi}/4}\alpha \\ e^{{-3i\pi}/4}\alpha \\ 0 \\ 0 \\ \end{pmatrix}
\label{phase_gate_t7_zero}
\end{equation}
\begin{equation}
|\Phi\rangle{t_7} =  \begin{pmatrix} e^{{-2i\pi}/4}\beta \\ e^{{-2i\pi}/4}\beta \\ 0 \\ 0 \\ \end{pmatrix}
\label{phase_gate_t7_one}
\end{equation}
\end{subequations}

-- {\large{\bf Time step $t_8$}}. 
\begin{subequations}
\begin{equation}
|\Psi\rangle_{t_8} = \begin{pmatrix} 0 \\ 0 \\ e^{{-4i\pi}/4}\alpha \\ e^{{-4i\pi}/4}\alpha \\ \end{pmatrix}
\label{phase_gate_t8_zero}
\end{equation}
\begin{equation}
|\Phi\rangle{t_8} =  \begin{pmatrix} 0 \\ 0 \\ e^{{-3i\pi}/4}\beta \\ e^{{-3i\pi}/4}\beta \\ \end{pmatrix}
\label{phase_gate_t8_one}
\end{equation}
\end{subequations}

-- {\large{\bf Time step $t_9$}}. 
\begin{subequations}
\begin{equation}
|\Psi\rangle_{t_9} = \begin{pmatrix} e^{{-4i\pi}/4}\alpha \\ e^{{-4i\pi}/4}\alpha \\ 0 \\ 0 \\ \end{pmatrix}
\label{phase_gate_t9_zero}
\end{equation}
\begin{equation}
|\Phi\rangle{t_9} =  \begin{pmatrix} e^{{-3i\pi}/4}\beta \\ e^{{-3i\pi}/4}\beta \\ 0 \\ 0 \\ \end{pmatrix}
\label{phase_gate_t9_one}
\end{equation}
\end{subequations}

-- {\large{\bf Time step $t_{10}$}}. 
\begin{subequations}
\begin{equation}
|\Psi\rangle_{t_{10}} = \begin{pmatrix} 0 \\ 0 \\ e^{{-5i\pi}/4}\alpha \\ e^{{-5i\pi}/4}\alpha \\ \end{pmatrix}
\label{phase_gate_t10_zero}
\end{equation}
\begin{equation}
|\Phi\rangle{t_{10}} =  \begin{pmatrix} 0 \\ 0 \\ e^{{-4i\pi}/4}\beta \\ e^{{-4i\pi}/4}\beta \\ \end{pmatrix}
\label{phase_gate_t10_one}
\end{equation}
\end{subequations}

-- {\large{\bf Time step $t_{11}$}}. 
\begin{subequations}
\begin{equation}
|\Psi\rangle_{t_{11}} = \begin{pmatrix} e^{{-5i\pi}/4}\alpha \\ e^{{-5i\pi}/4}\alpha \\ 0 \\ 0 \\ \end{pmatrix}
\label{phase_gate_t11_zero}
\end{equation}
\begin{equation}
|\Phi\rangle{t_{11}} =  \begin{pmatrix} e^{{-4i\pi}/4}\beta \\ e^{{-4i\pi}/4}\beta \\ 0 \\ 0 \\ \end{pmatrix}
\label{phase_gate_t11_one}
\end{equation}
\end{subequations}

So, at time $t_{11}$, the $|0\rangle$ wire has a phase equal to $e^{{-5i\pi}/4}$ while the $|1\rangle$ wire has a phase equal to $e^{{-4i\pi}/4}$, i.e. the $|1\rangle$ wire has a relative phase of $e^{{i\pi}/4}$
with respect to the $|0\rangle$ wire.

Finally, let us find out how to construct the Hadamard gate according to \cite{lovett10}. Please note that the graph structure proposed in \cite{lovett10} for the Hadamard gate (Fig. (\ref{graphs_for_quantum_gates_lovett}.c)) is divided into three parts:

\begin{itemize}

\item
As in the previous gates, the Hadamard gate (Fig. (\ref{graphs_for_quantum_gates_lovett}.c)) has as input states 

For $|0\rangle_{a,b}$
\begin{equation}
|\Psi\rangle_{t_1} = \begin{pmatrix} \alpha \\ \alpha \\ 0 \\0 \\ \end{pmatrix}
\label{phase_gate_t1_zero} 
\end{equation}

For $|1\rangle_{a,b}$ 
\begin{equation}
|\Phi\rangle_{t_1} = \begin{pmatrix} \beta \\ \beta \\ 0 \\0 \\ \end{pmatrix}
\label{phase_gate_t1_one} 
\end{equation}

\item

Part (a) of (Fig. (\ref{graphs_for_quantum_gates_lovett}.c)) adds a total phase of $e^{{-9i\pi}/4}$ to the $|0\rangle$ wire and a phase of $e^{{-7i\pi}/4}$ to the $|1\rangle$. We can see that from the number of $d=4$ nodes that the quantum walks is propagated through from the beginning to the very entrance of $G^{8}$: nine nodes for $|0\rangle$ and seven nodes for $|1\rangle$. Thus, states for part (a) of (Fig. (\ref{graphs_for_quantum_gates_lovett}.c)) are:

\begin{equation}
|\Psi\rangle_{t_A} = \begin{pmatrix} e^{{-9i\pi}/4}\alpha \\ e^{{-9i\pi}/4}\alpha \\ 0 \\0 \\ \end{pmatrix}
\label{hadamard_a_psi} 
\end{equation}

\begin{equation}
|\Phi\rangle_{t_A} = \begin{pmatrix} e^{{-7i\pi}/4}\beta \\ e^{{-7i\pi}/4}\beta \\ 0 \\0 \\ \end{pmatrix}
\label{hadamard_a_phi} 
\end{equation}

The same rationale applies to the phase applied to $|0\rangle$ and $|1\rangle$ wires on part (c) of (Fig. (\ref{graphs_for_quantum_gates_lovett}.c)). Thus, the total phase added to the $|0\rangle$ wire is $e^{{-18i\pi}/4}$ and to  the $|1\rangle$ wire is $e^{{-14i\pi}/4}$, i.e. there is a relative  phase of $e^{{-4i\pi}/4}=e^{{-i\pi}}=\cos\pi - i\sin\pi=-1$ on $|1\rangle$.

Of course, $|\Psi\rangle_{t_A}$ and $|\Phi\rangle_{t_A}$ are also the input states of Part B.

\item
According to \cite{lovett10}, Part B of (Fig. (\ref{graphs_for_quantum_gates_lovett}.c)) is composed of a $d=8$ graph that has two effects on Eqs. (\ref{hadamard_a_psi},\ref{hadamard_a_phi}): to combine the two inputs from $|0\rangle$ and $|1\rangle$ wires as well as to add a global phase of $3\pi/4$ to both wires. Applying Euler's identity as before we can see that $e^{{-3i\pi}/4}=\cos({-3i\pi}/4) + i\sin({-3i\pi}/4)=-1/\sqrt{2}-i/\sqrt{2}$, hence the factor $1/\sqrt{2}$ needed for the Hadamard operator (the number $-1-i$ is a global phase that would be experimentally irrelevant.)

\end{itemize}

Lovett {\it et al} finish by explaining how to build quantum circuits using the graphs and methods exposed in \cite{lovett10}, which is very similar to the method proposed in \cite{childs09}:  for the actual implementation of a general quantum gate as well as a discrete-time quantum-walk algorithm, we would only need to connect corresponding graphs using basis-state quantum wires.
\\\\
c) {\bf Universal computation by discontinuous quantum walk \cite{underwood10}}
\\

\begin{figure}
\begin{center}
\scalebox{0.5}{\includegraphics{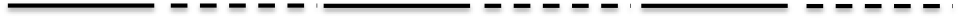}}
\end{center}
\caption{{\small Alternating wires (solid and dashed) on which the quantum walk propagates via perfect state transfer. Solid and dashed lines are turned on and off alternatively.}}
\label{underwood_lines}
\end{figure}

Based on an eclectic analysis of \cite{childs09} and \cite{lovett10}, Underwood and Feder \cite{underwood10} have proposed a hybrid quantum walk for realizing universal computation, consisting of propagating a quantum walker via perfect state transfer under continuous evolution. The quantum walk propagates on a line (quantum wire) which is actually composed of two alternating lines (Fig. (\ref{underwood_lines}).) The walker begins walking on the solid line of the graph LHS long enough to perfectly transfer to the end of the first solid line segment. Then, the solid line is turned off and, simultaneously, the dashed line is turned on, enabling then the walker to transfer to the end of the first dashed line segment. As in \cite{childs09,lovett10}, Underwood and Feder \cite{underwood10} have proposed a universal gate set (phase, identity and rotation graphs) as well as a method for building general unitary quantum gates and quantum circuits as a combination of basis state quantum wires and phase, identity and rotation graphs.

\paragraph{}
\cite{childs09,lovett10,underwood10}, together with the computational  equivalence proofs of several other models of quantum computations, provide a rich \lq toolbox'  for computer scientists interested in quantum computation, for they will be free to choose  from several models of quantum computation those that particularly suit their academic background  and interests.

\section{Conclusions}

In this paper we have reviewed theoretical advances on the foundations of both discrete- and continuous-time quantum walks, together with  the role that randomness plays in quantum walks, the connections between the mathematical models of coined discrete quantum walks and continuous quantum walks, the quantumness of quantum walks and a brief summary of papers published on discrete quantum walks and entanglement as well as a succinct review of experimental proposals and realizations of discrete-time quantum walks. Moreover, we have reviewed several algorithms based on quantum walks as well as a most important result: the computational universality of both continuous- and discrete-time quantum walks. 

Fortunately, quantum walks is now a solid field of research of quantum computation full of exciting open problems for physicists, computer scientists and engineers.  This review, which is meant to be situated as a contribution within the field of quantum walks from the perspective of a computer scientist, will best serve the scientific community if it encourages quantum scientists and quantum engineers to further advance on this discipline.

\section*{Acknowledgments}
I start by gratefully thanking my family for unconditionally supporting me during the holidays I spent working on this manuscript. I am also indebted to Professor Y. Shikano for his kind invitation, patience and support. Additionally, I acknowledge the financial support of ITESM-CEM, CONACyT (SNI member number 41594), and Texia. I thank Professor F.A. Gr{\"u}nbaum,  Professor A. Joye, Professor C. Liu, Professor M.A. Martin-Delgado, Professor A. P\'erez, Professor C. A. Rodr\'iguez-Rosario, Professor E. Rold\'an, Professor S. Salimi, Professor Y. Shikano, and the anonymous reviewers of this paper for their criticisms and useful comments. Finally, I thank Dr A. Aceves-Gaona for his kind help on artwork.

\end{document}